\documentclass[aps,prl,amsmath,amssymb,letterpaper,showpacs,reprint,floatfix]{revtex4-1}
\usepackage{placeins,color}
\usepackage[caption=false]{subfig}
\usepackage[allcolors=blue,colorlinks=true,breaklinks=true,bookmarks=false]{hyperref}
\usepackage{graphicx}

\newcommand{\ifw}[0]{\affiliation{Leibniz Institute for Solid State and Materials Research (IFW) Dresden, Dresden D-01069, Germany}}
\newcommand{\ifp}[0]{\affiliation{Institute for Solid State Physics, Dresden University of Technology, D-01069 Dresden, Germany}}
\newcommand{\lmu}[0]{\affiliation{Laboratory for Muon-Spin Spectroscopy, Paul Scherrer Institute, CH-5232 Villigen PSI, Switzerland}}
\newcommand{\kip}[0]{\affiliation{Kirchhoff Institute for Physics, University of Heidelberg, Heidelberg D-69120, Germany}}
\newcommand{\bessy}[0]{\affiliation{Helmholtz-Zentrum Berlin GmbH f\"ur Materialien und Energie, BESSY, D-12489 Berlin, Germany}}

\begin{document}

\title{Structural and electronic phase diagrams of CeFeAsO$_{1-x}$F$_x$ and SmFeAsO$_{1-x}$F$_x$}

\author{Hemke Maeter} \email{h.maeter@physik.tu-dresden.de} \ifp{}
\author{Til Goltz} \ifp{}
\author{Johannes Spehling} \ifp{}
\author{Andrej Kwadrin} \ifp{}
\author{Hans-Henning Klauss} \ifp{}

\author{Jorge Enrique Hamann Borrero} \ifw{}
\author{Agnieszka Kondrat} \ifw{}
\author{Louis Veyrat} \ifw{}
\author{Guillaume Lang} \ifw{}
\author{Hans-Joachim Grafe} \ifw{}
\author{Christian Hess} \ifw{}
\author{G\"unter Behr} \ifw{}
\author{Bernd B\"uchner} \ifw{}

\author{Hubertus Luetkens} \lmu{}
\author{Chris Baines} \lmu{}
\author{Alex Amato} \lmu{}

\author{Norman Leps} \kip{}
\author{R\"udiger Klingeler} \kip{}

\author{Ralf Feyerherm} \bessy{}
\author{Dimitri Argyriou} \bessy{}

\date{\today}

\begin{abstract}
We have studied the structural and electronic phase diagrams of CeFeAsO$_{1-x}$F$_x$ and SmFeAsO$_{1-x}$F$_x$ by a detailed analysis of muon spin relaxation experiments, synchrotron X-ray diffraction, M\"ossbauer spectroscopy, electrical resistivity, specific heat, and magnetic susceptibility measurements.
In these systems structural and magnetic degrees of freedom are strongly coupled and magnetism competes with superconductivity for the ground state.
The \textit{bulk} long range magnetic phase transition is always preceded by a tetragonal to orthorhombic structural phase transition that are both suppressed on doping.
However, in the Sm system we find small volumes that show long range magnetic order already $\approx$10~K above the structural transition temperature.
In both systems the structural transition is absent in superconducting materials.
We find a mixed phase of superconductivity and short range magnetic order for one composition in each system.
In view of available literature we conclude that it remains unclear whether this is electronic phase separation on a nanometer length scale or microscopic coexistence.
Our muon spin relaxation experiments reveal short range magnetic order in a broad temperature range for all magnetic materials.
For the Ce system, short range order develops into long range order close to a percolation of magnetically ordered clusters.
Whereas for the Sm system, the magnetically ordered clusters already show coherent muon spin precession, an indication for long range order. In both systems also the structural transition is broad, i.e.,
it is preceded by an increase of the Bragg peak width over a broad temperature range.
We quantitatively analyze the broadening of the magnetic transition.
In the superconducting state, the temperature dependence of the magnetic penetration depth is,
in all studied cases, compatible with a single nodeless s-wave gap.
We compare the magnetic and structural transition temperatures determined from direct ($\mu$SR and synchrotron X-ray diffraction) and indirect (electrical resistivity) measurements.
\end{abstract}



\maketitle

\section{Introduction}
Two intensely studied phenomena in correlated electron systems are superconductivity (SC) and quantum criticality (QC). The latter arises from the suppression of a continuous phase transition to zero temperature at a quantum critical point (QCP).
In its vicinity critical fluctuations play an important role for the electronic properties of solids.
Superconductivity is often found in the vicinity of a magnetic quantum critical point, i.e., near the suppression of a magnetic phase, e.g., by doping or pressure. 
It is widely believed, that magnetic fluctuations near a magnetic QCP may play a crucial role in the pairing mechanism of Cooper pairs in correlated electron systems.
Both in heavy fermion superconductors and the copper oxide high-$T_c$ superconductors this interplay is studied. 
CePd$_2$Si$_2$\cite{mathur98} is an example for a heavy-Fermion system that is antiferromagnetic at ambient pressure, and becomes superconducting under pressure near the QCP of the antiferromagnetic order.
In the cuprate superconductor La$_{2-x}$Sr$_x$CuO$_{4-y}$ quantum criticality arises from the suppression of the antiferromagnetic order by hole doping via substitution of La by Sr (for a review see Ref.~\onlinecite{pines05} and references therein). 
The emergence of low energy spin fluctuations are believed to be crucial not only for the pairing mechanism of superconductivity but also for the electronic properties over a wide range of its phase diagram\cite{cuprates1}. 
This general behavior under hole doping is found in other cuprate materials as well.

After the discovery of the ferropnictide superconductor LaFeAsO$_{1-x}$F$_x$ a plethora of different phase diagrams were investigated. 
Among the first were the F doped group of rare earth \textit{R}=La, Ce, and Sm containing \textit{R}FeAsO$_{1-x}$F$_x$ \cite{zhao08,drew09,luetkens09}. 
The undoped systems are semi metals and show a tetragonal to orthorhombic structural phase transition on lowering the temperature followed by a transition to a spin density wave phase (SDW). 
Upon doping both phases are suppressed and superconductivity emerges with a maximal $T_c$$\approx$56~K. 

The phase transition from magnetic order to superconductivity as a function of doping for \textit{R}=La is abrupt, with only a small decrease of the magnetic ordering temperature before superconductivity emerges and magnetic order is fully destroyed\cite{luetkens09}.
The available experimental reports for $R$=Ce, Sm are contradictory concerning the nature of the phase transition from magnetic order to superconductivity as a function of doping.
The phase diagram of the Ce system reported by \citet{zhao08} allows both, a magnetic and a possibly structural QCP, whereas \citet{sanna10} and \citet{shiroka11} claim ``nanoscopic'' coexistence of magnetic order and superconductivity (phase separation with nanometer sized domains) which indicates the absence of a magnetic QCP (due to a first order phase transition as a function of doping).
The latter situation agrees with the claim of (nanoscopic) phase separation between magnetism and superconductivity in the Sm system\cite{sanna09}. 
Alternatively the $\mu$SR data presented by Sanna, and Shiroka and co-workers\cite{sanna09,sanna10,shiroka11} may be explained by microscopic coexistence of magnetic order and superconductivity, i.e., they coexist (spatially) homogeneous and form a single thermodynamic phase that is both superconducting and magnetically ordered.
This implies that both order parameters are coupled, or, in different words, the same electrons contribute to superconductivity and magnetic order\cite{fernandes10}. 

The structural phase transition for $R$=Sm is discussed controversially:
\citet{margadonna} reported a structural phase transition with a strongly suppressed phase transition temperature $T_S$ in underdoped superconducting SmFeAsO$_{1-x}$F$_x$, whereas \citet{martinelli11} find that $T_S$ is almost doping independent even beyond the optimally doped regime. 
In the present work we want to clarify these contradicting results with new experimental data and a comparison with available literature. 
We will present a thorough analysis of our experimental data that reveal, new details of the phase diagrams of the $R$FeAsO$_{1-x}$F$_x$ ($R$=Sm, Ce) superconductors.

The intricate interplay between structure, magnetism and superconductivity requires the combination of several experimental techniques. 
No single experiment exists that can detect all relevant aspects. 
It is of utmost importance to realize the limitations and advantages of each experimental technique. 
For example, diffraction experiments with neutrons or X-rays are only sensitive to order with long correlation lengths. 
This is an essential limitation. 
For example, when discussing microscopic coexistence and phase separation of SC and small moment, or short range magnetic order. 
Local probes such as M\"ossbauer spectroscopy and muon spin relaxation ($\mu$SR) on the other hand, are very sensitive to disordered magnetism.
$\mu$SR is also an excellent tool to determine the absolute values of the magnetic penetration depth of superconductors.

Here we report the results of our synchrotron X-ray diffraction (XRD), muon spin relaxation and rotation ($\mu$SR), M\"ossbauer spectroscopy, and electrical resistivity experiments of CeFeAsO$_{1-x}$F$_x$ for $x$=0, 0.042(2), 0.048(3), 0.063(2), 0.145(20), 0.150(20), and SmFeAsO$_{1-x}$F$_x$ with $x$=0, 0.02, 0.04, 0.06, 0.08, 0.10.
We also present specific heat measurements of the undoped Sm system and of $x$=0.06.
The results we obtain for the Ce system clarify the above mentioned features of the phase diagram.
We unambiguously show that magnetic order and the tetragonal to orthorhombic structural transition are suppressed with doping but only short range magnetic order (SRO) is found together with superconductivity for $x$=0.063(2) in the Ce, and $x$=0.06 in the Sm system, whereas the structural transition disappears as superconductivity emerges with doping.

Our experiments confirm a mixed phase of superconductivity and magnetic order in both systems but the indications for phase separation are not sufficient to rule out a magnetic quantum critical point that possibly emerges as a function of F doping.
We determined the structural phase transition temperature from a detailed analysis of the lattice distortion of the orthorhombic phase.
In several publications\cite{zhao08,martinelli11} the broadening of Bragg peaks was used to obtain the structural transition temperature.
Our results show that this procedure can be inaccurate in the systems at hand.
In fact, this kind of analysis should always be supported by additional experimental evidence such as thermal expansion or specific heat to verify the presence of a phase transition.
To study the order parameter of superconductivity we employed transverse field $\mu$SR.
We find that the temperature dependence of the superfluid density $n_s$$\propto$$1/\lambda^2$ is consistent with a nodeless s-wave gap in both systems.
This is in agreement with measurements of the penetration depth $\lambda$ that find an exponential decrease as a function of temperature \cite{hashimoto09,malone09}.
We discuss our results in terms of the effects of random stress, and disorder on the magnetic, and structural phase transition. We also study the doping dependence of the rare earth magnetic order.

\section{Experimental Methods}
\label{sec.exp}
Polycrystalline samples of SmFeAsO$_{1-x}$F$_x$ and CeFeAsO$_{1-x}$F$_x$ were synthesized using the two step solid state reaction route described by \citet{kondrat} and annealed all samples in vacuum.
The crystal structure was verified by powder Mo $K_\alpha$ X-ray diffraction \cite{kamihara,margadonna,nomura,clarina,kondrat} and the FullProf software\cite{fullprof1}.
Electrical resistivity was measured using a standard 4-point geometry with an alternating dc current.
Synchrotron powder X-ray diffraction data was collected ($\lambda$=1.07813 \AA{}) for the (2,2,0)$\rm_T$ and (0,0,6)$\rm_T$ Bragg peaks (subscript T refers to tetragonal), in a temperature range between 7~K and 200~K, at the beam line for Resonant Magnetic Scattering and High-Resolution Diffraction (MAGS) at the 7 T Wiggler at BESSY in Berlin, Germany.
Muon spin relaxation and rotation ($\mu$SR) experiments were conducted at the Paul Scherrer Institute (PSI) using standard $^4$He flow cryostats, a $^3$He-$^4$He dilution cryostat at the GPS, Dolly, and LTF instruments.
M\"ossbauer spectroscopy was conducted using a $^4$He flow cryostat with a $^{57}$Co-in-Rh matrix gamma radiation source kept at room temperature (emission line half-width-at-half-maximum of 0.130(2)~mm/s).
From X-ray diffraction, less than 5~mol\% impurity phases are inferred.
From M\"ossbauer spectroscopy less than 1~atom\% of Fe bearing impurities are inferred for CeFeAsO$_{1-x}$F$_x$ with $x$=0, 0.48(2), and 0.63(3); $x$=0.145(20) contains $\approx$5 weight\% of Fe$_2$As.
\citet{panarina10} studied samples from the same batch of SmFeAsO$_{1-x}$F$_x$ for $x$$\geq$0.06 and reported evidence for bulk superconductivity from a carefully analysis of susceptibility measurements.
We determined the doping levels of the Sm samples using wavelength dispersive X-ray spectroscopy.
To determine the F contents of the Ce samples we used nuclear quadrupole resonance, with the corresponding study of the electronic 
properties to be published separately\cite{lang12}.

\section{S\lowercase{m}F\lowercase{e}A\lowercase{s}O$_{1-x}$F$_x$}
In this section we present the results of our electrical resistivity (Sec.~\ref{sec.sm-rho}), synchrotron X-ray diffraction (Sec.~\ref{sec.sm-xrd}), and muon spin relaxation measurements (Sec.~\ref{sec.sm-mu}).
We also present specific heat measurements in Sec.~\ref{sec.sm-sm}, and \ref{sec.sm-fluct} to support the results of our $\mu$SR measurements.
The structural transition and the Fe magnetic transition are both broadened. $\mu$SR enables us to disentangle the temperature dependence of the magnetic order parameter and the magnetically ordered volume fraction---this is, in general, not possible with diffraction experiments.
In Sec.~\ref{sec.sm-fe-model} we use this knowledge to construct a phenomenological model that correlates all static experimental parameters of the $\mu$SR in a self consistent way.
In Sec.~\ref{sec.sm-sm} we discuss the evolution of the Sm magnetic order as a function of doping and in Sec.~\ref{sec.sm-fluct} the magnetic fluctuations unique to SmFeAsO$_{1-x}$F$_{x}$.
In Sec.~\ref{sec.sm-sum} we summarize our main results and present our phase diagram of SmFeAsO$_{1-x}$F$_{x}$.
Phase transition temperatures are shown in Tab.~\ref{tab.sm}, p.~\pageref{tab.sm}.

\begin{figure}[htb]
\includegraphics[width=8.5cm]{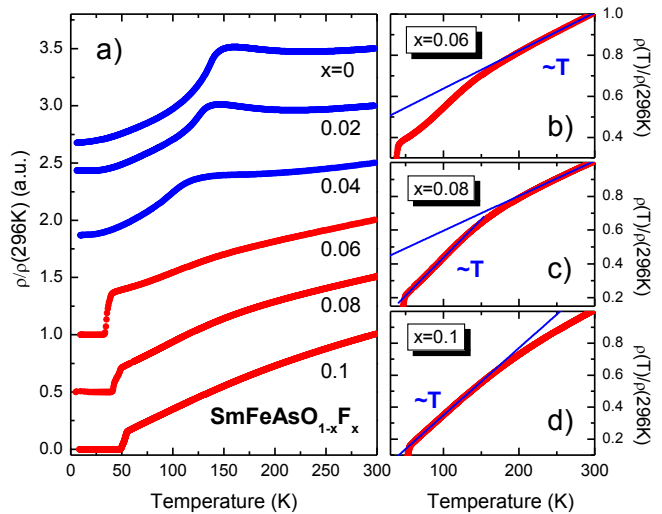}
\caption{Normalized electrical resistivity as a function of temperature of SmFeAsO$_{1-x}$F$_{x}$.
(Blue) Red curves refer to (non) superconducting compounds.
Curves are shifted vertically for clarity.}
\label{img.sm-rho}
\end{figure}

\subsection{Electrical resistivity measurements}
\label{sec.sm-rho}
We discussed the temperature dependence of the electrical resistivity $\rho(T)$ of SmFeAsO$_{1-x}$F$_x$ in Fig.~\ref {img.sm-rho} in detail in a previous work \cite{hess}.
Here we briefly mention the essential features for later comparison with the CeFeAsO$_{1-x}$F$_x$ system.
For $x$=0, and 0.02 we observe a cusp in the resistivity that has been used in the past to determine the structural and magnetic phase transition temperatures at the maximum and the inflection point, respectively \cite{hess,klauss}.
We find the maximum of the resistivity at $T_{max}^{\rho}$=160, 146, and 135~K for $x$=0, 0.02, and 0.04, respectively (see Ref.~\onlinecite{hess} for details).
For the inflection point we find $T_i^\rho$=136~K, 125~K, 97~K for $x$=0, 0.02, and 0.04, respectively.
Upon doping, this anomaly is slightly suppressed to lower temperatures.
It is absent in the \emph{superconducting} compounds ($x$$\geq$0.06).
Instead, $\rho(T)$ is linear for $T$$>$150~K and exhibits a pronounced change of slope at ~150~K (see Fig.~\ref{img.sm-rho}(b)-(d)) which has been attributed as to the spin fluctuations related to proximate
spin density wave state.
We find $T_c$=36.2, 44.5, and 52.1~K for $x$=0.06, 0.08, and 0.10, respectively.

\begin{figure*}[htb]
\begin{minipage}{8.6cm}
\includegraphics[width=8.6cm]{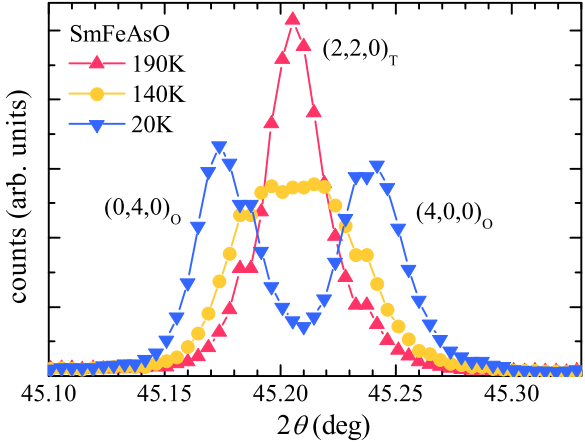}
\end{minipage}
\begin{minipage}{8.6cm}
\includegraphics[width=8.6cm]{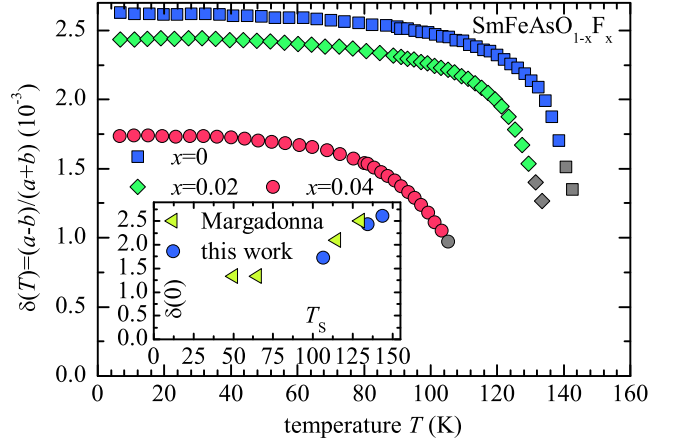}
\end{minipage}
\caption{\textit{Left}: representative X-ray diffracted patterns of the (2,2,0)$_{\rm T}$ tetragonal and the (4,0,0)$\rm_O$ and (0,4,0)$\rm_O$ orthorhombic Bragg peaks at different temperatures for SmFeAsO.
\textit{Right}: orthorhombic distortion $\delta(T)$ as a function of temperature.
Colored symbols indicate the temperature range in which $\delta(T)$ is concave, whereas the gray symbols indicate that $\delta(T)$ is convex.
\textit{Inset}: $\delta(T\to 0)$ as a function of the bulk critical temperature $T_S$ from this work and from \citet{margadonna}}
\label{img.sm-ortho}
\end{figure*}

\begin{figure}[htb]
\includegraphics[width=8.6cm]{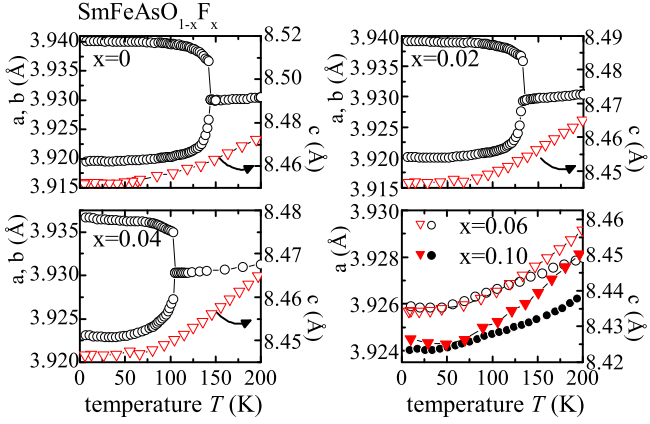}
\caption{Lattice constants as a function of temperature for SmFeAsO$_{1-x}$F$_{x}$.
The black circles refer to the $a$ and $b$ lattice parameters and the red triangles to $c$.
Below the tetragonal to orthorhombic transition, $a$ and $b$ constants are divided by $\sqrt{2}$ for comparison.}
\label{img.sm-lattice}
\end{figure}

\subsection{X-ray diffraction measurements}
\label{sec.sm-xrd}
To study the tetragonal to orthorhombic structural transition, we measured the tetragonal (2,2,0)$\rm_T$ Bragg peak (space group $P4/nmm$) as a function of temperature.
In Fig.~\ref{img.sm-ortho} we show representative diffraction patterns of SmFeAsO.
When we lower the temperature below the structural transition temperature $T_{S}$=143.7($+$1$-$5)~K, 134.5($+$1$-$5)~K, 106.3($+$1$-$5)~K for $x$=0, 0.02, and 0.04, respectively, the (2,2,0)$\rm_T$ peak splits into the (4,0,0)$\rm_O$ and (0,4,0)$\rm_O$ orthorhombic Bragg peaks (space group $Cmma$). 

The order parameter of this transition is the orthorhombic distortion $\delta(T)$ in Fig.~\ref{img.sm-ortho}.
It is the normalized splitting of the Bragg peaks $\delta(T)$=$(a-b)/(a+b)$ where $a$ and $b$ are the lattice constants.
We determined the temperature dependence of the lattice constants shown in Fig.~\ref{img.sm-lattice} from the Bragg peak positions.
The structural transition of $x$=0, 0.02, and 0.04 is indicated by $a(T)$$\neq$$b(T)$ below the transition temperature $T_S$.
We observe no such feature for $x$=0.06, and 0.10 (we did not study $x$=0.08 with synchrotron X-ray diffraction).
The temperature dependence of the specific heat $c_p(T)$ of $x$=0.06 in Fig.~\ref{img.sm-cp}, p.~\pageref{img.sm-cp} also shows no indication of a structural transition.
This indicates that for $x$$\geq$0.06 the structure remains tetragonal down to the lowest temperature $T$$\approx$10~K, that we measured in this study.

\begin{figure}[htb]
\includegraphics[width=8.6cm]{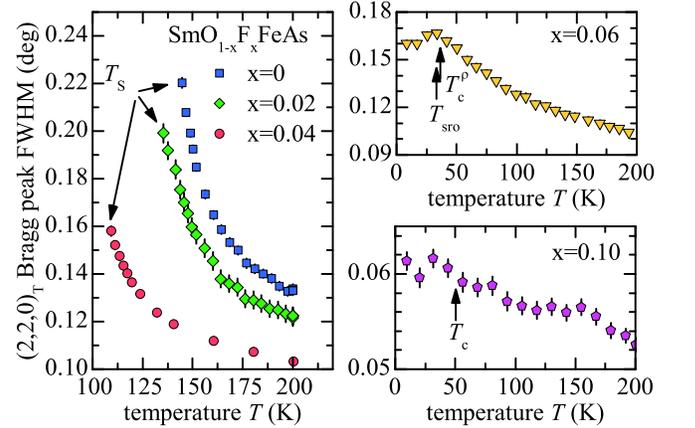}
\caption{FWHM of the (2,2,0)$_{\rm T}$ Bragg peak (left) above the structural transition, and (right) in the superconducting superconducting regime of SmFeAsO$_{1-x}$F$_x$.
Arrows indicate the superconducting transition temperature $T_c$, and the onset of magnetic short range order at $T_{sro}$. Please note the small increase of the FWHM for $x$=0.10 below 200~K that is followed by a plateau between 150~K and $\approx$100~K (see text for a detailed discussion of this anomaly).}
\label{img.sm-fwhm}
\end{figure}


In the following we present a detailed analysis of the structural transition.
When we approach the structural transition from high temperatures the tetragonal (2,2,0)$\rm_T$ Bragg peak broadens (see Fig.~\ref{img.sm-fwhm}) by up to $\approx$70\% before we can observe a clear splitting at $T_{S}$ shown in Fig.~\ref{img.sm-ortho}.
\textit{A priori} it is unknown whether this is due to an unresolved splitting or an intrinsic broadening of the (2,2,0)$\rm_T$ Bragg peak or a combination of both.
When we approach the structural transition from low temperatures the splitting of the orthorhombic (4,0,0)$\rm_O$ and (0,4,0)$\rm_O$ Bragg peaks decreases and becomes comparable to the full-width-at-half-maximum (FWHM) of the Bragg peaks.
In this temperature range the Bragg peak positions and their FWHM become correlated fit parameters.
To lift this correlation we keep the FWHM at the constant value $w_0$.
The value of $\delta(T)$ for $T$$>$$T_S$ close to the transition depends on the choice of $w_0$ wheras at lower temperatures $T$$\leq$$T_S$, $\delta(T)$ is independent of $w_0$.
Here we set $w_0$ to the value that we obtain for $T$$\approx$10~K.
For the discussion of the order parameter we therefore limit the analysis to data for which the splitting is larger than $w_0$.
This coincides with the temperature range for which a description of the data with a single Bragg peak is not sufficient anymore.

The so obtained orthorhombic distortion is shown in Fig.~\ref{img.sm-ortho}.
In a first step we attempt to analyze it with a general power law $\delta(T)$$\propto$$(1-(T/T_S)^\alpha)^\beta$.
Here $\alpha$ controls the temperature dependence of the order parameter at low temperatures and the critical exponent $\beta$ the steepness close to $T_S$.
Below $T_S$ the bulk orthorhombic distortion follows the above power law.
However, this analysis does not yield reliable estimates of the critical exponents.

\begin{figure}[htb]
\includegraphics[width=8.6cm]{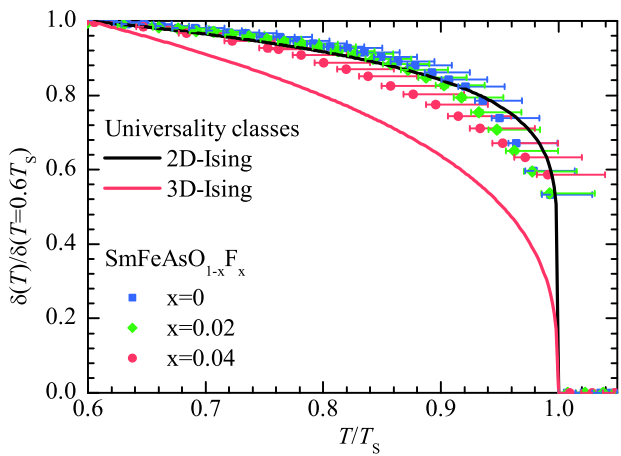}
\caption{The normalized orthorhombic distortion of SmFeAsO$_{1-x}$F$_x$ compared with the temperature dependence of a 2D-Ising, and 3D-Ising type order parameter.
Close to the transition it is not clear whether $\delta(T)$ is the true bulk order parameter.
Therefore the temperature error is large and asymmetric (see text).
Qualitatively all three order parameters are closer to the 2D-Ising than the 3D-Ising type.
However, $\delta(T)$ tends towards the 3D-Ising type but within the error we can make no clear classification.}
\label{img.sm-norm-ortho}
\end{figure}

Nevertheless, we are able to identify trends of the order parameter graphically.
We normalize the order parameter to the value at $T$=0.6$T_S$ and the temperature axis with the estimated transition temperature $T_S$ in Fig.~\ref{img.sm-norm-ortho} and compare it with the 2D-Ising $\delta(T)$$\propto$$(1-T/T_S)^\beta$ with $\beta$=0.125, and the 3D-Ising $\beta$=0.325.
In this analysis we have to take into account the large uncertainty of $T_S$.
For all three doping levels, the order parameter is close to the 2D-Ising universality class.
However, $\delta(T)$ of $x$=0.04 tends towards the 3D-Ising case.
A clear classification is not possible due to the large error of $T_S$.

This result is similar to that obtained by \citet{wilson} for the \textit{magnetic} transition.
They concluded a change of dimensionality from 2D- to 3D-Ising of the \textit{magnetic} phase transition under doping by an analysis of available experimental data.
We find a similar trend for the \textit{structural transition} of the Sm system.
In section~\ref{sec.ce-xrd}, p.~\pageref{sec.ce-xrd}, we show that this trend is more pronounced in the Ce system (due to the slightly higher doping level).

In the inset of Fig.~\ref{img.sm-ortho} we show $\delta(T$$\to$0) as a function of $T_S$ from our, and the work by \citet{margadonna}.
We find, that for low doping levels for which $T_S$$>$100~K, $\delta(T$$\to$0) depends linearly on $T_S$ and a fit yields $\delta(T_S)$=$(0.024(1) \textrm{K}^{-1} T_S-0.8(1))\cdot 10^{-3}$.
The two data points with $T_S$$<$70~K in the inset of Fig.~\ref{img.sm-ortho} that deviate from this linear dependence are from superconducting materials studied by \citet{margadonna} This indicates that in a narrow part of the superconducting regime of SmFeAsO$_{1-x}$F$_x$ the orthorhombic distortion is stabilized while the critical temperature is suppressed by doping.

\begin{figure}[htb]
\includegraphics[width=8.6cm]{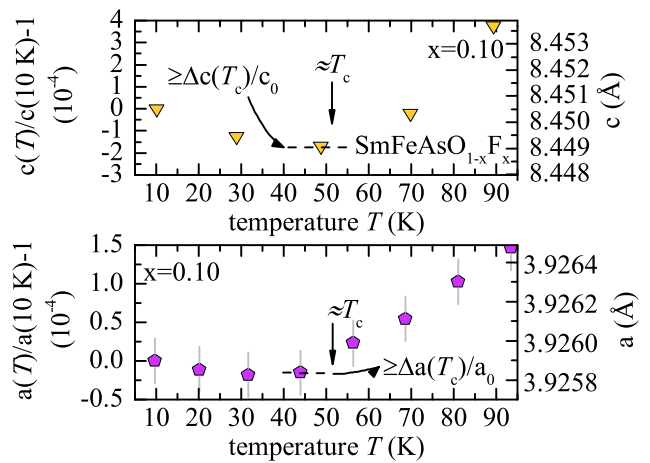}
\caption{The relative change of the lattice constants in the proximity of $T_c$ of SmFeAsO$_{1-x}$F$_x$ with $x$=0.10.
The change of slope from negative ($T$$<$$T_c$) to positive ($T$$>$$T_c$) indicates that $T_c$ decreases under pressure.
The steeper slope of $c(T)$ suggests that $T_c$ is more sensitive to uniaxial pressure along the $c$-axis than to in-plane pressure.
We were not able to resolve the effect of superconductivity for $x$=0.06---close to $T_c$, its lattice parameters are constant within the experimental error (not shown).}
\label{img.sm-a-c}
\end{figure}

For $x$=0.10 we observe superconductivity with $T_c$$=$52.1~K (from electrical resistivity).
The structural transition is absent for this composition.
But we find an upturn of the $a$, and $c$ lattice constants below $T$$\approx$$T_c$.
This upturn is shown in Fig.~\ref{img.sm-a-c}.
The relative changes of the lattice constants at $T$$\approx$$T_c$ with respect to the value at $T$=10~K are $\Delta c(T_c)/c_0$$\leq$$c(T_c)/c($10~K)$-$1=$-$1.7(2)$\cdot$10$^{-4}$, and $\Delta a(T_c)/a_0$$\leq$$a(T_c)/a($10~K)$-$1=$-$0.15(2)$\cdot$10$^{-4}$.

The origin of this anomaly of the temperature dependence of the lattice constants may be either magnetic order or superconductivity.
But for $x$=0.10 we can rule out a magnetic origin because our $\mu$SR measurements show the absence of any magnetic order in this temperature range.
It is more likely that the pressure dependence of $T_c$ governs the lattice contraction.
According to an Ehrenfest relation, the jump $\Delta \alpha_i$ of the thermal expansion coefficient $\alpha_i$ at $T_c$, the derivative of the thermal expansion along a crystallographic direction $i$, is proportional to the pressure dependence $dT_c/dp_i$ of $T_c$ (see e.g.
Ref.~\onlinecite{barron1999heat}).
The contraction of the lattice upon heating towards $T_c$ and subsequent expansion suggests a jump from $\alpha$$<$0 below $T_c$ to $\alpha$$>$0 above $T_c$, $\Delta \alpha$$\propto$$\Delta C_p$$dT_c/dp$ should then be negative.
Since $c(T)$ is steeper than $a(T)$, the jump $\Delta \alpha_c$$\propto$$dT_c/dp_c$ is larger than for the $a$-axis.
This indicates that the pressure dependence of $T_c$ is dominated by uniaxial pressure along the $c$-axis.
This anisotropy of the pressure dependence should be kept in mind when comparing pressure experiments with varying degrees of hydrostaticity and different pressure cell geometries\cite{chu09}.

Qualitatively, this result agrees with the decrease of $T_c$ of nearly optimally doped SmFeAsO$_{1-x}$F$_x$ under high pressure\cite{takahashi09}.
Quantitatively, $c(T_c)/c_0$$\leq$$-$1.7(2)$\cdot$10$^{-4}$, and $\Delta a(T_c)/a_0$$\leq$$-$0.15(2)$\cdot$10$^{-4}$ are consistent with the results of thermal expansion experiments of BaFe$_{2-x}$Co$_x$As$_2$ by \citet{budko09}.

For $x$=0.06, we were not able to resolve any effect on the lattice at or below $T_c$.
The jump of the thermal expansion coefficients $\Delta \alpha$$\propto$$\Delta C_p$$dT_c/dp$ is proportional to the specific heat jump at the transition, which should be proportional to $T_c$\cite{bcs57}.
This suggests a 30\% smaller effect on the lattice constants of $x$=0.06 compared to $x$=0.10 based solely on the different $T_c$.
In addition, absolute value of $dT_c/dp$$\approx$-2 to -6~K/GPa for nearly optimally doped SmFeAsO$_{1-x}$F$_x$ and SmFeAsO$_{1-\delta}$\cite{takabayashi08,yi08,dong10} is reduced by a factor of approximately two to $dT_c/dp$$\approx$+1 to +2.6~K/GPa for underdoped SmFeAsO$_{1-x}$F$_x$\cite{lorenz08,takabayashi08}.
These two factors equate to a reduction of the effect on the lattice constants (the jump $\Delta \alpha$) by $\approx$70$\%$.
Overall, this should make the effect of superconductivity on the lattice constants of $x$=0.06 undetectable in our X-ray diffraction study.

Fig.~\ref{img.sm-fwhm} shows the full-width-at-half-maximum (FWHM) of the (2,2,0)$\rm_T$ peak for $x$=0.06.
When cooling, it increases continuously until $T_c$ and then decreases slightly.
For $x$=0.10, it is reduced by 50-70\% compared to $x$=0.06 and increases continuously without any sharp anomalies.

The FWHM is a measure of the structural disorder and its inverse is related to the correlation length of the structural long range order.
For the sample with $x$=0.06, lattice fluctuations are enhanced upon cooling until we reach $T_c$.
In contrast, structural fluctuations are significantly reduced for $x$=0.10 but also show a similar increase towards lower temperatures, but no anomaly at $T_c$.
The behavior of the FWHM in figure \ref{img.sm-fwhm}, is reminiscent of the evidence of spin fluctuations observed in superconducting LaFeAsO$_{1-x}$F$_x$ by nuclear magnetic resonance \cite{nakai,nakai-njp}.
The importance of magneto-elastic coupling has been discussed also in other theoretical and experimental studies\cite{yildirim09,cano10,nandi10,johnston10}.
This should play an important role in the temperature dependence of the FWHM.
\citet{tropeano} and \citet{zhu-nernst} suggested that magnetic fluctuations observed above $T_c$ are suppressed in favor of superconductivity.

A suppression of spin fluctuations by superconductivity may also cause a decrease of the FWHM below $T_c$.
We observe a decrease of the FWHM below $T_c$ only for $x$=0.06.
However, for $x$$=$0.10, the system may be sufficiently far away from a spin density wave instability and for this mechanism to be relevant

Alternatively, stresses and disorder may similarly broaden the structural transition and cause a temperature dependence of the width of the tetragonal Bragg peak in a wide temperature range\cite{fernandes10-nematic,fernandes11b,cano12,fisher11,ran11}.

%

\citet{margadonna} reported that superconductivity can be found in the orthorhombic phase of SmFeAsO$_{1-x}$F$_x$ in a narrow doping range.
In Fig.~\ref{img.deltats} we show the orthorhombic distortion as a function of the structural transition temperature from this work and from \citet{margadonna}.
It is evident that both datasets are consistent.

\label{para.marti} In contrast, \citet{martinelli11} reported evidence for a tetragonal to orthorhombic structural transition with $T_S$$\approx$170~K in superconducting SmFeAsO$_{1-x}$F$_x$ with $x$=0.10 .
They obtained this result by a microstrain analysis that showed an anisotropic broadening of the (1,0,0)$\rm_T$ Bragg peak.
They concluded an unresolved splitting of the Bragg peak due to the structural transition.
This study was limited to temperatures above 90~K.
The authors suggested that at temperatures lower than 90~K a splitting of the Bragg peak should appear.
We also observed a significant broadening of the tetragonal (2,2,0)$\rm_T$ Bragg peak of the superconducting materials over a wide temperature range, as shown in Fig.~\ref{img.sm-fwhm}.
The Bragg peak FWHM for $x$=0.10 even shows a plateau like feature between 90~K and 150~K with a small drop of the FWHM at higher temperatures.
This resembles the anomaly observed by \citet{martinelli11} that they interpreted as an indication for a structural transition.
But we find no splitting of the Bragg peak down to $\approx$10~K that would indicate a structural transition.
Considering that F doping suppresses $T_S$ to 107~K already for $x$=0.06 it seems unlikely that for $x$=0.10 the transition should occur at 170~K as suggested by \citet{martinelli11}.
Here we point out that in the Fe based superconductors an increase of the FWHM of a Bragg peak alone cannot be taken as sufficient evidence for a structural transition.
This becomes even more evident when comparing the structural transition temperature obtained from the analysis of the FWHM for the non superconducting materials.
With this analysis \citet{martinelli11}
find only a small decrease of $T_S$ from $\approx$175~K in the undoped material to $\approx$170~K for $x$=0.1 whereas we clearly observe a significant suppression of $T_S$ from $\approx$144~K in the undoped material to $\approx$105~K for $x$=0.06 by the analysis of the orthorhombic distortion.
Please take note that the analysis of the FWHM in Ref.~\onlinecite{martinelli11} yields transition temperatures that are up to $\approx$60~K too high.

\subsection{Muon spin relaxation measurements}
\label{sec.sm-mu}
\subsubsection{Introduction}
For the $\mu$SR experiments a continuous beam of $S$=1/2, nearly 100\% spin polarized positve muons is directed onto the sample.
Through inelastic Coulomb scattering with the sample the muons completely lose their kinetic energy and finally stop at interstitial lattice sites.
They will usually stop in the sample at depths of 0.2--0.3~mm.
At the final interstitial lattice site the muon spin precesses in the local magnetic field.
For a single measurement several million positrons, generated by the muon decay, are recorded in time histograms by a set of positron detectors.
Properly normalized, these histograms yield the time evolution of the muon spin polarization $P(t)$.
This is possible because, during the muon decay the positron is emitted preferentially along the direction of the muon spin.
The analysis of $P(t)$ in zero, longitudinal, or transverse magnetic fields can reveal properties of, e.g.
magnetic order, magnetic fluctuations, and superconductivity, respectively.

In a previous work we calculated possible muon sites.
In agreement with experiment we found two muon sites (see below and Ref.~\onlinecite{maeter09}). In a recent DFT study, \citet{derenzi12} calculated similar muon sites.
Accordingly we chose a model for commensurate magnetic order to describe the zero field $\mu$SR time spectra that we deduced from the following general model with the two muon sites $A$ and $B$:
\begin{equation}
\begin{split}
P(t)=&V_{mag}(T)\bigl[ a (\frac{2}{3}e^{-\lambda_{T,A} t} \cos 2\pi f_A t +\frac{1}{3}e^{-\lambda_{L,A}t})\bigr.\\
&+(1-a) (\frac{2}{3}e^{-\lambda_{T,B} t} \cos2\pi f_B t+\frac{1}{3}e^{-\lambda_{L,B}t})\bigr]\\
&+(1-V_{mag}(T))\bigl[a(G(t,\sigma_A)e^{-\lambda_{nm,A}}\bigr.\\
&+\bigl.(1-a)G(t,\sigma_B)e^{-\lambda_{nm,B}}\bigr].
\end{split}
\label{eq.zf1}
\end{equation}
The first term models the signal from muons that come to rest in the volume of the sample that is magnetically ordered.
Its volume fraction is the magnetic volume fraction $V_{mag}(T)$.
The second term is due to muons that stop in the volume of the sample that does \textit{not} show long range magnetic order.
Its volume fraction is $(1$$-$$V_{mag}(T))$.
Each term contains two muon sites $A$ and $B$ with average muon occupation $a$ and $(1-a)$, respectively.
The last term models the paramagnetic part of the sample were muon-spin relaxation is caused by nuclear magnetic dipoles, described by the Gaussian-Kubo-Toyabe relaxation function $G(t,\sigma)$, and electron spin fluctuations, described by the exponential term\cite{hayano1979}.

The model (\ref{eq.zf1}) contains many free parameters - not all can be determined accurately.
It is often impossible to distinguish between the dynamic relaxation rates $\lambda_L$ and $\lambda_{nm}$ in the magnetically ordered, and the paramagnetic phase.
Therefore, we set $\lambda_{L,A}$=$\lambda_{nm,A}$ and $\lambda_{L,B}$=$\lambda_{nm,B}$.
The same applies to the rates that characterize the nuclear dipole field distribution, i.e.,
we set $\sigma$=$\sigma_A$=$\sigma_B$.
In case of $x$=0.06 we only observe one strongly damped signal without a frequency but two dynamic rates $\lambda_{L,A}$ and $\lambda_{L,B}$ we cannot distinguish $\lambda_{T,A}$, and $\lambda_{T,B}$, therefore we set $\lambda_T$=$\lambda_{T,A}$=$\lambda_{T,B}$.
For $x$=0, 0.02, and 0.04, above the rare earth magnetic ordering temperature for the Sm system we never observed more than one frequency and we set $a$=1.
In summary model (\ref{eq.zf1}) reduces to
\begin{equation}
\begin{split}
P(t)=V_{mag}(T)&\bigl[ \frac{2}{3}e^{-\lambda_{T} t} \cos 2\pi f_\mu t\\ &+\frac{1}{3}(a e^{-\lambda_{L,A}t}+(1-a)e^{-\lambda_{L,B}t})\bigr]\\
+(1-V_{mag}(T))G(t,\sigma)&\bigl[a e^{-\lambda_{L,A}} + (1-a) e^{-\lambda_{L,B}} \bigr].
\end{split}
\label{eq.zf2}
\end{equation}

Above model describes the signal of the majority phase in the sample. Small amounts of magnetic impurity phases cause relaxation of a few percent of the total signal during the dead time of the detectors for $x$$\geq$0.06, this reduces $P(t=0)$ below 1.

In Ref.~\onlinecite{maeter09} we have presented a calculation of the muon spin precession frequency based on our muon site calculation and, at the time available estimates of the Fe ordered moment from neutron diffraction.
For LaFeAsO, \citet{clarina} reported an Fe moment of $\approx$0.36~$\mu_B$.
Using this value we calculated the dipole fields $B_{loc}$ and the muon frequencies $f$=$\gamma_\mu/(2\pi) B_{loc}$ at the two muon sites.
However, this magnetic moment yields frequencies that are by a factor of 1.86 smaller than the experimentally reported frequencies of $f_A$=23~MHz and $f_B$=3~MHz\cite{klauss,luetkens09}.
Only recently, \citet{qureshi} reported a neutron diffraction study of LaFeAsO with different samples.
They find a much larger Fe magnetic moment of 0.63(1)$\mu_B$.
Using this value, the muon frequencies we calculate increase by factor of 1.75 to $f_A$=21.7(3)~MHz and $f_B$=2.3(4)~MHz where the error is estimated from the error of the magnetic moment given by \citet{qureshi}.
Within this error, the calculated and measured muon frequencies deviate by only $\approx$10--20\%.

\label{sec.sm-fe}
\begin{figure*}[htb]
\includegraphics[width=\textwidth]{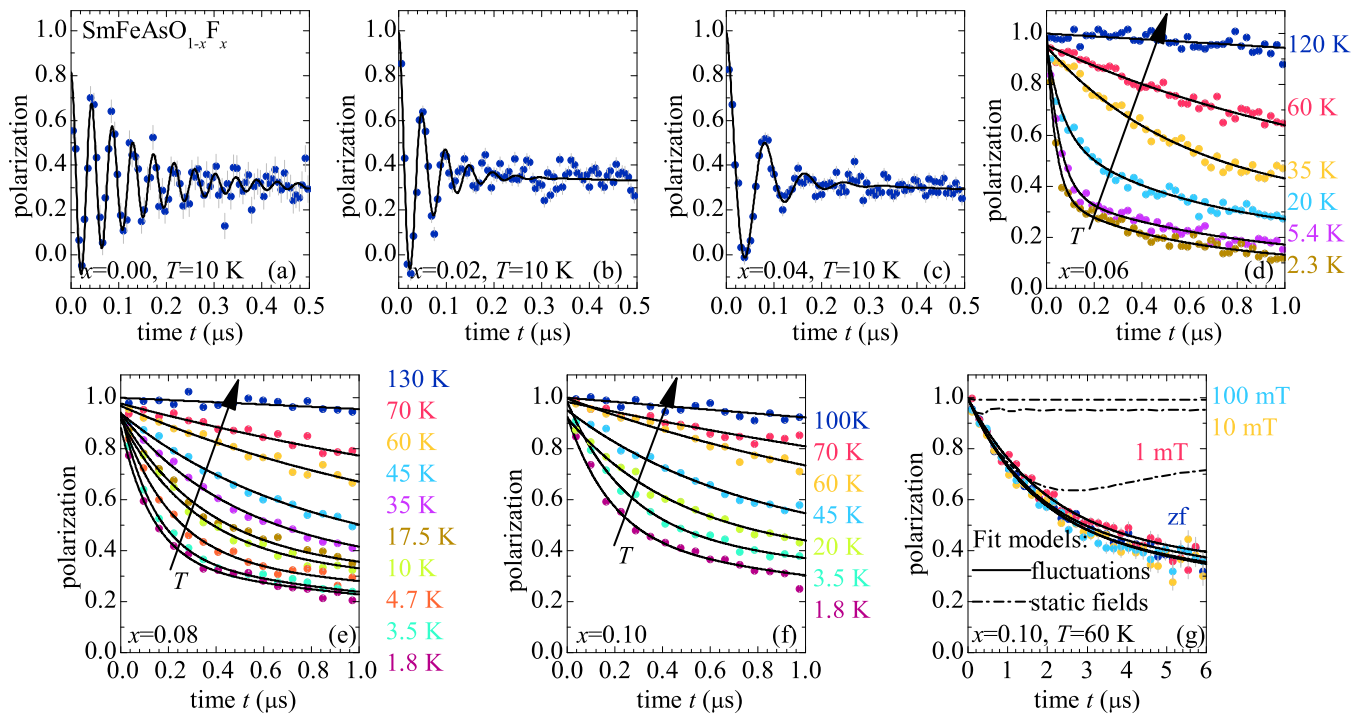}
\caption{Panels (a) through (f) show the zero field $\mu$SR time spectra characteristic for the SmFeAsO$_{1-x}$F$_x$ compounds we study here.
For $x$$\leq$0.04 we observe coherent muon spin precession which proofs long range magnetic order.
For $x$=0.06 a growing fraction of the signal shows fast relaxation below $T_{sro}$=28(2)~K which indicates short range magnetic order.
Panel (g) shows longitudinal field $\mu$SR measurements for $x$=0.10.
Small magnetic fields of up to 100~mT do not have any effect on the time spectra.
This indicates that only time dependent magnetic fields, i.e.,
spin fluctuations cause the relaxation and not static magnetic moments.}
\label{img.sm-histo}
\end{figure*}

This result has several implications:
\begin{enumerate}
\item{The coordinates of the calculated muon sites are very close to the real muon sites.}
\item{The interaction of the muon with its surrounding, i.e.,
electrostatic polarization, magnetic polarization, and local deformation of the crystal lattice have negligible effects on the $\mu$SR measurements of the studied ferropnictides, i.e., with $\mu$SR the intrinsic magnetic properties are measured.}
\item{No Fermi-contact hyperfine field contributes to the local magnetic fields at the interstitial muon lattice sites, because we can explain them fully by magnetic dipole fields.}
\end{enumerate}
These conclusions are in broad agreement with the results of \citet{derenzi12}: Based on their calculation of the muon site, they determined a Fe magnetic moment of 0.68~$\mu_B$, which is only 10\% larger than the value determined by neutron scattering\cite{qureshi}. They also conclude that Fermi contact hyperfine fields, by symmetry, do not contribute to the local magnetic fields at the muon sites\cite{derenzi12}. In addition they find a third muon site that may be relevant for muon diffusion, which takes place at temperatures larger than, for ferropnictides typical magnetic and structural phase transition temperatures\cite{ohishi11}.

The magnetically ordered phase may have a magnetic correlation length that is shorter than a few nanometers or may be broken up into small and isolated clusters.
In theses cases the relaxation rate $\lambda_T$ can be much larger than the precession frequency $f_\mu$.
As a result only a fast decay of the muon polarization without coherent muon spin precession can be observed.

For SmFeAsO$_{1-x}$F$_x$ we observed both situations, long range order with, and short range order without muon spin precession: a well defined muon spin precession, shown in Fig.~\ref{img.sm-histo} due to long range magnetic order for $x$$\leq$0.04, and for $x$=0.06 only a fast relaxation indicating short range magnetic order.
For higher doping levels static magnetic order is absent.
In the following we will discuss the results of the measurements in detail.

\subsubsection{Fe magnetic order}
In Fig.~\ref{img.sm-histo} we show our $\mu$SR time histograms of the Sm compounds. From the data we deduce, that the magnetic transition in the non-superconducting SmFeAsO$_{1-x}$F$_x$ with $x$=0, 0.02, 0.04 occurs in two steps.
We illustrate this process for SmFeAsO in Fig.~\ref{img.sm-undoped} with a detailed series of histograms at different temperatures:

\textit{Firstly}, the data display coherent muon spin precession already at high temperatures where only 5--10\% of the volume are magnetically ordered.
In the temperature range between $\approx$150~K and $\approx$140~K the magnetic volume fraction $V_{mag}(T)$ in Fig.~\ref{img.sm-vmag} increases gradually.
However, the temperature dependence of the order parameter, the muon spin precession frequency $f_\mu(T)$, shown in Fig.~\ref{img.sm-mag} does not follow the typical temperature dependence of an order parameter $f_\mu(T)$$\propto$$(1-(T/T_N)^\alpha)^\beta$---this is indicated by the symbols colored in gray in Fig.~\ref{img.sm-mag}.
Instead, $f_\mu(T)$ remains almost constant or decreases with decreasing temperature.
As the precession amplitude increases, the damping rate $\lambda_T(T)$ of the muon spin precession increases to $\approx$20~MHz, see Fig.~\ref{img.sm-mag} and Fig.~\ref{img.sm-undoped}.

\textit{Secondly}, only when 80--90\% of the sample volume are ordered, the damping rate $\lambda_T(T)$ decreases and $f_\mu(T)$ displays the typical temperature dependence expected for an order parameter $f_\mu(T)$$\propto$$(1-(T/T_N)^\alpha)^\beta$.

\begin{figure}[htb]
\includegraphics[width=8.6cm]{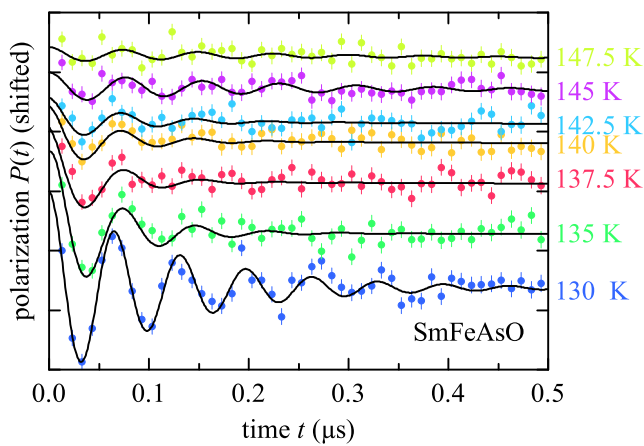}
\caption{Zero field muon spin relaxation time spectra of SmFeAsO (curves are shifted vertically for clarity).
We find muon spin precession, the indicator for long range magnetic order, already at temperatures for which only 5-10\% of the material show magnetic long range order.
This indicates that more and more clusters become ordered as we lower the temperature.
Only close to percolation $f_\mu(T)$ follows the typical temperature dependence of an order parameter (see Fig.~\ref{img.sm-mag}).}
\label{img.sm-undoped}
\end{figure}

Qualitatively the transition follows the same pattern for $x$=0, 0.02, and 0.04: A broad magnetic transition with a gradual increase of $V_{mag}(T)$ (see Fig.~\ref{img.sm-vmag}) with an unusual temperature dependence of $f_\mu(T)$ and a peak in $\lambda_T(T)$ (see Fig.~\ref{img.sm-mag}).
We will present a phenomenological interpretation of this broad transition in the next section (see p.~\pageref{sec.sm-fe-model}).

We also find a broad magnetic transition for $x$=0.06. magnetic fluctuations of the Sm 4$f$ moments (see Sec.~\ref{sec.sm-fluct}, p.~\pageref{sec.sm-fluct}). 
At high temperatures, the relaxation of $P(t)$ is due to the magnetic fluctuations of the Sm 4$f$ moments (see Sec.~\ref{sec.sm-fluct}, p.~\pageref{sec.sm-fluct}).
The magnetic order occurs below $T_{sro}$=33(1) and causes, as we lower the temperature, a gradually increasing part of $P(t)$ to relax during the first 0.2~$\mu$s of the time histograms in Fig.~\ref{img.sm-histo}.
This is indicated by the gradual increase of $V_{mag}(T)$ that reaches $\approx$100\% below $T_{100}$=6(2)~K, and $\lambda_T(T)$ below $T_{sro}$ (see Figs.~\ref{img.sm-vmag}, and ~\ref{img.sm-mag}). 

However, down to a temperature of 1.8~K we find no coherent muon spin precession---an indication for short range magnetic order. Since the susceptibility measurements\cite{panarina10} also show bulk superconductivity, we conclude that below $T_{100}$ the material is 100\% magnetically ordered and 100\% superconducting. Sanna, Shiroka and co-workers\cite{sanna09,sanna10,shiroka11} take this as evidence for nanoscopic coexistence, i.e., phase separation on length scales of a few nanometers.
However, this would only be clear evidence for phase separation if the $\mu$SR signals (volume fractions) from both phases were clearly separable, as e.g.
in Ba$_{1-x}$K$_x$Fe$_2$As$_2$\cite{aczel08,goko09,park09,julien09}.
Clear evidence for microscopic coexistences would be coupled order parameters as e.g.
in BaFe$_{2-x}$Co$_x$As$_2$\cite{pratt09}.
Considering the available evidence for the present case, we suggest that the situation is consistent with both situations.

At higher doping levels $x$=0.08, and 0.10 we observe no sign of Fe magnetic order. Instead, the relaxation of $P(t)$ in zero field is mostly due to magnetic fluctuations of the Sm 4$f$ moments (see Sec.~\ref{sec.sm-fluct}, p.~\pageref{sec.sm-fluct}).

\begin{figure*}[htb]
\includegraphics[width=\textwidth]{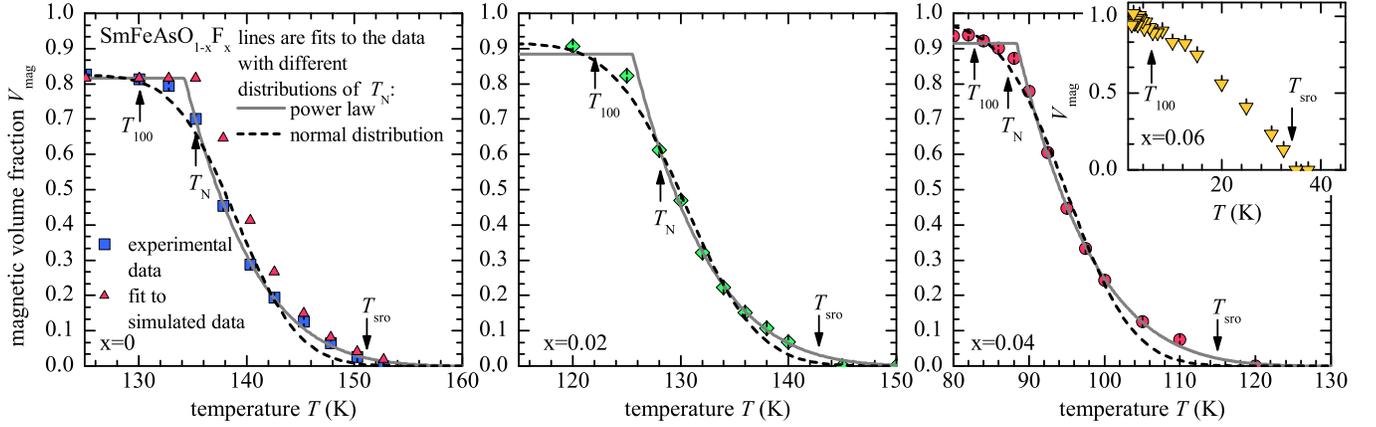}
\caption{The magnetic volume fraction $V_{mag}$ with fits according to power law, and normal distributed transition temperatures (see Eq.~(\ref{eq.distribution})).
A small but non-zero $V_{mag}$ indicates magnetic order in small volumes of the sample that appear below $T_{sro}$.
$V_{mag}(T)$ saturates at $T_{100}$.
For $x$=0 we also show the result of a fit to simulated data (red triangles, see Sec.~\ref{sec.sm-fe-model}).}
\label{img.sm-vmag}
\end{figure*}

\begin{figure*}[htb]
\includegraphics[height=6cm]{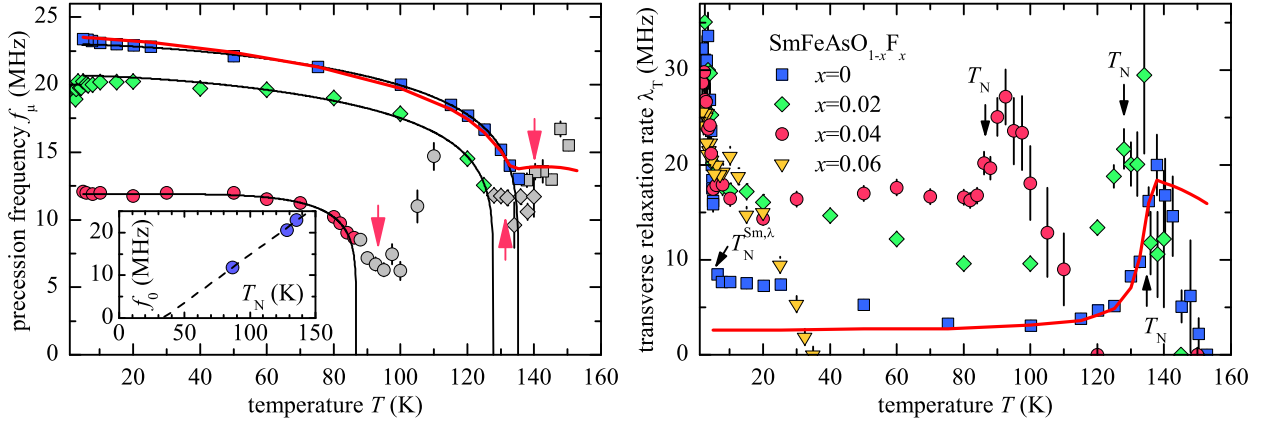}
\caption{\textit{Left}: The temperature dependence of the muon spin precession frequency $f_\mu(T)$ of SmFeAsO$_{1-x}$F$_x$.
It is proportional to the order parameter of the magnetic phase and indicates long range magnetic order.
The gray symbols indicate that the precession signal is mainly due to isolated clusters and does not reflect the bulk order parameter.
The black lines are fits to the data with the general power law described in the text---it was used to determine $T_N$.
The red arrows indicate the temperature of the maximum of $\lambda_T$.
\textit{Inset}: The low temperature limit of the muon spin precession frequency $f_0$=$f_\mu(T$$\to$$0)$ as a function the N\'eel temperature $T_N$.
The line is a linear fit to the data $f_0(T_N)$=0.224~MHz$/$K$\cdot$$T_N$-7.52~MHz.
\textit{Right}: The damping rate $\lambda_T$ of the muon spin precession signal is due to spatial and time dependent fluctuations of the ordered magnetic moment as well as quenched disorder.
However, within our model we can explain the peak at the transition with a large distribution of order parameters that narrows rapidly.
The red lines shown in the left and right panel were generated with the phenomenological model that we present in the text.}
\label{img.sm-mag}
\end{figure*}

Considering this broad magnetic phase transitions, it is not straight forward to define a phase transition temperature.
The temperature for which $V_{mag}(T)$ becomes non-zero is $T_{sro}$=151(1)~K, 143(2)~K, 115(5)~K, 33(1)~K for $x$=0, 0.02, 0.04, and 0.06, respectively.
We define $T_{100}$ as the temperature below which the majority of the sample is magnetically ordered and $V_{mag}(T)$ saturates.
We find $T_{100}$=130(2)~K, 122(2)~K, 83(1)~K, 6(2)~K for $x$=0, 0.02, 0.04, and 0.06, respectively.
We choose the N\'eel temperature $T_N$ so that $f_\mu(T)$ follows a power law $f_\mu(T)$$\propto$$(1-(T/T_N)^\alpha)^\beta$ for $T$$\leq$$T_N$=135.3(2)~K, 128.6(10)~K, 86.7(22)~K for $x$=0, 0.02, and 0.04.

The fits with this general power law are shown in Fig.~\ref{img.sm-mag}.
The exponent $\alpha$=1.49(4), 3.0(3), 5.5(21) for $x$=0, 0.02, and 0.04, respectively, controls the saturation at low temperatures.
The increase of $\alpha$ with doping indicates that the ordered magnetic moment saturates faster as a function of temperature the higher the F content.
The critical exponent $\beta$=0.143(2), 0.19(2), 0.15(8) for $x$=0, 0.02, and 0.04, respectively, controls the steepness of $f_\mu(T)$ close to $T_N$ and indicates the universality class of the phase transition.
Due to the unusual temperature dependence of $f_\mu(T)$ in the vicinity of $T_N$ the uncertainty of $\beta$ is too large to draw definite conclusions.
The values of $\beta$ are between $\beta$=0.125 expected for a 2D-Ising transition and $\beta$=0.325 of a 3D-Ising transition.
This is consistent with previously reported values (see Ref.~\onlinecite{wilson} and references therein).

The saturation value of the precession frequency $f_0$=23.07(4)~MHz, 20.06(4)~MHz, 11.91(7)~MHz for $x$=0, 0.02, and 0.04, respectively, is proportional to the ordered magnetic moment.
In Ref.~\onlinecite{maeter09} we showed that the magnetic moment of the $R$FeAsO ($R$=rare earth) is practically independent of the rare earth.
The ordered Fe magnetic moment of 0.61~$\mu_B$ of LaFeAsO\cite{qureshi} corresponds to a frequency of $f_0$=23~MHz\cite{klauss,luetkens09}.
This indicates that the measured frequencies in the Sm compounds correspond to an ordered Fe magnetic moment of 0.61~$\mu_B$, 0.53~$\mu_B$, and 0.32~$\mu_B$ for $x$=0, 0.02, and 0.04, respectively.

In the inset of Fig.~\ref{img.sm-mag} we plot $f_0=f_\mu(T=0)$ as a function of $T_N$.
We find that $f_0$ is linear in $T_N$ and a fit to the data yields $f_0(T_N)$=0.224~MHz$/$K$\cdot$$T_N$-7.52~MHz.
This seems to be a general feature of the ferropnictides: \citet{fernandes10-order}
analyzed magnetic neutron scattering data of BaFe$_{2-x}$Co$_x$As$_2$ with a simple two band model and showed that, in agreement with experiment, the magnetic moment decreases nearly linear as a function of the ordering temperature.
However, it is not a general consequence of their two band model, but rather of the particular detuning of the Fermi surface pockets by doping\cite{fernandes10-order}.

The gradual increase of $V_{mag}(T)$ suggests that magnetic order occurs in clusters that show already the hallmark of long range magnetic order: a non-zero muon spin precession frequency $f_\mu(T)$ shown in Fig.~\ref{img.sm-mag}.
The fact that close to $T_N$, $f_\mu(T)$ does not follow the typical temperature dependence of an order parameter indicates that the amount of ordered clusters has the main effect on $f_\mu(T)$ and not the temperature dependence of the bulk order parameter.
This is consistent with the peak in $\lambda_T(T)$.
It suggests that near the transition the ordered clusters have a broad distribution of ordered moments, which dominates $\lambda_T(T)$ and narrows as the ordered moments saturate at lower temperatures.
In the next section we present a phenomenological model that quantifies the above considerations.

This type of cluster formation could indicate a first order transition.
However, specific heat measurements of the $R$FeAsO type ferropnictides show no clear sign of a delta-function like anomaly or a temperature hysteresis at the phase transition that would indicate a first order phase transition \cite{ding:180510,mcguire,mcguire09}.

\subsubsection{A phenomenological model for the Fe magnetic phase transition}
\label{sec.sm-fe-model}

In the following we present a phenomenological model that will help to understand the broad transition that manifests in the slow increase of $V_{mag}(T)$, the peak of $\lambda_T(T)$ and the unusual temperature dependence of $f_\mu(T)$ for $T$$>$$T_N$.

The basis for this model is the following picture based on disorder.
The investigated samples are polycrystals and each grain of a polycrystal may have a local N\'eel temperature $T_N^\prime$ (as opposed to our estimation of the bulk $T_N$ presented in the preceding section) which depends on density of defects and dopant atoms, and possibly the size and direction of stress.
Therefore, we expect a probability distribution $P(T_N^\prime)$ of local N\'eel temperatures: at a temperature $T$ all grains with $T_N^\prime$$\geq$$T$ constitute the magnetic volume fraction:
\begin{equation}
V_{mag}(T)=\int_T^\infty P(T_N^\prime) dT_N^\prime.
\label{eq.vmag}
\end{equation}
\textit{A priori} it is not clear what functional form $P(T_N^\prime)$ should have.
Authors of previous works chose the normal distribution\cite{wilson,sanna09}
\begin{equation}
P(T_N^\prime)=\frac{1}{\sqrt{2\pi\sigma}}e^{-\frac{(T_N^\prime-T_0)^2}{2\sigma^2}}.
\end{equation}
We show fits with a normal distribution and Eq.~(\ref{eq.vmag}) in Fig.~\ref{img.sm-vmag} as dashed lines.
These fits systematically underestimate the data close to $T_{sro}$.
To improve the fit close to $T_{sro}$ we construct a distribution based on a power law: 
\begin{multline}
P(T_N^\prime)=\\
\left\{
\begin{aligned}
\frac{b+1}{2\sigma^{b+1}} (T_0-T)^b&\textnormal{, for } T_0\leq T\leq T_0-\sigma 2^{1/(b+1)} \\
0 &\textnormal{, else}
\end{aligned}
\right..
\label{eq.distribution}
\end{multline}
Here $b$$>$0 is an exponent and $T_0$ the upper limit of the distribution.
The prefactor is chosen so that $\sigma$ is a measure of the width of the distribution with $V_{mag}(T$$=$$T_0$$-$$\sigma)$$=$0.5.
The distribution is cut-off towards lower temperatures so that $V_{mag}(T)$$\leq$1.
This distribution yields the fits shown as gray lines in Fig.~\ref{img.sm-vmag}.
It describes the data well except close to $T_{100}$, the temperature at which $V_{mag}(T)$ saturates.
The parameters we obtain from the fits are $T_0$=167(2)~K, 161(3)~K, 139(4)~K, and $\sigma$=29(2)~K, 31(3)~K, 44(4)~K for $x$=0,0.02, and 0.04, respectively.
Within the error $b$=4.0(5) is independent of doping and was optimized simultaneously for all three doping levels.
The exponent $b$=4.0(5) may depend on the microscopic details of the disorder.
It is close to $b$=3.1 predicted for the mean field McCoy-Wu model\cite{berche98}.
It describes a disordered layered planar Ising magnet where the disorder is correlated perpendicular to the planes.

In the following we will use the above power law distribution to also describe $f_\mu(T)$ and $\lambda_T(T)$.
We do this by a simulation of zero field muon spin relaxation time spectra.
$P(t)$ is proportional to the real part of the Fourier transform of the probability distribution $P(B)$ of the local magnetic field $B$ at the muon stopping site:
\begin{equation}
\begin{split}
P(t)&=\int_{-\infty}^\infty \cos(\gamma_\mu B t) P(B) dB\\
&=\int_T^\infty \cos(2\pi f_\mu(T,T_N^\prime) t) P(T_N^\prime) dT_N^\prime.
\end{split}
\label{eq.spectrum}
\end{equation}
Here, $f_\mu(T)$=$f_0(T_N^\prime)(1-(T/T_N^\prime)^\alpha)^\beta$ where $\alpha$ and $\beta$ are variable fit parameters, and $f_0(T_N^\prime)$=0.224~MHz$/$K$\cdot$$T_N^\prime$-7.52~MHz as discussed above and shown in Fig.~\ref{img.sm-mag}.

\begin{figure*}[htb]
\includegraphics[width=\textwidth]{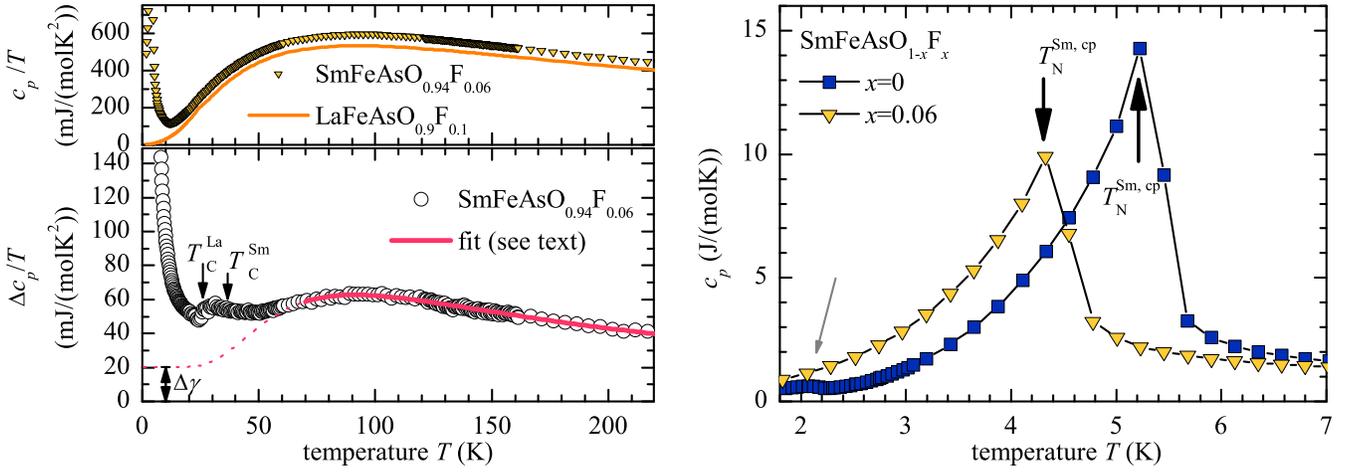}
\caption{\textit{Left}: The temperature dependence of the specific heat $c_p(T)/T$ of SmFeAsO$_{0.94}$F$_{0.06}$ (upper panel).
This material is superconducting and does not show the anomaly associated with the structural or magnetic phase transition\cite{baker09}.
$c_p(T)/T$ shows a broad maximum at $T$$\approx$75~K similarly to LaFeAsO$_{0.9}$F$_{0.1}$.
To extract the contribution of the Sm CEF levels we follow the procedure described by \citet{baker09} and calculate the difference $\Delta c_p(T)/T$ of the specific heat of SmFeAsO$_{0.96}$F$_{0.06}$ and LaFeAsO$_{0.9}$F$_{0.1}$ (lower panel).
$\Delta c_p(T)/T$ displays a broad Schottky-type anomaly due to the thermal population of Sm CEF levels and a difference in Sommerfeld coefficients $\Delta \gamma$=20.3(4)~mJ/(molK$^2$).
Below $\approx$50~K the onset of the superconductivity in the La system ($T_c^{La}$) and Sm system ($T_c^{Sm}$), and the magnetic order in the Sm system limit this analysis.
\textit{Right}: Low temperature $c_p(T)$ of SmFeAsO and SmFeAsO$_{0.96}$F$_{0.06}$.
The $\lambda$-anomaly indicates a second order phase transition which is attributed to the Sm magnetic order.
This anomaly is only slightly suppressed by doping.
A small Schottky-type hump in the specific heat of SmFeAsO centered at 2~K (marked with the small, gray arrow) was also observed in Ref.~\onlinecite{baker09} and is most likely related to an impurity.}
\label{img.sm-cp}
\end{figure*}

We fit the spectra simulated by Eq.~\eqref{eq.spectrum} with a damped oscillation $P(t,T)$=$P_0(T)\cos(2\pi f_\mu(T) t)\exp(-\lambda_T(T) t)$ to obtain the frequency $f_\mu(T)$, the damping rate $\lambda_T(T)$, and the magnetic volume fraction which is just $P_0(T)$.
We used this procedure to optimize the critical exponents $\alpha$ and $\beta$ for temperatures lower than $T_N$=135.3~K.
For $x$=0 we find that $\alpha$=1.35(3) and $\beta$=0.170(3) yield the best description of $f_\mu(T)$ for $T$$<$$T_N$=135.3~K.
This fit is shown as a red line in Fig.~\ref{img.sm-mag}.
Although we only optimized the critical exponents for $T$$<$$T_N$ the model gives good quantitative agreement also for $T_N$=135.3(2)~K$<$$T$$<$$T_{sro}$=151(1)~K.

This procedure also yields quantitative agreement with relaxation rate $\lambda_T(T)$ shown in Fig.~\ref{img.sm-mag}.
In particular, it reproduces the peak of $\lambda_T(T)$ near the phase transition.
Note that $\alpha$ and $\beta$ were only optimized with respect to $f_\mu(T)$.
The quantitative agreement with $\lambda_T(T)$ indicates that this simple model captures most features of the magnetic phase transition correctly.
In particular it describes the unusual temperature dependence of $f_\mu(T)$ for $T$$>$$T_N$ and the decay of $\lambda_T(T)$ for $T$$<$$T_N$.

In summary, we can describe $f_\mu(T)$ and $\lambda_T(T)$ by a broad and asymmetric probability distribution of magnetic ordering temperatures.
We suggest that the formation of an inhomogeneous magnetic state with paramagnetic and magnetically ordered domains is an \textit{intrinsic} response of the magnetic system triggered by \textit{extrinsic} disorder.
Its systematic study by experimental and theoretical methods may reveal important properties of the system without disorder.

\begin{figure}[htb]
\includegraphics[width=8.6cm]{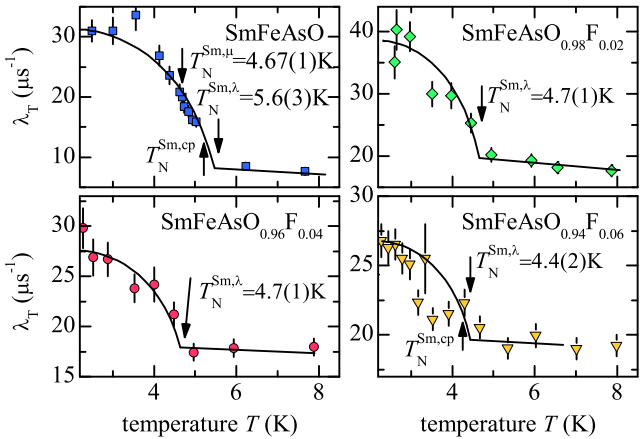}
\caption{The temperature dependence of the damping rate $\lambda_T$ of the muon spin precession.
Its increase at low temperatures marks the onset of the Sm magnetic order.
Lines are guides to the eye.}
\label{img.sm-lamt-lt}
\end{figure}


\subsubsection{Sm magnetic order}
\label{sec.sm-sm}
Additional muon spin precession signals in the undoped material mark the onset of the Sm magnetic order below $T_N^{Sm,\mu}$=4.66(1)~K\cite{maeter09}.
In the ordered phase of the Sm moments for $x$=0.02 we also observe additional spin precession signals, but the onset temperature is not as clear.

The specific heat $c_p(T)$ probes the onset of Sm magnetic order very sensitively.
We have measured $c_p(T)$ for $x$=0, and 0.06 at low temperatures, see the right-hand side panel of Fig.~\ref{img.sm-cp}.
We find a $\lambda$-anomaly with a maximum at the transition temperature $T_N^{Sm,cp}$=5.2(1)~K, and 4.3(1)~K for $x$=0, and 0.06, respectively.
From a general point of view we can expected that $T_N^{Sm,\mu}$$\leq$$T_N^{Sm,cp}$ because, although the Sm magnetic order parameter is non-zero for $T$$\leq$$T_N^{Sm,cp}$ it may be undetectable with $\mu$SR because \textit{(a)} the order parameter is too small, \textit{(b)} the local magnetic field is a vector sum of the fields created by the Fe, and Sm magnetic order, which reduces the effect of the Sm order on the muon spin precession frequency, \textit{(c)} the (spatial) fluctuations of the order parameter destroy a coherent muon spin precession, or \textit{(d)} the magnetic field of the ordered magnetic moment cancels out at the muon site.

The $\lambda$-anomaly is \textit{not} accompanied by a divergence of the dynamic relaxation rate $\lambda_L(T)$, see Fig.~\ref{img.sm-laml}.
This indicates that the hyperfine coupling of the muon is too small or the critical fluctuations are too fast to be detected by $\mu$SR.
Thus, $T_N^{Sm}$ cannot be determined directly from $\lambda_L(T)$ and its temperature dependence is not related to the critical fluctuations expected for this second order magnetic transition (a detailed discussion follows in the next section).
\textit{A posteriori} we can identify $T_N^{Sm}$ with the increase of the damping rate $\lambda_{T}(T)$.
In Fig.~\ref{img.sm-lamt-lt} we show $\lambda_{T}(T)$ for low temperatures.
The increase of $\lambda_{T}(T)$ correctly indicates the onset of the broadening of the spin precession frequency spectrum due to the Sm magnetic order.
It coincides with the maximum of the $\lambda$-anomaly of $c_p(T)$.
It enables us to identify the ordering temperature of the Sm moments using $\mu$SR by the increase of $\lambda_{T}(T)$.
We call the ordering temperature of the Sm moments that we determine in this way $T_N^{Sm,\lambda}$=5.6(3)~K, 4.7(1)~K, 4.7(1)~K, and 4.4(2)~K for $x$=0, 0.02, 0.04, and 0.06, respectively.

This approach is not possible for $x$=0.08, and 0.10 because the Sm magnetic order causes no additional static relaxation.
Instead the relaxation is fully due to spin fluctuations.
We verified the dynamic character of the relaxation by measuring in a longitudinal magnetic field, see Fig.~\ref{img.sm-histo}(g).
The absence of any sizable decoupling indicates the dynamic nature of the relaxation.
The increase of $\lambda_{L,(A/B)}$ below $\approx$5~K could be caused by the Sm order (see next section and Fig.~\ref{img.sm-laml}).
Because this feature is present for all doping levels it is likely that Sm magnetic order persists also for $x$=0.08, and 0.10 with an estimated $T_N^{Sm}$$\approx$3.5(10)K.
However, this only indicates a possible phase transition and should not be taken as proof for Sm magnetic order for $x$=0.08, and 0.10.

\subsubsection{Spin fluctuations}
\label{sec.sm-fluct}

In the preceding sections we discussed the Fe and the Sm magnetic order.
In this section we focus on the magnetic fluctuations characterized by $a$, $\lambda_{L,A}$ and $\lambda_{L,B}$.
$\lambda_{L,(A/B)}$ are the muon spin-lattice relaxation rates.
In SmFeAsO$_{1-x}$F$_x$ and CeFeAsO$_{1-x}$F$_x$ fluctuations the rare earth and the Fe magnetic moment could contribute to the spin-lattice relaxation rate.

Several nuclear magnetic resonance (NMR) studies show an enhancement of magnetic fluctuations in the paramagnetic state above the Fe magnetic transition, and the superconducting transition\cite{ahilan,nakai,nakai-njp,oka11}.
However, we, and other groups, find no such fluctuations in the $R$FeAsO$_{1-x}$F$_x$ materials with $\mu$SR\cite{luetkens08,luetkens09,klauss,sanna10,carlo09,ohishi11,shiroka11}, indicating that the associated muon spin-lattice relaxation rate is too small to be measured in the time window of $\mu$SR (several $\mu$s).

For the $R$FeAsO$_{1-x}$F$_x$ materials with a magnetic rare earth $R$=Ce, Pr, and Nd no sizable spin-lattice relaxation in $\mu$SR has been reported\cite{sanna10,maeter09,carlo09}---with the exception of $R$=Sm\cite{maeter09,khasanov08,drew09,drew,sanna09}.
It is not immediately clear where these spin fluctuations originate.
In superconducting SmFeAsO$_{1-x}$F$_x$ the Sm spin fluctuations lead to $1/T_1$$\propto$$T^{-0.6}$ (from NMR\cite{prando10}).
The $\mu$SR spin-lattice relaxation rates $\lambda_{L,(A/B)}$ contain additional contributions that are temperature activated,\cite{maeter09,khasanov08,drew,sanna09} see Fig.~\ref{img.sm-laml}.
\citet{drew} suggested that they are related to spin fluctuations in the vicinity of the superconducting transition.

In Fig.~\ref{img.sm-laml}, \textit{all} materials show a temperature activated (step-like) temperature dependence of $\lambda_{L,(A/B)}$ close to $T$$\approx$50~K followed by a plateau or slight increase and a steep increase below $\approx$5~K.
In materials with Fe magnetic order ($x$$\leq$0.06) an additional cusp is present in the vicinity of $T_N$ which is due to the enhanced spin fluctuations at the phase transition.
Apart from this cusp, the temperature dependence of the spin-lattice relaxation rate is qualitatively independent of doping.
This suggests that the enhancement of $\lambda_{L,(A/B)}$ at $\approx$50~K is a general feature of all SmFeAsO$_{1-x}$F$_x$ compounds and related neither to superconductivity nor to Fe magnetic order but rather to Sm spin fluctuations.

\begin{figure}[ht]
\includegraphics[width=8.25cm]{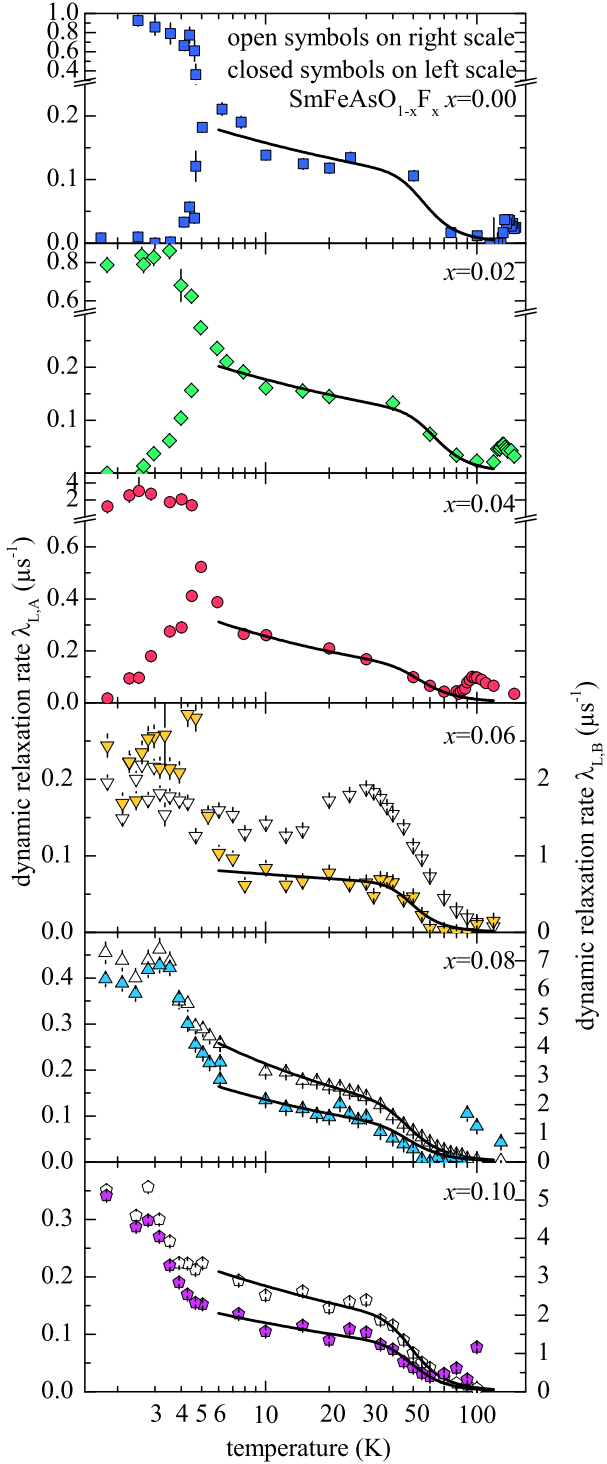}
\caption{The longitudinal relaxation rates of SmFeAsO$_{1-x}$F$_{x}$.
The open symbols mark the additional relaxation rates that we find for $x$$\geq$0.6.
The Fe magnetic order causes a small cusp at high temperatures.
It is followed by a temperature activated contribution that saturates below $\approx$30~K and a power law increase down to $\approx$5~K.
At even lower temperatures the Sm magnetic order could be responsible for the steep increase below $\approx$5~K.}
\label{img.sm-laml}
\end{figure}

For $x$=0, 0.02, and 0.04 the onset of the Sm magnetic order is accompanied by the appearance of additional precession frequencies, and two distinct longitudinal relaxation rates $\lambda_{L,A}$ and $\lambda_{L,B}$ ($\lambda_{L,A}=\lambda_{L,B}$ for $T>T_N^{Sm}$).

The second relaxation rate $\lambda_{L,B}$ for $x$$\geq$0.06 for $T>T_N^{Sm}$ is most likely related to the population of the second muon site in the Sm-O layer\cite{maeter09}.
The muons that stop in the Sm-O layer have a much larger hyperfine coupling to the Sm spin fluctuations which leads to an enhanced spin-lattice relaxation rate $\lambda_{L,B}$$<$$\lambda_{L,A}$.
The probability for a muon to stop in the Fe-As layer is $a$ in (\ref{eq.zf2}), and the probability to stop in the Sm-O layer is 1-$a$.
We find that $a$ is temperature independent for $T$$<$150~K and $a$=0.37(1), 0.38(1), 0.46(1) for $x$=0.06, 0.08, and 0.10, respectively.

A a double-logarithmic plot of $\lambda_{L,B}(T)$ for SmFeAsO$_{0.92}$F$_{0.08}$ is shown in Fig.~\ref{img.sm-laml-det}.
The temperature activated contribution (green dashed line) used previously\cite{khasanov08} increases down to $T$$\approx$30~K and then saturates.
$\lambda_{L,B}(T)$, however, continues to increase.
For 6--20~K we can describe it with a power law $\lambda_{L,B}(T)$=$8.9(4)T^{-0.42(2)}$.
This exponent is consistent with a recent $^{19}$F NMR work\cite{prando10}.
Therein, \citet{prando10} report a $^{19}$F spin-lattice relaxation rate $1/T_1$$\propto$$T^{-0.6(1)}$.
They interpreted $1/T_1$ to be due to critical fluctuations close to the magnetic phase transition of Sm, which should lead to a maximum of $1/T_1$ for $T=T_N^{Sm}$.
Clearly, $\lambda_{L,(A/B)}(T)$ does not show a maximum at $T=T_N^{Sm}\approx 5~K$ (see Fig.~\ref{img.sm-laml}).
It follows that $\mu$SR is not sensitive to the itinerant Sm spin fluctuations sensed by NMR, but rather to the local fluctuations, i.e., we have to consider the fluctuations of the Sm moment between different crystal electric field levels.

According to Orbach\cite{orbach61}, the temperature dependence of the fluctuation rate between CEF levels (at high enough temperatures) due to two-phonon scattering leads to a finite lifetime of the CEF ground state given by
\begin{equation}
\tau^{phonon}\propto 1/\Gamma^{phonon}\propto e^{\Delta/(k_BT)}-1,
\end{equation}
where $\Gamma^{phonon}$ is the line width of the quasi-elastic CEF excitation, $k_B$ is the Boltzman constant.
The transition between the degenerate spin states proceeds via the first excited CEF level with the energy $\Delta$ (the so-called resonant-Raman or Orbach process).
Orbach's treatment\cite{orbach61} neglects any kind of interaction of the rare earth moments with the conduction band electrons.
In first order, interaction with the conduction band electrons causes a Korringa-type broadening of the quasi-elastic CEF excitations\cite{goetze74} $\Gamma^{3d-4f}\propto 1/\tau^{3d-4f}\propto T$.
\citet{goetze74} considered corrections to the linear temperature dependence due to magnetic exchange interaction (valid at low temperatures) that lead to $\Gamma^{3d-4f}\propto T^\alpha$, where $\alpha<1$ depends on the density of states at the Fermi level and the hybridization (valid for $T\gg T_K$, where $T_K$ is the Kondo temperature).
Such power laws are often found by NMR measurements of the nuclear spin-lattice relaxation rate in heavy-fermion compounds, e.g., in CeFePO \cite{bruning08}.

Assuming that both processes are independent, the overall line width of the quasi-elastic CEF excitation is just the sum $\Gamma=\Gamma^{phonon}+\Gamma^{3d-4f}$, i.e.,
\begin{equation}
\begin{split}
\lambda_L(T)\propto\tau\propto&\left(\frac{1}{\tau^{phonon}}+\frac{b}{\tau^{3d-4f}}\right)^{-1}\\
&=\left(\frac{1}{e^{\Delta/(k_BT)}-1}+b T^\alpha\right)^{-1}
\end{split}
\label{eq.cef}
\end{equation}
where $b$ scales the two contributions to the quasi-elastic line width, and $\lambda_L$ is the muon spin-lattice relaxation rate (assuming the fast fluctuation limit).

\begin{figure}[htb]
\includegraphics[width=8.6cm]{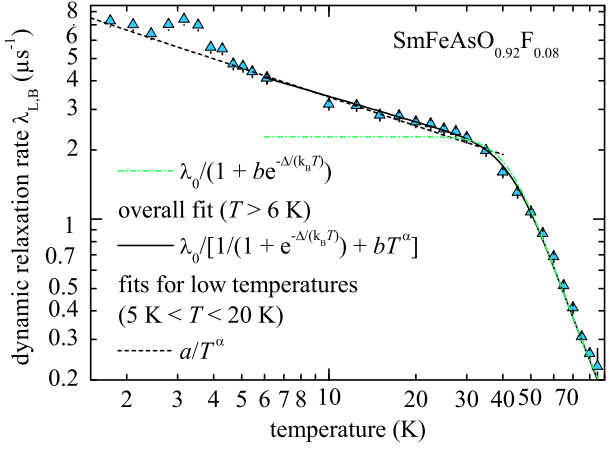}
\caption{The temperature dependence of spin-lattice relaxation rate $\lambda_{L,B}(T)$ for SmFeAsO$_{0.92}$F$_{0.08}$.
This rate belongs to the muon site in the Sm-O layer.
The temperature activated behavior is indicated by the green dashed-dot line.
Below $\approx$30~K as a power law, below $\approx$5~K it increases steeper but not as steep as expected for critical fluctuations at the Sm magnetic phase transition.}
\label{img.sm-laml-det}
\end{figure}

\begin{figure}[htb]
\includegraphics[width=8.5cm]{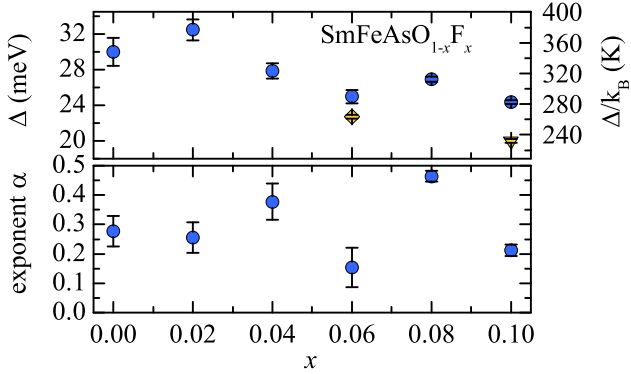}
\caption{Parameters with best fit of equations (\ref{eq.cef}) to the data.
The fits are shown in Fig.~\ref{img.sm-laml} and \ref{img.sm-laml-det}.
We have added the energy of the lowest CEF excitation determined by $c_p(T)$: $\Delta_1$=20.0(2)~meV for $x$=0.10 by \citet{baker09} (diamond), and $\Delta_1$=22.0(2)~meV for $x$=0.06 in this work (triangle).} 
\label{img.sm-cef}
\end{figure}

Within the error, $b=10.8(4)\cdot 10^{-4}$ is independent of doping, $\Delta$ and the exponent $\alpha$ is shown in Fig.~\ref{img.sm-cef}.
The two relaxation rates for $x$$\geq$0.06 have (within error) identical temperature dependencies, therefore we optimized $\alpha$ and $\Delta$ simultaneously for both rates.
The data is well described by the fits of Eq.~\eqref{eq.cef} (see Fig.~\ref{img.sm-laml}), however on cooling below $T\approx 5$~K the data increases steeper than expected, and reaches a cusp for $T\approx 3$~K.
As discussed in the previous section, this is not related to critical fluctuations close to the Sm magnetic transition.
This deviation suggests that magnetic Sm-Sm interaction should be taken into account for a description of the data for $T<5$~K (see e.g. Ref.~\onlinecite{fulde85})---such an analysis is beyond the scope of this work.

$\Delta$ determined above is consistent with the CEF levels determined from the specific heat $c_p(T)$ (see Fig.~\ref{img.sm-cp}):
To determine the CEF excitation energies of SmFeAsO$_{0.94}$F$_{0.06}$ we used $c_p(T)$ of LaFeAsO$_{0.09}$F$_{0.1}$ as a reference and proceeded as \citet{baker09}.
They reported three crystal electric field doublets for SmFeAsO$_{0.90}$F$_{0.10}$ with energies $\Delta_{1,2}$=20.0(2), 45(1)~meV of the first and second excited levels, respectively.
For $x$=0.06 we find $\Delta_{1,2}$=22.7(2), 52.7(9)~meV.
We plot $\Delta_1$ as a function of doping together with $\Delta$ (from $\mu$SR) in Fig.~\ref{img.sm-cef}.
We find that $\Delta$ is systematically $\approx$10\% larger than $\Delta_1$.
Part of the deviation should be due to the uncertainties of the analysis of both the specific heat and the $\mu$SR data.
In particular our model \eqref{eq.cef} neglects the second CEF excitation, one-phonon scattering, and the contribution of Sm-Sm magnetic interactions to the line width of the quasi-elastic CEF excitations.

In conclusion, we interpret $\lambda_{L,(A/B)}$ as the spin-lattice relaxations rates of muons that stop in the Fe-As layer ($\lambda_{L,A}$) and the Sm-O layer ($\lambda_{L,B}$).
Qualitatively, $\lambda_{L,(A/B)}(T)$ are doping independent and mainly probe magnetic fluctuations of the Sm 4$f$ states but with different hyperfine coupling strengths.
This is supported by the quantitative agreement of the lowest CEF excitation $\Delta_1$ determined by $\mu$SR and specific heat measurements in this, and Baker's study\cite{baker09}, see Fig.~\ref{img.sm-cef}.
Furthermore, the $^{19}$F NMR studies by \citet{prando10} show a power law divergence of $1/T_1$, that they attribute to the Sm 4$f$ states, similar to the power law increase we find between 5~K and 30~K, see Fig.~\ref{img.sm-laml-det}.
However, we suggest that $\mu$SR is sensitive to the local Sm fluctuations (due to phonons and magnetic exchange with the conduction band electrons), whereas $^{19}$F NMR senses the collective Sm spin fluctuations (due to Sm magnetic order at low temperatures).
The Sm order is not accompanied by critical fluctuations in the time window of $\mu$SR.
The upturn of $\lambda_{L,(A/B)}(T<5\textrm{~K})$ should be due to the broadening/splitting of the CEF ground state doublet by the Sm magnetic interaction/order.

\subsubsection{Superconductivity}
\label{sec.sm-tf}
To investigate the magnetic penetration depth of SmFeAsO$_{0.90}$F$_{0.10}$ in the superconducting phase we applied a transverse magnetic field $\mu_0H$=70~mT that causes muon spin precession.
In general we expect the relaxation above $T_c$ to be solely due to the spin fluctuations described in Sec.~\ref{sec.sm-fluct}.
For a powder sample of an anisotropic type-II superconductor, the flux line lattice causes an additional Gaussian relaxation rate $\sigma_{sc}(T)$=$\sqrt{\sigma(T)^2-\sigma_{nm}^2}$ where $\sigma_{nm}$ is determined above $T_c$ \cite{brandt88}.
$\sigma_{sc}$ is a measure of the superfluid density, or the inverse penetration depth $\sigma_{sc}$$\propto$$1/\lambda^2$$\propto$$ n_s/m^\ast$ \cite{brandt88}.
Both effects contribute to the damping of the precession signal, which is modeled by
\begin{equation}
A(t)=(a e^{-\lambda_{L,A} t} + (1-a)e^{-\lambda_{L,B} t}) e^{-\frac{t^2}{2\sigma^2}} \cos(\gamma_\mu B_{loc} t+\phi).
\end{equation}
In the limit of fast spin fluctuations, the longitudinal and transverse relaxation are identical to $\lambda_{L,(A/B)}$\cite{hayano1979}.
Both can be measured simultaneously at the GPS instrument of the Paul-Scherrer Institute, which improves the accuracy of $\sigma$.
For briefness we do not show $\lambda_{L,(A/B)}(T)$  (they are identical to the rates measured in zero field, see Fig.~\ref{img.sm-laml}).

Bulk superconductivity is indicated by the reduction of the internal field $B_{loc}$ and the increase of $\sigma_{sc}$ below $T_c$ shown in Fig.~\ref{img.sm-sig}.
The saturation of $\sigma_{sc}$ indicates a low density of states at the Fermi level characteristic for a nodeless gap.
A gap with nodes, as e.g.
for extended s-wave symmetry would lead to a linear decrease of $\lambda$ at low temperatures \cite{vorontsov09}.
$\sigma_{sc}(T)$ has been analyzed using the temperature dependence of $1/\lambda^2$ assuming a nodeless s-wave gap $\Delta(T)$ \cite{tinkham2004,*carrington03}.
The best fit, shown in Fig.~\ref{img.sm-sig}, yields $\Delta(T=0)$=10.0(5)~meV, $T_c$=50.6(4), and $\sigma_{sc}(0)$=1.88(5)~$\mu$s$^{-1}$.
For an anisotropic superconductor the in-plane penetration depth can be estimated to $\lambda_{ab}(0)[$nm$]$=248.85$(\sigma_{sc}(0)[$$\mu$ s$^{-1}])^{-0.5}$=180(2)~nm \cite{brandt88,fesenko91}.
Vortex lattice disorder can artificially reduce $\lambda_{ab}$, therefore 180(2)~nm is a lower limit for 180(2)~nm.
Nevertheless, this value is in agreement with the (more reliable)
results of \citet{weyeneth10}. They studied polycrystalline SmFeAsO0$_{85}$F$_{0.15}$ (nominal
composition, $T_c = 54$~K from electrical resistivity) by TF-$\mu$SR and found
$\lambda_{ab}=200$ nm by carefully measuring the field dependence of $\sigma_{sc}$ . This procedure
allowed them to exclude vortex lattice disorder due to pinning. They were able to determine
the sizes of both superconducting gaps $\Delta(0)=13.8, 5.3$~meV.
The resulting $\sigma_{sc}(T)$, shown as a dashed line in Fig.~\ref{img.sm-sig}, is compatible with our data.
However, within the error of our data, a fit with two gaps is not possible.

\begin{figure}[htb]
\includegraphics[width=8.6cm]{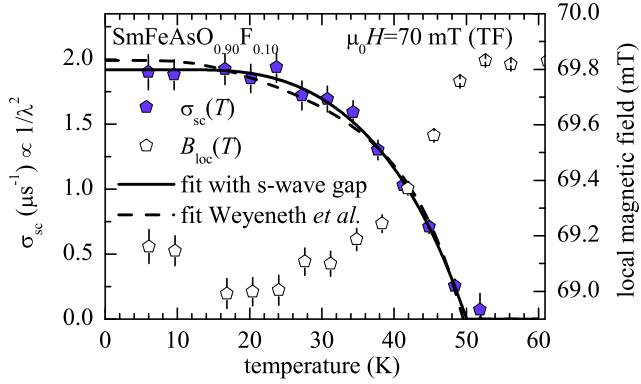}
\caption{Weak transverse field $\mu$SR of SmFeAsO$_{0.90}$F$_{0.10}$: The Gaussian relaxation rate $\sigma_{sc}$ due to the flux line lattice, and the reduction of the local magnetic field $B_{loc}$ due to the diamagnetic response below $T_c$ indicate bulk superconductivity.} 
\label{img.sm-sig}
\end{figure}

\subsection{Summary}
\label{sec.sm-sum}
In Tab.~\ref{tab.sm} we summarize the characteristic temperatures relevant for the structural and electronic phase transitions of SmFeAsO$_{1-x}$F$_x$ determined by muon spin relaxation, synchrotron X-ray diffraction, electrical resistivity, and specific heat measurements.

\paragraph*{General results} The phase diagram is shown in Fig.~\ref{img.sm-phase}.
As a result of the detailed examination of the (broadened) phase transitions with $\mu$SR and synchrotron X-ray diffraction the hierarchy of phase transition is not as clear as thermodynamic probes suggest (see e.g.
Refs.~\onlinecite{jesche08,jesche10}).
We observe long range magnetic order (lrmo) in clusters below $T_{sro}$$\approx$$T_{max}^\rho$$>$$T_S$ already above the structural transition but the order parameter follows the expected temperature dependence only within the orthorhombic phase ($T_{N}$$<$$T_S$).
Both phase transitions are suppressed by doping and only the magnetic order, albeit now truly short ranged, survives as superconductivity appears for $x$$\geq$0.06.
At higher doping levels we observe only superconductivity.
The magnetic penetration depth of the superconducting material $x$=0.10 is consistent with a single nodeless s-wave gap.
However, within the uncertainty of the analysis it is also possible that two gaps are present as suggested in several works\cite{malone09,gonnelli09,daghero09,weyeneth10}.
The magnetic transition of the Sm moments is almost unaffected by doping and is only reduced by $\approx$1~K over the whole phase diagram.
The spin fluctuations of the Sm are also present throughout the phase diagram and are almost unaffected by the disappearance of Fe magnetic order.

\begin{figure}[htbp]
\includegraphics[width=8.6cm]{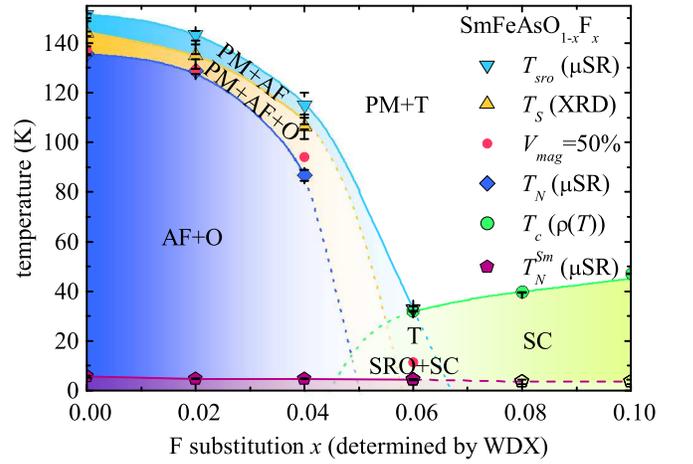}
\caption{The electronic phase diagram of SmFeAsO$_{1-x}$F$_x$.
The extend and shape of the region with a mixed phase with superconductivity, Fe magnetic order, and non-zero orthorhombicity is still unclear.
In this work we only find a mixed phase of superconductivity and Fe magnetic short range order with almost identical transition temperatures that has a tetragonal crystal structure.
This indicates that $x$=0.06 is close to a tetra critical point.
Ref.~\onlinecite{sanna09} indicates that a region with Fe magnetic long range order and superconductivity exists.
Refs.~\onlinecite{margadonna}, ~\onlinecite{drew} indicate that superconductivity and Fe magnetic (long range) order occur also in the orthorhombic phase.
All lines are guides to the eye.
Dotted lines are extrapolations of the phase boundaries indicated by solid lines.
For $x$=0.08, and 0.10, it is unclear whether Sm order occurs (see Sec.~\ref{sec.sm-sm}, p.~\pageref{sec.sm-sm}), this is indicated by the open symbols and the dashed line.
For $x$=0.06 the triangle that marks $T_{sro}$($\approx$$T_c$) was drawn open for clarity.}
\label{img.sm-phase}
\end{figure}

\paragraph*{Disorder} The structural phase transition is also not sharp, but occurs over a broad temperature interval, which may be caused by random or homogeneous stress, or order parameter fluctuations in a broad temperature interval.
In this regard it is worth mentioning, that also the possible spin-nematic phase transition can be broadened by stress\cite{hu12,fernandes11a}.
The broad magnetic transition may also be due to homogeneous\cite{cano12} or inhomogeneous stress\cite{hu12}.
\citet{vavilov11} showed in a theoretical study that disorder may have the same effect on the phase transitions as charge doping.
In light of this result we cannot dismiss quenched disorder (impurities, lattice defects) as a second possible source of the broad phase transitions.

\paragraph*{Structural transition} The tetragonal to orthorhombic phase transition does not occur for $x$=0.06 where short range magnetic order occurs together with superconductivity.
This is supported by the absence of any extra anomaly of the temperature dependence of the specific heat $c_p(T)$ (Fig.~\ref{img.sm-cp}, p.~\pageref{img.sm-cp}).
Also our X-ray diffraction study shows no indication for a structural phase transition.
It follows that the structural phase transition be absent for $x$$\geq$0.06.
Due to the large FWHM of the (2,2,0)$\rm_T$ Bragg peak, we are not able to resolve small orthorhombic distortions $\delta$$<$1$\cdot$10$^{-3}$ (the discussion of the resolution limit on p.~\pageref{para.resolution} also applies here).
We note that small changes of the FWHM \emph{alone} cannot be taken as proof for a structural phase transition, as this can lead to false phase transition temperatures (see p.~\pageref{para.marti}).

\begin{figure*}[htbp]
\includegraphics[width=\textwidth]{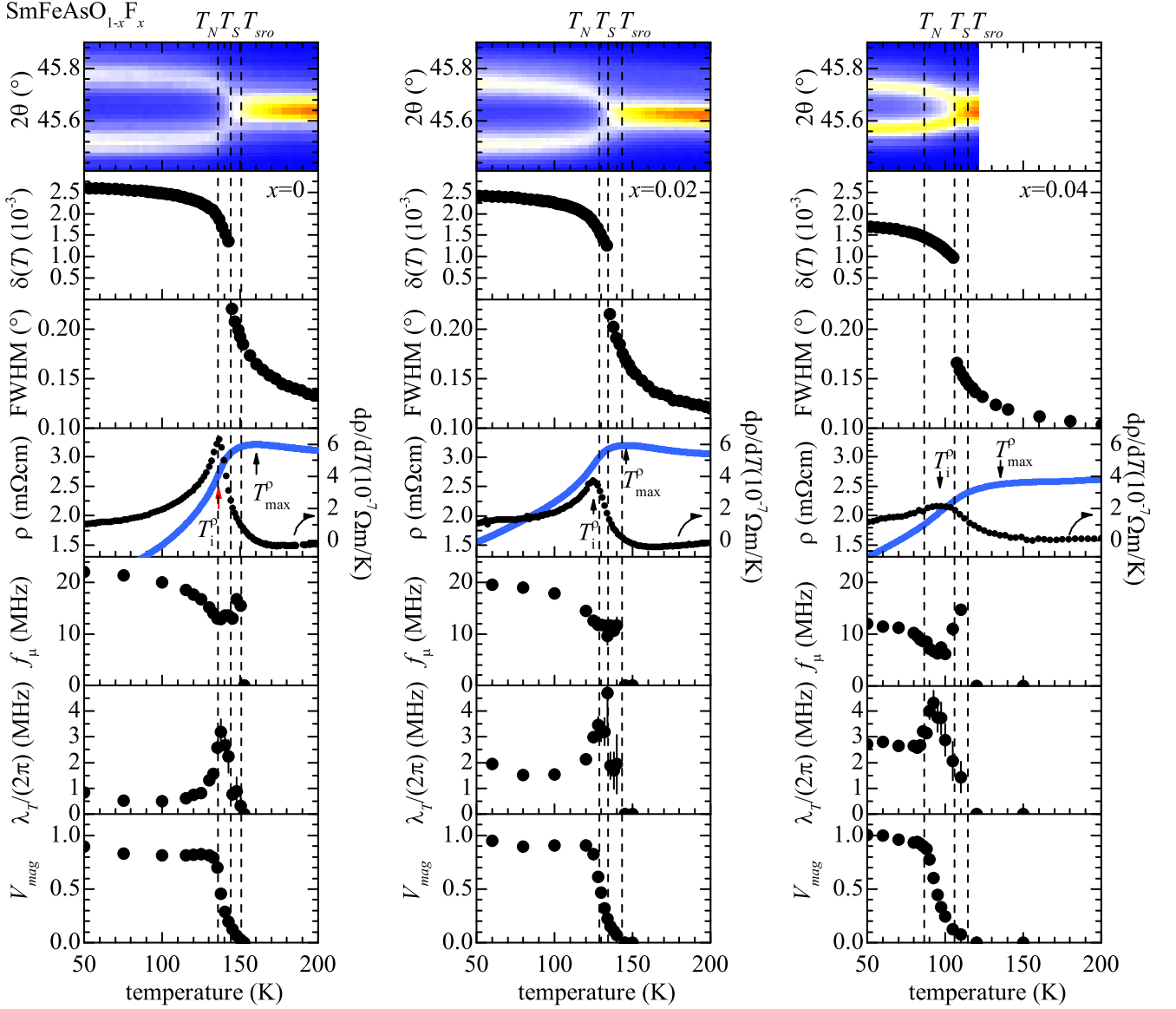}
\caption{A comparison of the X-ray diffraction patterns, orthorhombic distortion $\delta$ and the FWHM of the (2,2,0)$\rm_T$ Bragg peak, the electrical resistivity $\rho$ and its derivative, and the muon spin precession frequency $f_\mu$, transverse relaxation rate $\lambda_T/(2\pi)$ and magnetic volume fraction $V_{mag}$ of the Sm samples with a structural phase transition.
The vertical dashed lines indicate the phase transition temperatures.
Where applicable arrows indicate the maximum ($T_{max}^\rho$) and the inflection point ($T_i^\rho$) of $\rho(T)$.
This comparison shows that the phase transitions cause the anomalies of the resistivity, but no simple relationship exists between the phase transition temperatures and the shape, and position of the anomalies.
Except the color coding of the density plots, the $y$-axes scaling is identical for the different doping levels.}
\label{img.sm-comp}
\end{figure*}

\begin{table*}[htbp]
\begin{tabular}{c|ccccccccc}
$x$&$T_S$&$T_c$&$T_{sro}$&$T_N$&$T_{100}$&$T_N^{Sm}$&$T^\rho_{max}$&$T^\rho_i$\\
\hline
\hline
0.00&143.7+1-5&--&151(1)&135.3(2)&130(2)&5.2(1) ($c_p$), 5.6(3) ($\lambda_T$)&160&136\\
0.02&134.5+1-5&--&143(2)&128.6(10)&122(2)&4.7(1) ($\lambda_T$)&146&125\\
0.04&106.3+1-5&--&115(5)&86.7(22)&83(1)&4.7(1) ($\lambda_T$)&135&97\\
0.06&--&36.2 ($\rho(T)$)&33(1)&--&6(2)&4.3(1) ($c_p$), 4.4(2) ($\lambda_T$)&--&--\\
0.08&--&44.5(1) ($\rho(T)$)&--&--&--&3.5(10) ($\lambda_L$)$^\ast$&--&--\\
0.10&--&52.1(1) ($\rho(T)$), 50.6(4) ($\mu$SR)&--&--&--&3.5(10) ($\lambda_L$)$^\ast$&--&--
\end{tabular}
\caption{Phase transition temperatures for the structural transition ($T_{S}$, see Sec.~\ref{sec.sm-xrd}, p.~\pageref{sec.sm-xrd}), superconducting phase transition ($T_c$ obtained by $\mu$SR and electrical resistivity $\rho(T$$\leq$$T_c)$$\approx$0, see Sec.~\ref{sec.sm-tf}, p.~\pageref{sec.sm-tf} and \ref{sec.sm-rho}, p.~\pageref{sec.sm-rho}), onset of magnetic order ($T_{sro}$), bulk magnetic long range order ($T_N$), saturation of the magnetic volume fraction at $T$=$T_{100}$ ($T_N$, $T_{sro}$, $T_{100}$ obtained by $\mu$SR, see Sec.~\ref{sec.sm-fe}, p.~\pageref{sec.sm-fe}).
$T_N^{Sm}$ is the magnetic ordering transition temperature of the Sm moments obtained by specific heat measurements ($c_p$), from the $\mu$SR transverse relaxation rate ($\lambda_T$), or the $\mu$SR transverse relaxation rate ($\lambda_L$), see Sec.~\ref{sec.sm-sm}, p.~\pageref{sec.sm-sm}.
All temperatures are given in K.
$^\ast$The Sm transition temperatures determined from $\lambda_L(T)$ should not be considered as proof for a phase transition (see Sec.~\ref{sec.sm-sm}, p.~\pageref{sec.sm-sm} for details).}
\label{tab.sm}
\end{table*}

\paragraph*{Mixed phase} For $x$=0.06 we find magnetic short range order below $T_{sro}$=33(1)~K and superconductivity with $T_c$=32.0(1)~K.
Whether the two orders coexist microscopically or phase separated cannot be decided at this point---the relevant evidence are:
\begin{itemize}
\item bulk superconductivity (from $\chi(T)$\cite{panarina10}), and
\item bulk magnetic order (from $\mu$SR, see pp.~\pageref{img.sm-vmag}\textit{f}).
\end{itemize}
For microscopic coexistence both order parameters would be non-zero in the bulk and, possibly, we could observe a coupling of order parameters if $T_c$$<$$T_{sro}$\cite{fernandes10,wiesenmayer11,pratt09,marsik10}.
In case of phase separation the volume fractions should be coupled, i.e.,
one grows at the expense of the other\cite{aczel08,goko09,park09,julien09,bernhard09}.
We also have to consider nanoscopic phase separation that has been mentioned in recent $\mu$SR works\cite{sanna09,sanna10,shiroka11}.
In this case even $\mu$SR is not able to directly detect the phase separation.
The available experimental evidence are compatible with both situations!
Note, that concerning the question of nanoscopic phase separation vs.
microscopic coexistence, our experimental results are consistent with the results in Ref.~\onlinecite{sanna09}.
We merely point out that this issue cannot be resolved using the available experimental data.

\paragraph*{Relevance for a possible QCP} This ambiguous evidence leaves the question open whether a magnetic quantum critical point could occur in the phase diagram of the Sm system, as it is not clear whether a second or first order phase transition occurs between magnetic order and superconductivity as a function of doping.

\paragraph*{Electrical resistivity \& phase transitions} The relationship between the electrical resistivity and the structural, and magnetic phase transitions is illustrated in Fig.~\ref{img.sm-comp}.
Apparently, \emph{no simple} connection exists between the magnetic, and structural phase transitions and temperature dependence of the electrical resistivity.
The maximum of the resistivity at $T$=$T_{max}^\rho$ does not coincide with any phase transition studied in this work since it occurs at somewhat higher temperatures.
The maximum marks the onset of a reduction of the critical fluctuations prior to the structural/magnetic transitions, and thus indicates the development of short range structural/magnetic order.
The inflection point agrees with the measured $T_N$ for $x$$\leq$0.04 within errors.
The upon cooling reduced decrease of $\rho(T)$ signals an effective depletion of electronic states at the Fermi level which is expected for a true long-range ordered spin density wave.
The mismatch between $T_i^\rho$ and $T_N$ could be related to the width of the transition.

\section{C\lowercase{e}F\lowercase{e}A\lowercase{s}O$_{1-x}$F$_x$}
In this section we present the results of our electrical resistivity (Sec.~\ref{sec.ce-rho}), X-ray diffraction (Sec.~\ref{sec.ce-xrd}), muon spin relaxation measurements (Sec.~\ref{sec.ce-mu}), M\"ossbauer spectroscopy (Sec.~\ref{sec.ce-mos}), and magnetic susceptibility measurements of the materials that is both superconducting and magnetically ordered (Sec.~\ref{sec.ce-sc}).
Our experimental data show that this system is similar to the Sm system and in most parts we applied the same analysis strategies and draw similar conclusions as in the preceding section.
In Sec.~\ref{sec.ce-sum} we summarize our main results and present the electronic phase diagram of CeFeAsO$_{1-x}$F$_{x}$.
All phase transition temperatures are summarized in Tab.~\ref{tab.ce}.

\subsection{Electrical resistivity measurements}
CeFeAsO shows, similar to SmFeAsO, a characteristic cusp in the temperature dependence of the electrical resistivity $\rho(T)$ at $T_{max}^{\rho}$$\approx$150~K followed by an inflection point at $T_i^\rho$=135(2)~K in Fig.~\ref{img.ce-rho}.
For the doped compounds, this anomaly is not observed, presumably due to the higher doping level compared to the non-superconducting Sm-based compounds.
The suppression of the cusp is correlated with the reduction of $T_S$ and the low temperature limit of the orthorhombic distortion $\delta(T$$\to$$0)$ to values smaller than those we presented for SmFeAsO$_{0.96}$F$_{0.04}$.
Instead of the cusp, we observe a resistivity minimum at $T$$\approx$70~K.
Later we will show that $T_S$ is above, and $T_N$ close to the minimum of $\rho(T)$ (see Sec.~\ref{sec.dis}).
For $x$=0.048(3), $\rho(T)$ is similar to $x$=0.042(2).
However, the minimum is followed by a cusp at $T$$\approx$26~K.
At lower temperatures the resistivity drops only almost to zero which indicates spurious superconducting grains and not bulk superconductivity.

\label{sec.ce-rho}
\begin{figure}[htb]
\begin{center}
\includegraphics[width=8.6cm]{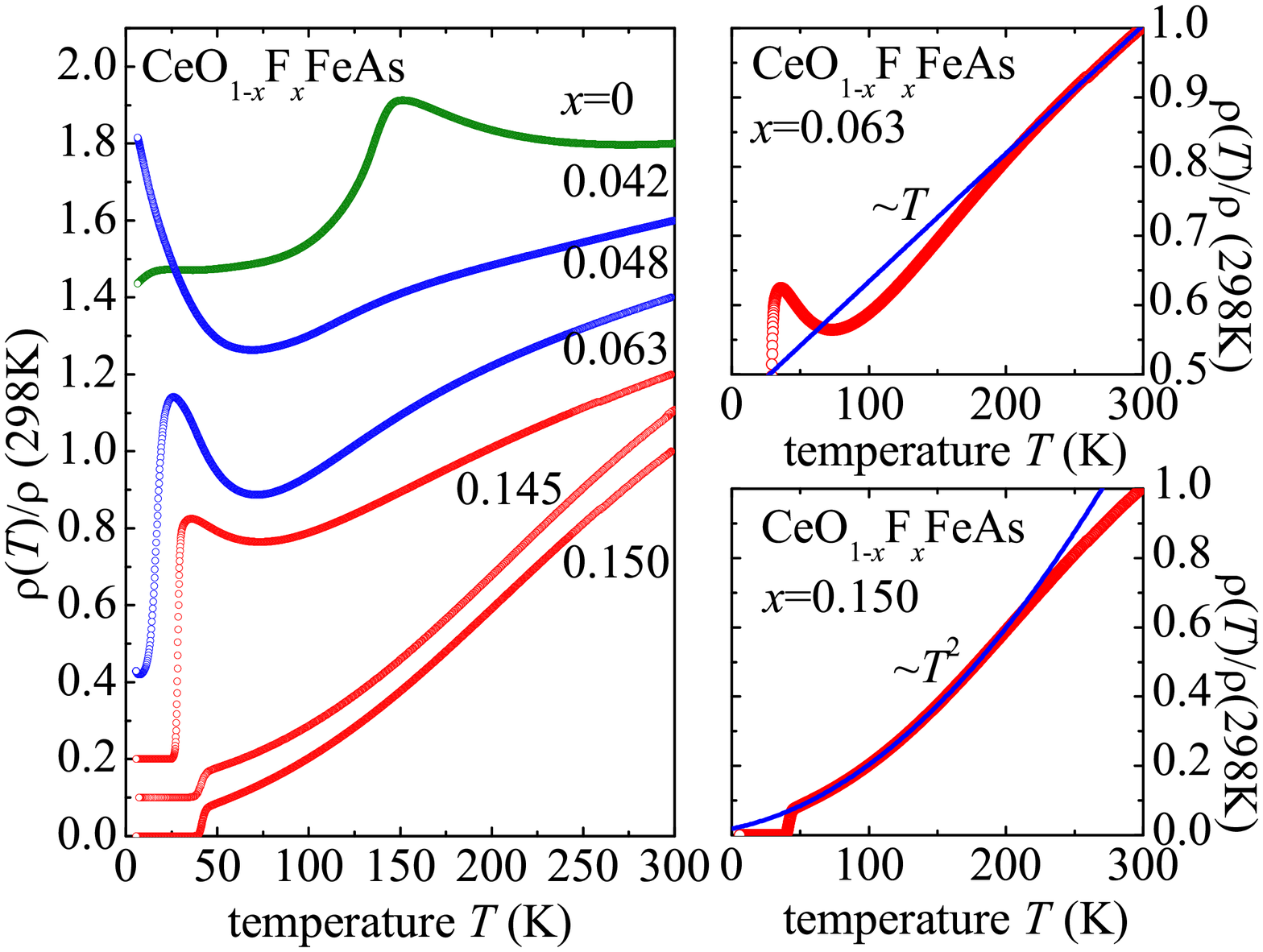}
\caption{Normalized resistivity as a function of temperature of CeFeAsO$_{1-x}$F$_{x}$.
(Blue and green) Red curves refer to (non-)superconducting compounds.
\textit{right:} b) the blue line is a linear fit above 200 K, below $T$=200 K resistivity differs from linearity similar to underdoped LaFeAsO$_{1-x}$F$_{x}$.
c) Quadratic temperature dependence of $\rho(T)$ below $T$$\approx$200~K in resemblance of overdoped LaFeAsO$_{1-x}$F$_{x}$.
Curves are shifted vertically for clarity.}
\label{img.ce-rho}
\end{center}
\end{figure}

We observe bulk superconductivity first for $x$=0.063(2) below $T_c$=29.5~K.
At higher temperatures $\rho(T)$ shows a cusp at $T$$\approx$35~K and a minimum at $T$$\approx$70~K much like for $x$=0.048(3).
Above $T$$\approx$240~K $\rho(T)$ becomes linear in $T$.
For $x$=0.145(20), and 0.150(20), we find superconductivity below $T_c$=43.8, 43.4~K, respectively.
At higher temperatures the resistivity shows neither a cusp nor a minimum but follows $\rho(T)$$\propto$$T^2$ for both materials.

In the superconducting regime, the resistivity of CeFeAsO$_{1-x}$F$_{x}$ shows similarities to LaFeAsO$_{1-x}$F$_{x}$, and SmFeAsO$_{1-x}$F$_{x}$.
In the optimally doped and overdoped regime ($x\geqslant 0.1$) LaFeAsO$_{1-x}$F$_{x}$ also shows linear $\rho(T)$ at high temperatures and $T^2$ behavior at low temperatures \cite{hess}.
In the underdoped regime of LaFeAsO$_{1-x}$F$_{x}$ ($x$$<$0.1), $\rho(T)$ is also proportional to temperature above T$\approx$200~K \cite{hess}.
At lower temperatures, $\rho(T)$ drops below the (linear) extrapolation of the high-tempearture linear behavior.
This resistivity drop, accompanied by a large Nernst effect anomaly, was interpreted as the signature of slow spin density wave fluctuations \cite{hess,kondrat11}.
The $x$=0.063(2) doped Ce sample shows the same feature in the resistivity which suggests a similar role of magnetic fluctuations.
In Sec.~\ref{sec.ce-fe} we will show that, contrary to the La system, these fluctuations slow down and eventually lead to magnetic order.

In the discussion in Sec.~\ref{sec.dis} we will compare the structural and magnetic phase transition as found by X-ray diffraction, $\mu$SR and M\"ossbauer spectroscopy to the features in the temperature dependence of the resistivity.

\subsection{Susceptibility measurements}
The temperature dependence of the magnetic susceptibility for $x$=0.063(2) in magnetic fields of 0.805(3)~mT, and 0.112(3)~mT cooled in field (fc) and in zero field (zfc) is shown in Fig.~\ref{img.ce-chi}.
Both fc and zfc measurements show strong diamagnetic signals in the superconducting state.
However, neither reaches 4$\pi\chi(T)$=$-1$ down to $T$$\approx$2~K.
The sizable fc and zfc diamagnetic susceptibilities indicate bulk superconductivity.
Later we present X-ray diffraction (via the jump thermal expansion coefficient) data that also support bulk superconductivity, and $\mu$SR data that is consistent with nearly 100\% superconductivity for $T_c$$\geq$$T$$>$20~K.
The granular, i.e., polycrystalline structure of the samples may cause the reduced shielding signal: 4$\pi\chi(T)$=$-1$ may only be observed if the weak Josephson-junctions between the grains provide superconducting paths for the Meissner shielding current across the whole sample, e.g., see Ref.~\onlinecite{klingeler10b}.
Hence, we conclude that the smallest applied field of about 0.1~mT, possible with our experimental setup already breaks some of the weak links.
However, the data clearly allow extracting $T_c$.
Note, that the onset of Ce magnetic order at $T_{N,\chi}^{Ce}$=2.7(1)~K is visible in the data by the slight upturn and cusp, too.
This is consistent with the measurements in Refs.~\onlinecite{sanna10,shiroka11}, and our $\mu$SR measurements (see below).
Also 4$\pi\chi(T)$$>$$-1$ is in agreement with results of \citet{sanna10} and \citet{shiroka11}
Based on $\mu$SR results, they also concluded bulk superconductivity in similarly doped materials with a comparable macroscopic susceptibility.
Important is also the almost constant susceptibility for $T$$<$20~K, it shows that the superconductivity is \textit{not}, and not even in parts, quenched by the low temperature Fe magnetic order present in this sample.


\subsection{X-ray diffraction measurements}
\label{sec.ce-xrd}
As for the Sm system, we study the structural phase transition of the Ce system using synchrotron X-ray diffraction.
We find that it is qualitatively identical in both systems.
The phase transition is also broad and preceded by lattice fluctuations.
Here we only summarize the features.
Please see Sec.~\ref{sec.sm-xrd}, p.~\pageref{sec.sm-xrd} for a detailed discussion of our analysis strategy.

\begin{figure}[htb]
\includegraphics[width=8.6cm]{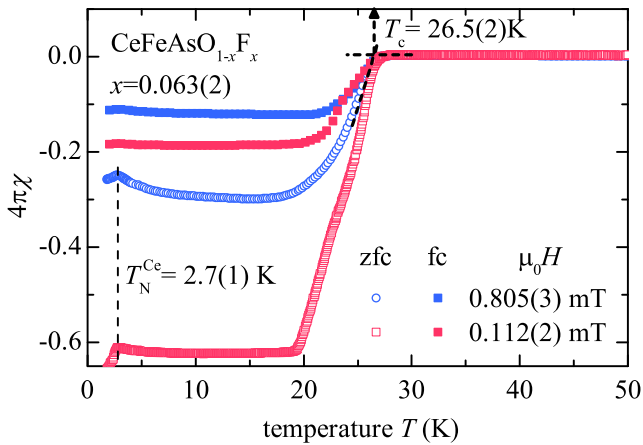}
\caption{The temperature dependence of the magnetic susceptibility 4$\pi\chi(T)$ of CeFeAsO$_{1-x}$F$_x$ with $x$=0.063(2).
Open symbols denote measurements upon heating after the sample was cooled in zero magnetic field (zfc) and closed symbols after cooling in the applied magnetic field (fc).}
\label{img.ce-chi}
\end{figure}

\begin{figure*}[htb]
\begin{minipage}{8.6cm}
\includegraphics[width=8.6cm]{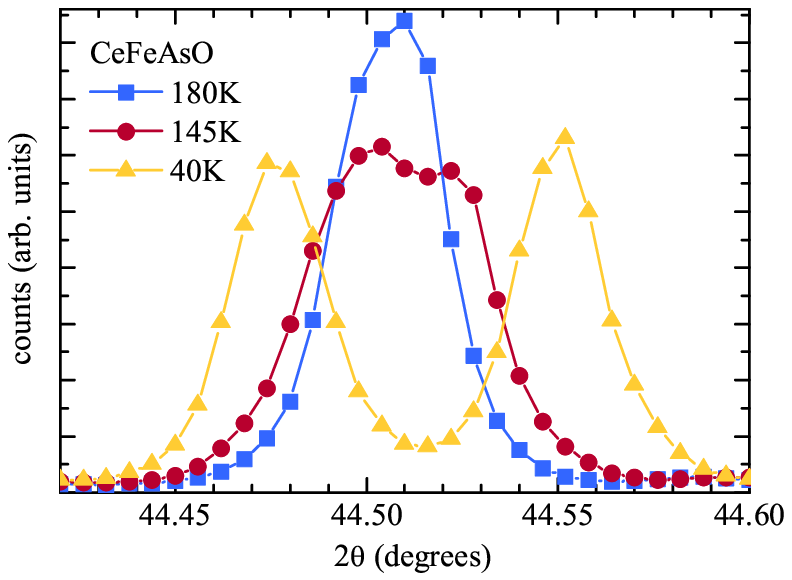}
\end{minipage}
\begin{minipage}{8.6cm}
\includegraphics[width=8.6cm]{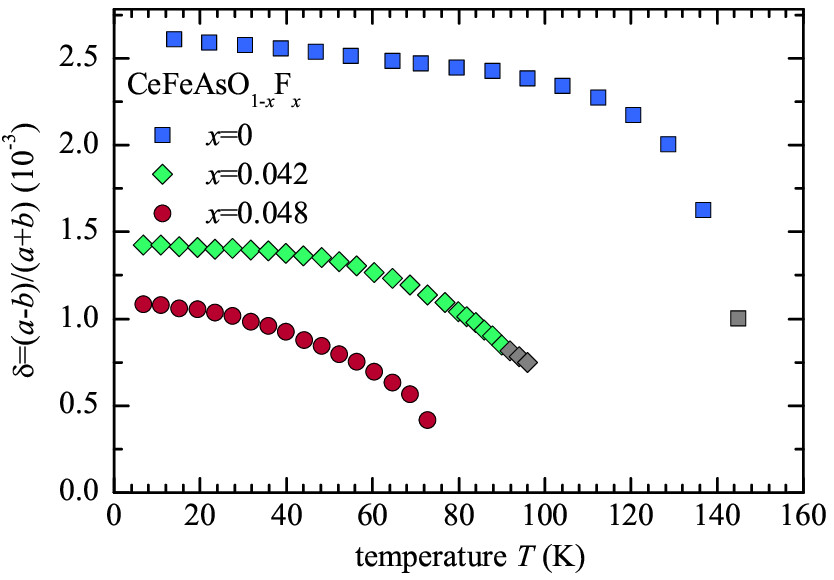}
\end{minipage}
\caption{\textit{Left}: representative X-ray diffracted patterns of the (2,2,0)$_{\rm T}$ tetragonal and the (4,0,0)$\rm_O$ and (0,4,0)$\rm_O$ orthorhombic Bragg peaks at different temperatures for CeFeAsO.
\textit{Right}: The orthorhombic distortion as a function of temperature and doping level of CeFeAsO$_{1-x}$F$_{x}$.
Gray data points indicate the temperature range for which $\delta(T)$ is convex (see Sec.~\ref{sec.sm-xrd}, p.~\pageref{sec.sm-xrd} for a detailed explanation).}
\label{img.ce-ortho}
\end{figure*}

\begin{figure}[htb]
\includegraphics[width=8.6cm]{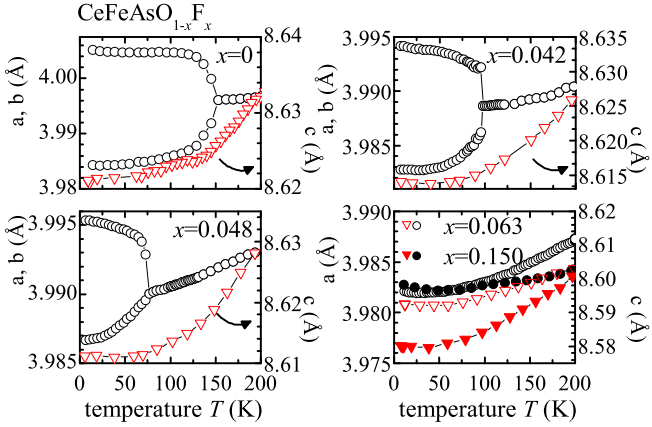}
\caption{The lattice constants as a function of temperature of CeFeAsO$_{1-x}$F$_{x}$.
The black circles refer to $a$ and $b$ lattice parameters and the red triangles to $c$.
Below the tetragonal to orthorhombic transition, $a$$\neq$$b$ are divided by $\sqrt{2}$ for comparison.}
\label{img.ce-lattice}
\end{figure}

\begin{figure}[htb]
\includegraphics[width=8.6cm]{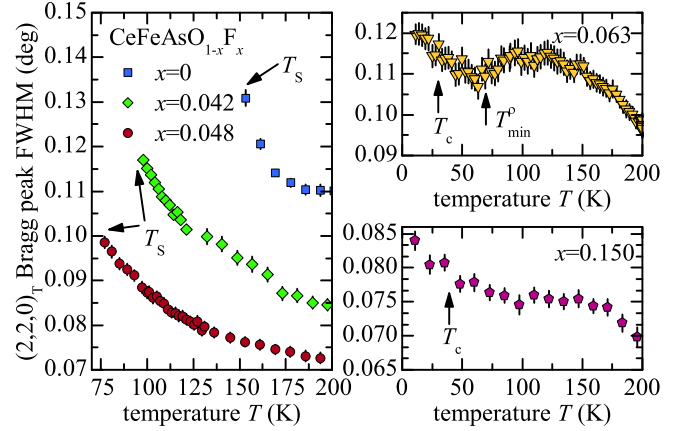}
\caption{FWHM of the (2,2,0)$_{\rm T}$ (left) above the structural transition, and (right) in the superconducting superconducting regime of CeFeAsO$_{1-x}$F$_x$ arrows indicate the structural and superconducting transition temperatures $T_S$ and $T_c$ and the minimum of the resistivity ($T_{min}^\rho$) for $x$=0.063(2).}
\label{img.ce-fwhm}
\end{figure}

Representative diffraction patterns of CeFeAsO, and the orthorhombic distortion are shown in Fig.~\ref{img.ce-ortho}.
We determined the lattice constants $a$, $b$, and $c$ in Fig.~\ref{img.ce-lattice} from the positions of the tetragonal (2,2,0)$\rm_T$ and (0,0,6)$\rm_T$, and the orthorhombic (4,0,0)$\rm_O$ and (0,4,0)$\rm_O$ Bragg peaks.
The temperature dependence the orthorhombic distortion in Fig.~\ref{img.ce-ortho}, the lattice constants in Fig.~\ref{img.ce-lattice}, and the FWHM of the tetragonal peak above the transition in Fig.~\ref{img.ce-fwhm} follow the same trends we reported for the Sm system.
The tetragonal peak also broadens significantly when we approach the structural transition from high temperatures.
We find $T_S$=148.9($+$2$-$8)~K, 97.0($+$1$-$7)~K, 74.7($+$1$-$5)~K for $x$=0, 0.042(2), and 0.048(3), respectively (the positive error is due to the temperature step of the measurement and the negative error is the temperature range for which $\delta(T)$ is convex).
The splitting is apparent in the raw data shown in Fig.~\ref{img.ce-ortho}.
It persists only for the non-superconducting compounds with $x$=0, 0.042(2), and 0.048(3), and is absent for the superconducting compounds with $x$=0.063(2),~and~0.150(20).


\begin{figure}[htb]
\includegraphics[width=8.6cm]{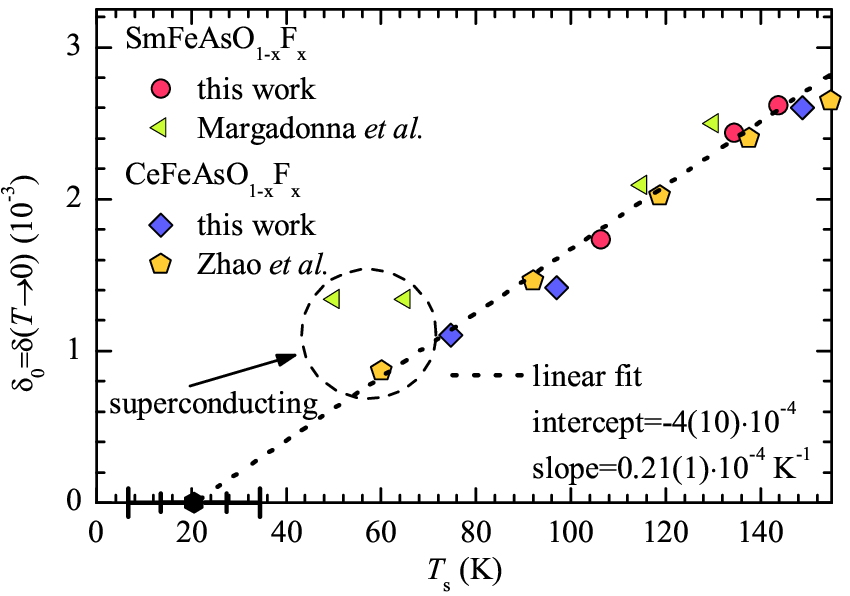}
\caption{The low temperature limit of the orthorhombic distortion $\delta(T$$\to$$0)$ as a function of the structural transition temperature $T_S$ for the Sm and the Ce system.
The dotted line is a linear fit and the fit parameters are given in the figure.
For comparison, we also show the data taken from the work of \citet{zhao08}, and \citet{margadonna}.
The marking on the $x$-axis indicates $T_S^0$ for which $\delta(T\to 0,T_S^0)=0$, and the bars indicate the error of $T_S^0$ (one, and two standard deviations).}
\label{img.deltats}
\end{figure}

At first glance, the structural transition temperature $T_S$ and the low temperature orthorhombic distortion $\delta(T$$\to$$0)$ are strongly reduced in the two doped CeFeAsO$_{1-x}$F$_x$ specimens as shown in Fig.~\ref{img.ce-ortho}.
However, Fig.~\ref{img.deltats} indicates that the reduction of $\delta(T$$\to$$0)$ is proportional to $T_S$, i.e.,
$\delta(T$$\to$$0,T_S)$$\propto$$T_S$ and that $\delta(T$$\to$$0,T_S)$ is quantitatively identical for the Sm and the Ce system.
In Fig.~\ref{img.deltats} we also show the data obtained by \citet{margadonna} for the Sm system, and \citet{zhao08} for the Ce system.
For the non-superconducting Sm and Ce materials $\delta(T$$\to$$0,T_S)$ follows the same linear dependence on $T_S$.
For the two superconducting specimens studied by \citet{margadonna} the orthorhombic distortion is higher than expected from the linear $\delta(T$$\to$$0,T_S)$ at lower doping levels, whereas the superconducting specimen studied by \citet{zhao08} follows the linear behavior.
At a quantum critical point the order parameter and the transition temperature should simultaneously become zero.
However, a linear extrapolation of $\delta(T$$\to$$0,T_S)$, shown in Fig.~\ref{img.deltats} indicates that transition temperature and order parameter \textit{do not} simultaneously become zero (excluding the two outlying data points taken from \citet{margadonna}).
On the other hand, the two outlying data points suggest that the decrease of the orthorhombic distortion slows down, while $T_S$ rapidly vanishes in the superconducting regime.
Also in this case, both parameters do not simultaneously vanish.
However, a structural QCP (or the related nematic QCP \cite{fang08}) may be recovered by a ``back bending'' of the structural phase transition line, which has been reported for doped BaFe$_2$As$_2$ \cite{fernandes10,nandi10,moon10}.

The analysis of the orthorhombic distortion with the general power law $\delta(T)$$\propto$$(1-(T/T_S)^\alpha)^\beta$ is, just as for the Sm system, not feasible due to the broadened transition.
But we can compare the temperature dependence of the normalized orthorhombic distortion with 2D-Ising and 3D-Ising order parameters (see Fig.~\ref{img.norm-ortho}).
Qualitatively, this indicates that the order parameter universality class changes from 2D-Ising ($x$=0) to 3D-Ising ($x$=0.048(3)) via an intermediate temperature dependence ($x$=0.042(2)) similar to the Sm system (see Fig.~\ref{img.sm-norm-ortho}).

For $x$=0.063(2), and 0.150(20) we observe an upturn of the $a$ lattice constant below $T$$\approx$$T_c$ in Fig.~\ref{img.ce-a} (\textit{cf}. Sec.~\ref{sec.sm-xrd}).
We can exclude a magnetic origin of the upturn (see next sections): for $x$=0.063(2) a sizable magnetic volume fraction develops only at lower temperatures, and $x$=0.150(20) is non-magnetic.
The expansion of the $a$ axis shown in Fig.~\ref{img.ce-a} is most likely due to superconductivity, and suggests $\Delta \alpha_a$$\propto$$dT_c/dp_a <0$ (the effect of superconductivity on the $c$ lattice constant could not be resolved).
More importantly, this finding for $x$=0.063(2) supports that a substantial part of the sample becomes superconducting.


\begin{figure}[htb]
\includegraphics[width=8.6cm]{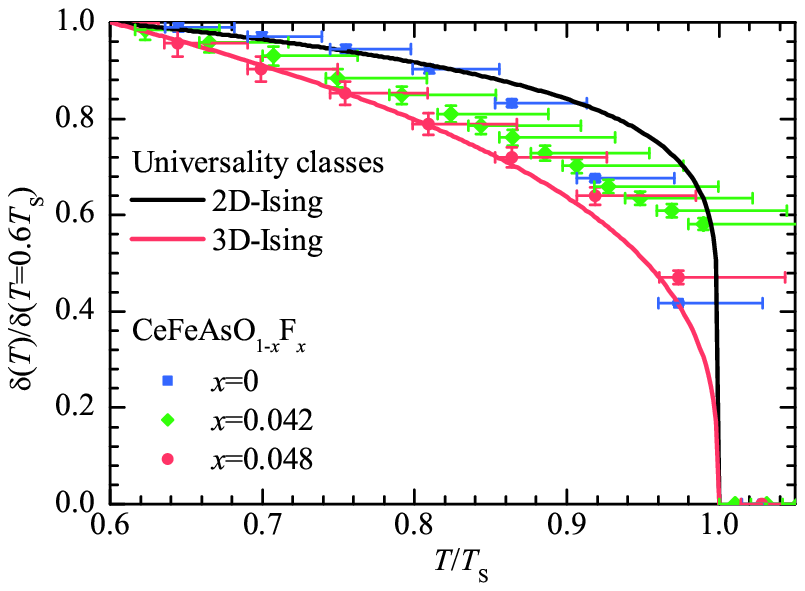}
\caption{The orthorhombic distortion $\delta(T)/\delta(T$$=$$0.6T_S)$ normalized to the value at $T$$=$$0.6T_S$ as a function of reduced temperature $T/T_S$ for CeFeAsO$_{1-x}$F$_x$.
The Sm and the Ce compounds show similar doping and temperature dependencies (compare with Fig.~\ref{img.sm-ortho}, p.~\pageref{img.sm-ortho}).
To illustrate the change of the universality class of the order parameter we also show the expected temperature dependence for an order parameter of the 2D-Ising ($\propto$$(1-T/T_S)^{0.125}$), and the 3D-Ising type ($\propto$$(1-T/T_S)^{0.325}$).
Due to the large asymmetric error of $T_S$ we could not determine the universality classes unambiguously.
Within the uncertainty of the analysis $x$=0.00 belongs to the 2D-Ising universality class, whereas $x$=0.048(3) belongs to the 3D-Ising class and $x$=0.042(2) lies between the two.
The change of dimensionality is more pronounced in the Ce than in the Sm system.
This is possibly due to the lower transition temperatures of the Ce materials.}
\label{img.norm-ortho}
\end{figure}

\begin{figure}[htb]
\includegraphics[width=8.6cm]{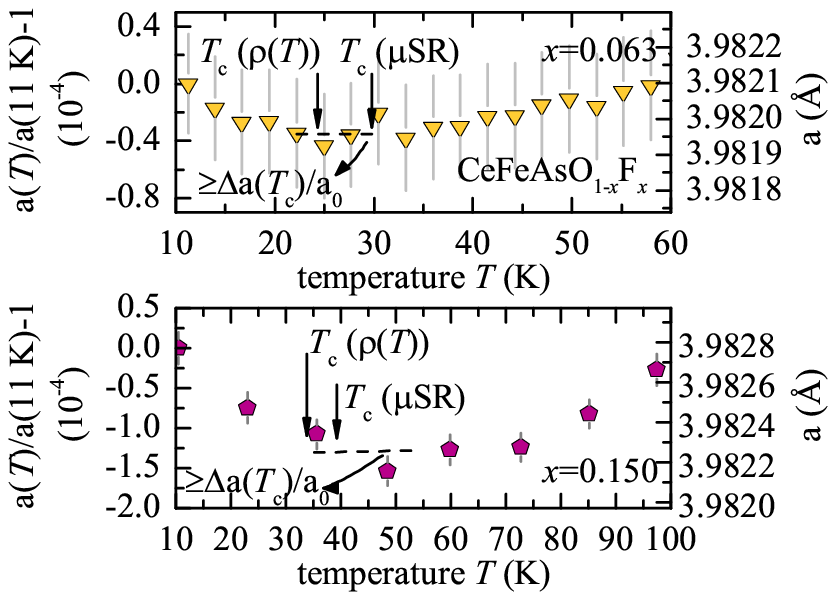}
\caption{The low temperature upturn of the $a$ lattice constant below $T$$\approx$$T_c$ of CeFeAsO$_{1-x}$F$_x$ with $x$=0.063(2) (top), and $x$=0.150(20) (bottom).
This is similar to the upturn of the $a$ and $c$ lattice constants of SmFeAsO$_{1-x}$F$_x$ with $x$=0.048(3) (see Fig.\ref{img.sm-a-c}, p.~\pageref{img.sm-a-c}).
It is due to the pressure dependence of $T_c$.
The change of slope from negative ($T$$<$$T_c$) to positive ($T$$>$$T_c$) suggests that $T_c$ would decrease under in-plane pressure.
We were not able to resolve the effect of superconductivity on the $c$ lattice constant with our X-ray diffraction experiments.
Because the dependence of $T_c$ on hydrostatic pressure depends both on in-plane (along $a$) and out-of-plane (along $c$) pressure we can draw no conclusion as to whether $T_c$ should increase or decrease under hydrostatic pressure.}
\label{img.ce-a}
\end{figure}

The FWHM (full-width-at-half-maximum) of the (2,2,0)$\rm_T$ peak in Fig.~\ref{img.ce-fwhm} for $x$=0.063(2) strongly increases at low temperatures.
It reaches a broad maximum at $T$$\approx$100~K.
This coincides with an inflection point in the electrical resistivity in Fig.~\ref{img.ce-rho}.
The minimum of $\rho(T)$ at $T$$\approx$70~K roughly coincides with the minimum of the FWHM of the (2,2,0)$\rm_T$ Bragg peak.
For $x$=0.150(20) the FWHM shows no sharp anomalies, and also $\rho(T)$ shows no pronounced features.
Similar to the Sm system, the absolute value of the FWHM is reduced by $\approx$50\% compared to $x$=0.063(2).
Contrary to the Sm system, we observe no anomalies at $T_c$ for both $x$=0.063(2), and 0.150(20) which speaks against suppression of lattice fluctuations by superconductivity.

The question arises whether the broadening of the tetragonal Bragg peak for $x$=0.063(2), and 0.150(20) in Fig.~\ref{img.ce-fwhm} could be an unresolved orthorhombic splitting.
The minimum and the following increase of $x$=0.063(2) for $T$$<$70~K could indicate an unresolved splitting (see Fig.~\ref{img.ce-fwhm}).
To answer this question, we discuss the resolution limit of our X-ray diffraction experiment below.
It is mainly limited by the intrinsic width of the (2,2,0)$\rm_T$ Bragg peak.
For $T$=70~K, the (2,2,0)$\rm_T$ Bragg peak is at 2$\theta$=45.02$^\circ$ with a FWHM of $\approx$0.1088$^\circ$.
The angle resolution of our experiment is $\approx$0.013$^\circ$ and the error of each intensity value is $\approx$3\%.
To estimate the resolution limit we simulate a series of diffraction patterns of a hypothethical orthorhombic phase with two Gaussian Bragg peaks ((0,4,0)$\rm_O$ and (4,0,0)$\rm_O$) at 2$\theta_\pm$=45.02$^\circ$$\pm$$0.5 \Delta$ with the FWHM of the tetragonal peak (0.1088$^\circ$) and the above angle resolution.
We fit each simulated pattern with a \textit{single} Gaussian Bragg peak.
It turns out, that for an orthorhombic distortion $\delta$=0.40$\cdot$10$^{-3}$ ($\Delta$=0.0381$^\circ$) the simulated diffraction pattern is indistinguishable from a single Bragg peak with a FWHM of $\approx$0.12$^\circ$, which corresponds to the experimental FWHM at $T$$\approx$10~K.
Therefore, $\delta$$\leq$0.40$\cdot$10$^{-3}$ is the upper limit of the orthorhombic distortion that is still consistent with the experimental data.
To unambiguously resolve the orthorhombic distortion, its error should be smaller than the orthorhombic distortion itself.
With the given FWHM and experimental resolution, this is satisfied for an orthorhombic distortion larger than $\delta$$\approx$1$\cdot$10$^{-3}$.

\label{para.resolution} Several studies of ferropnictides suggest\cite{delaCruz10,jesche08,prokes10} that the orthorhombic distortion is proportional to the ordered Fe magnetic moment.
A linear extrapolation using the orthorhombic distortion and the M\"ossbauer Fe hyperfine field, which is directly proportional to the ordered Fe moment (see Sec.~\ref{sec.ce-mos}) suggests an orthorhombic distortion of $\approx$0.5(1)$10^{-3}$ for $x$=0.063(2) at lowest temperatures.
This value is still compatible with the data but our experiment would not be able to resolve this distortion.

In summary, the resolution limit of our diffraction experiment is $\delta$$\approx$1$\cdot$10$^{-3}$ and the largest orthorhombic distortion that is still compatible with the data is $\delta$$\approx$0.4$\cdot$10$^{-3}$.
Considering the above estimates of $\delta(T\to 0)$ we cannot exclude a structural transition for $x$=0.063(2).
However, in that case, $\delta(T\to 0)$ should be smaller than $\approx$0.4$\cdot$10$^{-3}$.

\subsection{Muon spin relaxation measurements}
\label{sec.ce-mu}


\subsubsection{Fe magnetic order}
\label{sec.ce-fe}
\begin{figure*}[htb]
\includegraphics[width=\textwidth]{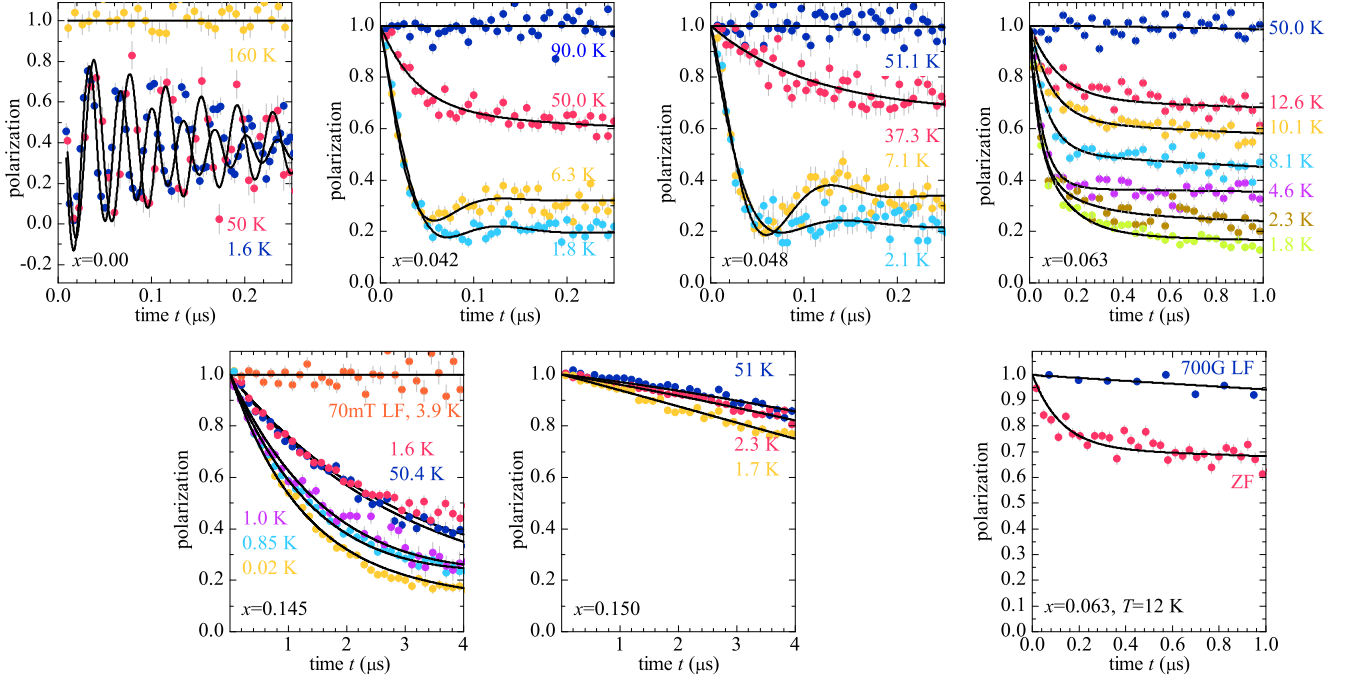}
\caption{Zero and longitudinal field muon spin relaxation time spectra of CeFeAsO$_{1-x}$F$_x$ and (bottom right) a weak longitudinal field decoupling time spectrum for $x$=0.063(2).
Note the different time scales for the various plots.}
\label{img.ce-histo}
\end{figure*}

\begin{figure}[htb]
\includegraphics[height=5cm]{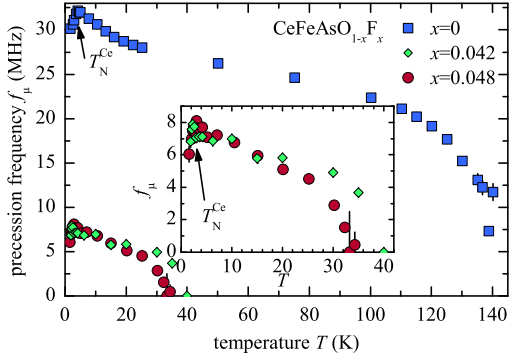}
\caption{The muon spin precession frequency $f_\mu$ is proportional to the AFM order parameter and indicates long range magnetic order.}
\label{fig.freq}
\end{figure}

\begin{figure*}[htb]
\includegraphics[width=0.9\textwidth]{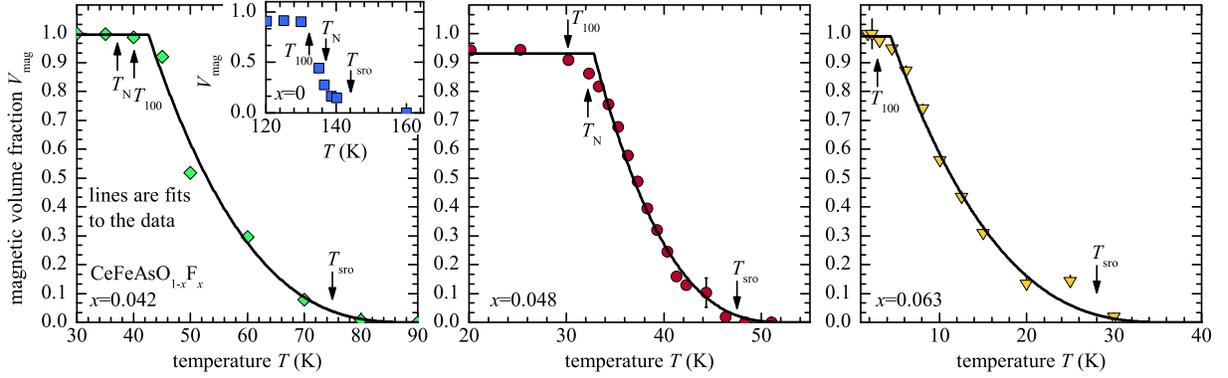}
\caption{The magnetic volume fraction $V_{mag}$ with fits according to power law distributed transition temperatures.
A small but non-zero $V_{mag}$ indicates magnetic short range order.}
\label{fig.vmag}
\end{figure*}

Zero field $\mu$SR data for $x$$\leq$0.063(2) are best described by a model for commensurate magnetic order:
\begin{equation}
\begin{split}
A(t)= &V_{mag}\bigl[(1-a_1) a_2 (\frac{2}{3}e^{-\lambda_T t} \cos\omega t +\frac{1}{3}e^{-\lambda_Lt})\bigr.
\\
&+ \bigl.(1-a_2) e^{-\lambda_{Ce}t}\bigr]\\
&+ (1-V_{mag})G(t,\sigma)e^{-\lambda_{nm}}.
\end{split}
\label{eq.zf}
\end{equation}
The first term models the commensurate Fe magnetic order.
The second term is non-zero ($a_2$$<$1) for the doped materials only below $\approx$4~K where Ce magnetic order causes an additional fast relaxation with rate $\lambda_{Ce}$ due to magnetic fluctuations evident from the reduction of $P(t)$ below 1/3.
The last term models the paramagnetic phase with weak Gaussian-Kubo-Toyabe relaxation $G(t,\sigma)$, and magnetic fluctuations that cause relaxation with rate $\lambda_{nm}$ \cite{hayano1979}.
The fluctuations close to the magnetic transition cause part of the signal to relax during the dead time of the detectors, this is modeled by the factor $a_1$ that is only non-zero and of the order of a few percent close to the magnetic transition.

When we cool the non-superconducting compounds with $x$$\leq$0.048(3) the magnetic order manifests in the $\mu$SR data in two steps: First magnetic short range order develops at $T_{sro}$=145(5)~K, 75(3)~K, 49(1)~K and the magnetically ordered phase volume $V_{mag}$ starts to increase gradually until it reaches 100\% at $T_{100}$=133(2)~K, 40(2)~K, 30.0(5)~K for $x$=0,0.042(2), and 0.048(3), respectively.
This can been seen by the absence of an oscillation and the fast exponential relaxation with rate $\lambda_T\sim20$~MHz of the muon spin polarization $A(t)$ for $t<0.5$~$\mu$s that reaches a signal fraction of $\approx2/3$ when the material is fully ordered (see Fig.~\ref{img.ce-histo}).
This indicates that magnetically ordered clusters with short coherence lengths form throughout the material below $T_{sro}$ that saturate at $T_{100}$.
We observe coherent long range magnetic order with a well defined order parameter, the muon spin precession frequency $f_{\mu}$ only close to $T_{100}$ with $T_N$=137(2)~K, 37(1)~K, 32.0(5)~K for $x$=0, 0.042(2), and 0.048(3), respectively.
This resembles the broadened magnetic phase transition of SmFeAsO$_{1-x}$F$_x$ that we reported in Sec.~\ref{sec.sm-fe}.
However, in the Ce system, we find no coherent precession signal above $T_{100}$ which suggests that the magnetically ordered clusters are too small or too disordered for long range magnetic order.
$f_\mu(T)$ shown in Fig.~\ref{fig.freq} is approximately four times smaller in the doped materials, similar to the observations\cite{drew09,sanna09} in SmFeAsO$_{1-x}$F$_x$ in Sec.~\ref{sec.sm-fe}.
This is different than in LaFeAsO$_{1-x}$F$_x$ where $f_\mu$ is only reduced by 10\% before superconductivity is induced by doping\cite{luetkens09}.

%
The width of the transition $\Delta T$=$T_{sro}$-$T_{100}$ in the undoped material is 6(2)~K and increases drastically in the doped materials up to 32(2)~K for $x$=0.042(2).
This broadening is not symmetric, $V_{mag}(T)$ follows the same shape that we reported for the Sm system (see Fig.~\ref{fig.vmag}, and Sec.~\ref{sec.sm-fe}).
To quantify the broadening we model the $V_{mag}(T)$ with the distribution of ordering temperatures (\ref{eq.distribution}) and (\ref{eq.vmag}).
Best fits are shown in Fig.~\ref{fig.vmag}.
Optimized parameters are $T_0$=90(8)~K, 53(4)~K, 37(6)~K, and $\sigma$=36(8)~K, 16(4)~K, 25(6)~K for $x$=0.042(2), 0.048(3), and 0.063(2), respectively ($\sigma$ indicates the width of the transition, $V_{mag}(T_0-\sigma)$=0.5).
The exponent $b$=1.8(7) is independent of doping.
For the Sm system we obtain a larger exponent $b$=4.0(5), interestingly this value is close to what \citet{berche98} determined for the McCoy-Wu model: $V_{mag}(T)$$\propto$$T^{4.1}$.
Within this work, we did not determine $V_{mag}(T)$ accurate enough close to the onset of magnetic order to accurately determine the exponent $b$.
In particular, we cannot exclude the possibility of exponential decrease of $V_{mag}(T)$ towards high temperatures, which would point to statistically rare unperturbed domains (for a review see Ref.~\onlinecite{tvoijta06}).
On the other hand, the exponent determined by our analysis of $V_{mag}(T)$ with a power law may be related to the type of disorder or the underlying magnetic interactions.

We also find a broad magnetic transition for $x$=0.063(2), as can be seen by a non-zero $V_{mag}$ below $T_{sro}=28(1)$~K in Fig.~\ref{fig.vmag}.
But, corresponding compound of the Sm system, we detect no coherent muon spin precession which indicates that this is short range order.
To exclude that the fast relaxation for $t<0.5$~$\mu$s in zero magnetic field is due to magnetic fluctuations we measured in an applied longitudinal magnetic field (LF).
It is well known, that a small LF will not have a significant effect on the muon spin relaxation rate if it is solely due to magnetic fluctuations.
Hence, the decoupling in LF seen in Fig.~\ref{img.ce-histo} is a clear indication for predominantly static magnetic order that develops in the material below $T_{sro}=28(1)$~K.

For $x$=0.145(20), we find, for a paramagnetic material, an unexpectedly fast relaxation of the muon spin polarization shown in Fig.~\ref{img.ce-histo}---but it is $\approx$100 times smaller than for any of the magnetically ordered materials.
This relaxation is temperature independent between 50~K and 1.6~K and can be decoupled in a longitudinal magnetic field of 70~mT, see Fig.~\ref{img.ce-histo}.
This indicates that it is caused by dilute magnetic impurities.
This is different from the relaxation in the superconducting Sm compounds where magnetic fluctuations of the Sm 4$f$ moments dominate (see Fig.~\ref{img.sm-laml}, p.~\pageref{img.sm-laml}).
This is consistent with the $\approx$5 wt.\% of Fe$_2$As impurities detected by M\"ossbauer spectroscopy (see next section)

For $x$=0.150(20) we find a weak relaxation of Gauss-Kubo-Toyabe\cite{hayano1979} form, typical for a non-magnetic material.
Only at lowest temperatures we observe an additional exponential relaxation, most likely due to the slowing down of Ce magnetic moment fluctuations, see Fig.~\ref{img.ce-histo}.

\subsubsection{Ce magnetic order}
\label{sec.ce}
In the undoped material the onset of long range Ce order is marked by a cusp in $f_{\mu}(T)$.
This is also visible for $x$=0.042(2), and 0.048(3) but less pronounced.
This suggests long range Ce magnetic order below $T_N^{Ce}$=4.4(3), 3.5(10), 3.5(10)~K for $x$=0, 0.042(2), and 0.048(3), respectively.
The larger error for the doped material stems from the large relaxation rate $\lambda_T$ which increases the error of $f_{\mu}$ and masks the onset of the Ce order.
It is accompanied by an extra signal fraction that is fully dynamic, i.e.,
due to rapid magnetic fluctuations ($a_2$$<$1 in (\ref{eq.zf})).
It can be seen in the time spectra in Fig.~\ref{img.ce-histo} by the lowering of the polarization for $t$$>$0.25~$\mu$s below 1/3 expected for static magnetic order.

For higher F content, long range Fe order is suppressed and the Ce order is weakened.
For $x$=0.063(2), the Ce order is marked by aforementioned dynamic signal fraction at $T_N^{Ce}$=2.5(10)~K which is accompanied by a cusp in the static relaxation rate $\lambda_T$, i.e.,
the width of the local magnetic field distribution.
For $x$=0.145(20) $T_N^{Ce}$=1.0(2)~K is further suppressed.
Above $T$=1~K impurities dominate the muon spin relaxation (see previous section).
Below $T=1$~K the relaxation rate increases by a factor of 3, marking the onset of Ce magnetic order.
For $x$=0.150(20) the relaxation rate is further suppressed to 0.01~$\mu$s$^{-1}$ which is of the order of the relaxation rate caused by randomly oriented nuclear magnetic dipoles.
This rate increases to 0.03~$\mu$s$^{-1}$ at $T$$\approx$2~K,
which is too small to be due to magnetic order.
This means that we do not observe static magnetic order of the Ce subsystem for the $x$=0.150(20) sample down to the lowest temperature of $T$=1.7~K measured in this experiment.

In summary, we observe no long range Ce magnetic order in the absence of Fe magnetic long range order.
The magnetic ordering of the Ce subsystem may stem from RKKY interaction between the localized Ce-4$f$ moments mediated by the Fe conduction band electrons.
\citet{akbari11} showed that the RKKY interaction is sizable in the ferropnictides and that doping, Fermi surface topology, and the SDW ordering of the Fe moments influence the RKKY interaction of local moments.
Additionally, disorder induced smearing of the conduction bands may also reduce the RKKY interaction strength and lead to a destruction of long range correlations and a reduction of the ordering temperature.

\subsubsection{Superconductivity}
\label{sec.ce-sc}
\begin{figure}[htb]
\includegraphics[width=8.6cm]{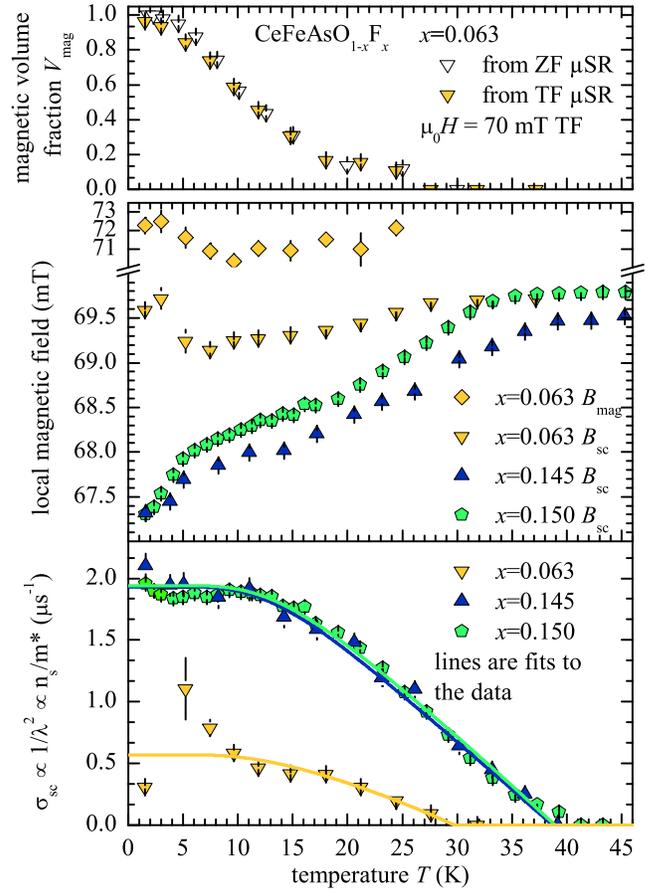}
\caption{The temperature dependence of the parameters of the SC phase, \textit{top} the volume fraction of the short range order of $x$=0.063(2), \textit{middle}: the internal field that is reduced by the diamagnetic response, \textit{bottom}: the SC order parameter $\sigma_{sc}$$\propto$$\lambda^{-1}$$\propto$$n_s/m^\ast$.
}
\label{img.ce-sc}
\end{figure}

As for the Sm system we use transverse magnetic field $\mu$SR to determine the magnetic penetration depth.
To account for the magnetic order for $x$=0.063(2), we introduce a corresponding signal to the fit function of the $\mu$SR time spectra:
\begin{equation}
\begin{split}
A(t)=&V_{mag} e^{-\lambda_{mag}} \cos(\gamma_{\mu} B_{mag} t)\\
&+(1-V_{mag}) e^{-\lambda_{tf} t}e^{-\frac{1}{2}(\sigma t)^2} \cos(\gamma_{\mu} B_{sc} t + \phi_0)
\end{split}
\end{equation}
The first term models the magnetically ordered phase volume, and the second term describes the  (purely) superconducting phase volume (\textit{cf}. Sec.~\ref{sec.sm-tf})
The close agreement between $V_{mag}(T)$ obtained from the weak transverse and zero field $\mu$SR in Fig.~\ref{img.ce-sc} indicate that the description of the $\mu$SR spectra is consistent.
$\sigma_{sc}(T)$=$\sqrt{\sigma(T)^2-\sigma_{nm}^2}$ is a measure of the superfluid density, or the inverse penetration depth $\sigma_{sc}$$\propto$$1/\lambda^2$$\propto$$ n_s/m^\ast$ \cite{brandt88}.
The relaxation rate $\lambda_{tf}$ accounts for relaxation due to static and dynamic magnetism in addition to the relaxation due to the flux-line lattice.

For $x$=0.063(2), the diamagnetic reduction of the local magnetic field, and the additional Gaussian relaxation shown in the middle and bottom panel of Fig.~\ref{img.ce-sc} indicate bulk superconductivity with $T_c$=29.8(2).
Close to $T_c$, the magnetic volume fraction $V_{mag}(T)$ does not grow larger than $\approx$10\% for $T$$>$17~K, see top panel of Fig.~\ref{img.ce-sc}.

For $x$=0.145(20), static magnetism, due to dilute magnetic impurities is already present above for $T$$>$$T_c$ and in zero field the relaxation rate does not change between 50~K and 1.6~K (see Fig.~\ref{img.ce-histo}, p.~\pageref{img.ce-histo}).
In the applied transverse field $\mu_0H$=70~mT, $\lambda_{tf}(T)$ increases linearly from 0.7~$\mu$s$^{-1}$ at $T$=50~K to 1~$\mu$s$^{-1}$ at $T$=3~K.
For $x$=0.150(20), the transverse relaxation can be described fully by setting $V_{mag}$=0.
This indicates a homogeneous superconducting phase with dilute magnetic impurities.
The bulk superconductivity is confirmed by the reduction of $B_{scc}$ and the increase of $\sigma_{sc}$ for $T<T_c$=39.0(9).

For $x$=0.150(20), in addition to the Gaussian relaxation we observe a temperature independent slow relaxation with $\lambda_{tf}(T)$=0.05(1)~$\mu$s$^{-1}$ due to fluctuations (see previous sections).
They are enhanced by applied magnetic fields $>$0.1~T and also present above $T_c$.
The reduction of $B_{sc}$ and the increase of $\sigma_{sc}$ at $T_c$=39.0(9) indicate bulk superconductivity, as above.

The saturation of $\sigma_{sc}$ for $x$=0.2, and 0.150(20) indicates a low density of state at the Fermi level characteristic for a nodeless gap.
A gap with nodes, as e.g.
for extended s-wave symmetry would lead to a linear decrease of $\lambda$ at low temperatures \cite{vorontsov09}.
For all three superconducting materials $\sigma_{sc}$$\propto$$1/\lambda^2$$\propto$$n_s$ has been analyzed using the temperature dependence of $1/\lambda^2$ assuming a nodeless s-wave gap $\Delta(T)$ \cite{tinkham2004,*carrington03}.
Fits are shown in Fig.~\ref{img.ce-sc}.
Best agreement has been found with the fit parameters shown in Tab.~\ref{tab.sc} along with the in-plane magnetic penetration depth $\lambda_{ab}(0)$.
This closely agrees with the roughly linear scaling of $T_c$ with the superfluid density $n_s$$\propto$$1/\lambda^2$ observed in many cuprate and ferropnictide superconductors (see Fig.~\ref{fig.uemura}).
Since we performed our measurements on powder samples in which vortex lattice disorder can result in an additional damping of the $\mu$SR signal, which in principal cannot
be disentangled from $\sigma_{sc}$, the values listed in Tab.~\ref{tab.sc} have to be regarded as lower limits of the penetration depth.

%
\begin{table}[htb]
\small
\begin{tabular}{c|ccccc} 
$x$&$T_c$(K)&$\Delta(0)$(meV)&$\frac{2\Delta(0)}{k_BT_c}$&$\sigma_{sc}(0)$($\mu$s$^{-1}$)&$\lambda_{ab}(0)$(nm)\\
	\hline
	\hline
	0.063(2)&29.8(2)&4.5(1)&3.5(1)&0.54(1)&339(5)\\
	0.145(20)&39.0(9)&5.2(2)&3.1(1)&1.94(4)&179(5)\\
	0.150(20)&38.7(1)&5.06(2)&3.04(1)&1.93(1)&179(1)
\end{tabular}
\caption{Best fit parameters of $\sigma_{sc}(T)\propto1/\lambda(T)^2$ for a nodeless s-wave gap.}
\label{tab.sc}
\end{table}

\subsection{M\"ossbauer spectroscopy measurements}
\label{sec.ce-mos}
\begin{figure*}[htb]
\begin{minipage}{\textwidth}
\includegraphics[height=5cm]{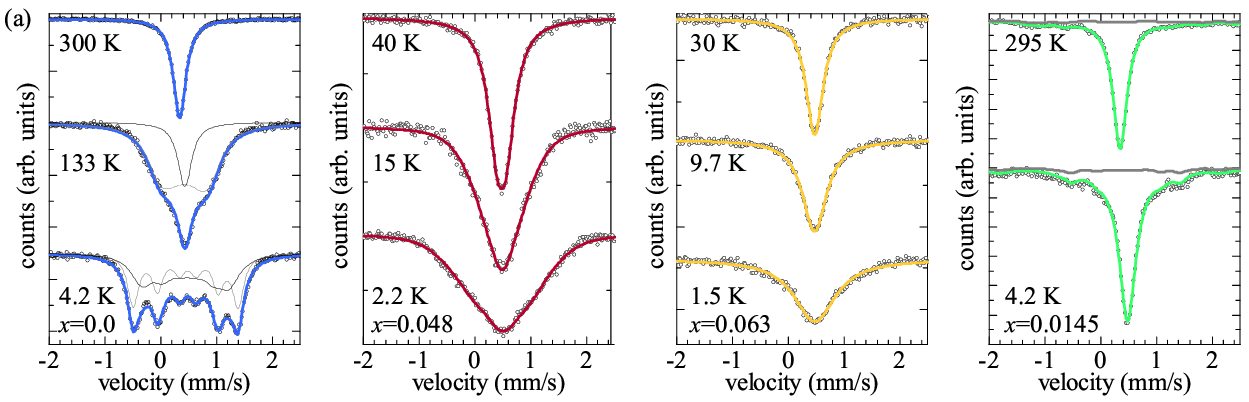}
\includegraphics[height=6cm]{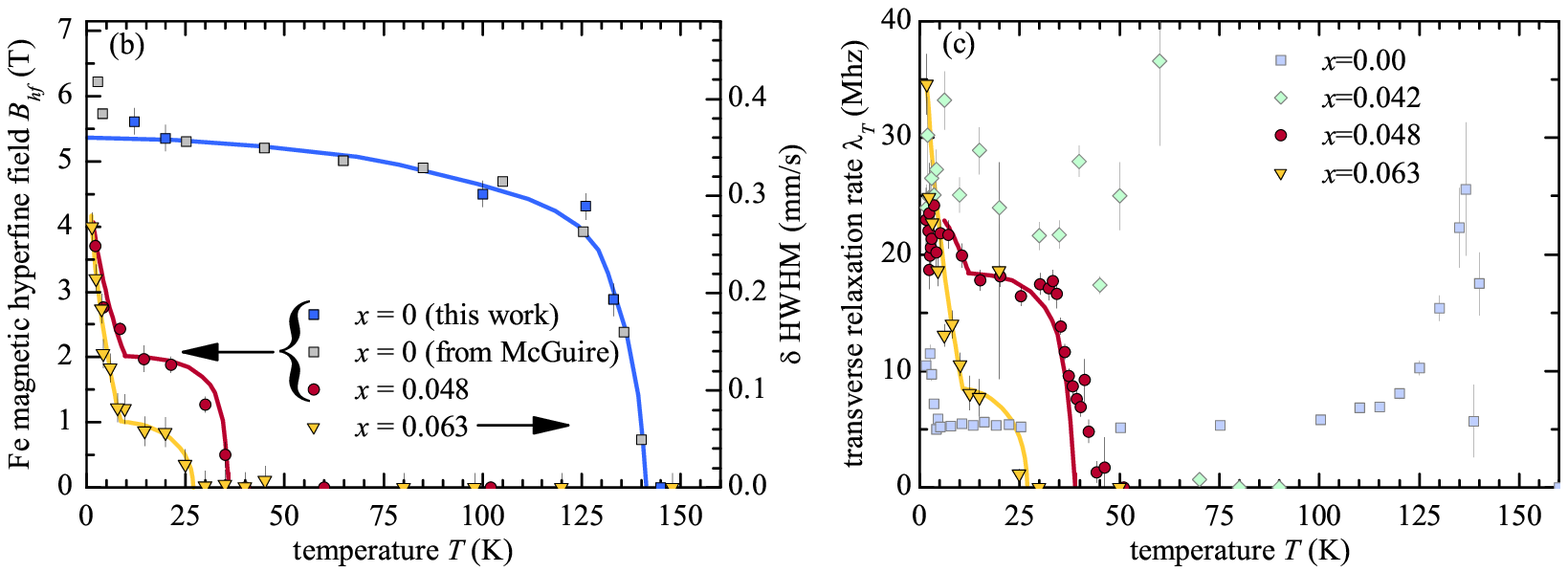}
\end{minipage}
\caption{(\textit{a}): Typical M\"ossbauer spectra.
The full lines are fits to the data and corresponding subspectra (see text).
(\textit{b}): M\"ossbauer Fe magnetic hyperfine field for $x$=0, and 0.048(3) and the magnetic contribution $\delta$HWHM to the absorption line width for $x$=0.063(2).
(\textit{c}): $\mu$SR transverse relaxation rate $\lambda_T$ ($x$=0.048(3), 0.063(2) are emphasized for clarity).
Lines are guides to the eye only.
Note the consistency of data produced with $\mu$SR and M\"ossbauer spectroscopy in (\textit{b}) and (\textit{c}).}
\label{img.ce-mos}
\end{figure*}

In the following, we present our $^{57}$Fe M\"ossbauer spectroscopy experiments.
The results corroborate the results of the $\mu$SR experiments presented above.
Typical M\"ossbauer absorption spectra are shown in Fig.~\ref{img.ce-mos}(a).
We recorded the spectra using a $^{57}$Co/Rh gamma radiation source (emission line half-width-at-half-maximum $\Gamma$=0.130(2)~mm/s) and a silicon drift diode detector.
We calibrated the spectrometer to $\alpha$-Fe at room temperature.

At room temperature, all spectra of CeFeAsO$_{1-x}$F$_x$ show a single sharp Lorentzian absorption line with a HWHM $\Gamma$=0.15(1)~mm/s and an isomer shift of $\delta^{\text{IS}}$=0.45(1)~mm/s with respect to $\alpha$-Fe.
This is consistent with the results of other M\"ossbauer studies of $R$FeAsO$_{1-x}$F$_x$\cite{klauss,mcguire09}.

Magnetically ordered Fe bearing impurity phases can be detected with M\"ossbauer spectroscopy if the concentration of the impurity is larger than $\approx$1~atom\%.
Only the spectra of superconducting material with $x$=0.145(20) show absorption lines in addition to the main line.
The additional absorption lines are consistent with the Fe hyperfine field\cite{ray76} of Fe$_2$As and their spectral weights are equivalent to $\approx$5~wt.\% of Fe$_2$As impurities.
The spectra of $x$=0, 0.048(3), and 0.063(2) show no additional absorption lines at room temperature, which indicates that these samples do not contain any iron-bearing impurities.

The spectra of $x$=0.00 show a distinct broadening of the singlet pattern below 135~K that develops into a resolved sextet at low temperatures due to the Fe magnetic order consistent with other studies of $R$FeAsO\cite{klauss,mcguire09}.
We analyzed the spectra of $x$=0.00 like in Ref.~\onlinecite{klauss} by diagonalizing the hyperfine Hamiltonian including electric quadrupole and static magnetic hyperfine interaction.
To account for the small broadening we used two sextets to describe the absorption spectra.
The temperature dependency of the magnetic hyperfine field $B_{\text{hf}}$ (see Fig.~\ref{img.ce-mos}(b)) and the low temperature values of the isomer shift $\delta^{\text{IS}}$=0.58(1)~mm/s, the quadrupole shift {$\epsilon$=-0.04(1)}~mm/s and the Fe hyperfine field $B_{\text{hf}}$=5.6(1)T at $T$=12~K are consistent with the results of \citet{mcguire09}.
In addition to the hyperfine field of the Fe order they found that the magnetic order of the Ce moments increases the measured Fe hyperfine field slightly below $\approx$5~K due to a small transferred hyperfine field.

At low temperatures, the absorption lines of the doped compounds with $x$=0.048(3), and 0.063(2) broaden significantly at low temperatures.
Down to $\approx$4~K we observe no resolved sextet as for CeFeAsO.
The broad absorption lines indicate Fe magnetic order with a small ordered moment and a distribution of ordered moments.

\begin{figure}[htbp]
\includegraphics[width=8.6cm]{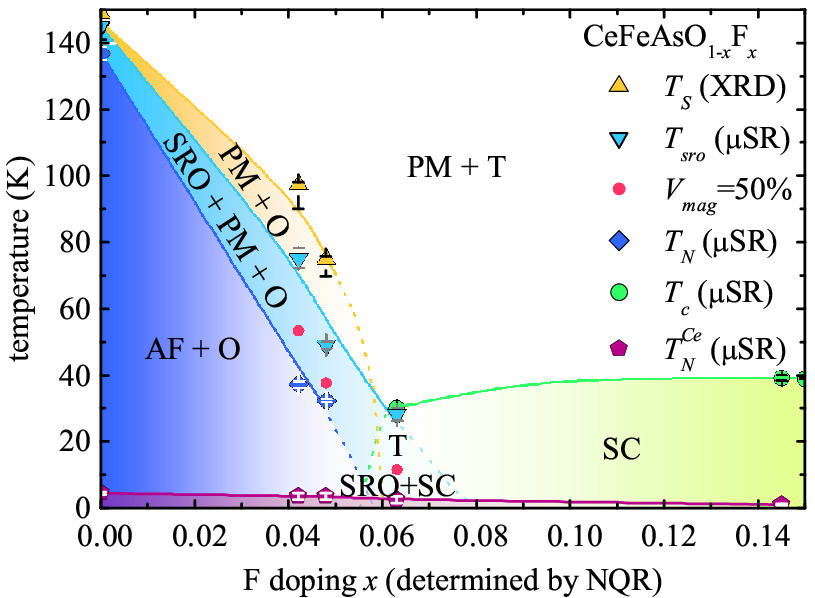}
\caption{The electronic phase diagram of CeFeAsO$_{1-x}$F$_{x}$ with a region where magnetic short range order ({\sc sro}) and superconductivity occur at the same time in the tetragonal structure, denoted as ``{\sc sro+sc+t}'' At present it is unclear whether this is phase separation or microscopic coexistence.
The close proximity of the transition temperatures $T_{sro}$ and $T_c$ for $x$=0.063(2) indicates that this composition is close to the tetra critical point of the phase diagram.
All lines are guides to the eye.
Dotted lines are extrapolations of the phase boundaries indicated by solid lines.
For $x$=0.150(20) we did not study the magnetic properties for $T$$<$1.6~K, therefore it remains unclear, if the Ce magnetic order survives for this composition.}
\label{fig.phase}
\end{figure}

We analyzed the spectra of $x$=0.048(3) with a Gaussian distribution of hyperfine fields $B_{\text{hf}}$ with width $\sigma$ and mean $\left\langle B_{\text{hf}}\right\rangle$, shown in Fig.~\ref{img.ce-mos}.
It turns out that the ratio $\left\langle B_{\text{hf}}\right\rangle$/$\sigma$=1.41 is temperature independent.

For $x$=0.063(2) this analysis is not possible because in this case the fit parameters $\sigma$, and $\left\langle B_{\text{hf}}\right\rangle$ are too strongly correlated.
Instead we fit a single Lorentzian line to the absorption spectra and extract the magnetic contribution $\delta$HWHM(T)=0.5($\Gamma(T)$-$\Gamma(300$~K$)$) in Fig.~\ref{img.ce-mos} to the total line width $\Gamma(T)$ by subtracting the line width at room temperature $\Gamma(300$~K$)$.
For a quantitative comparison of $\delta$HWHM with the hyperfine field of $x$=0, and 0.048(3) we additionally analyzed the spectra of $x$=0.048(3) with a single Lorentzian line.
It turns out, that for $x$=0.048(3) we recover the ratio $\left\langle B_{\text{hf}}\right\rangle$/$\delta$HWHM=14.7~T/(mm/s), which is specific to the $^{57}$Fe nucleus, from our fits for all temperature.
This proves the validity of our analysis and allows us to compare $\delta$HWHM of $x$=0.063(2) with the Fe hyperfine field of $x$=0, and 0.048(3)---the $y$-axes in Fig.~\ref{img.ce-mos}(b) are already scaled with this factor.

The spectra of $x$=0.145(20) show a single Lorentzian absorption line with a temperature independent line width in addition to the impurity lines (see above).
This confirms the $\mu$SR results that show the absence of magnetic order of $x$=0.145(20) for $T$$\geq$1.6~K.
Due to experimental limitations we did not study the Ce magnetic order with M\"ossbauer spectroscopy because it occurs at even lower temperatures (see Sec.~\ref{sec.ce}, p.~\pageref{sec.ce}).

The low temperature value $B_{hf}(12\rm K)$=$5.6(1)$~T in the undoped compound is reduced to 2.0(1)~T for $x$=0.042(2) and to 1.0(1)~T for $x$=0.063(2) (see Fig~\ref{img.ce-mos}).
Qualitatively, this is consistent with the reduction of the low temperature value of the muon spin precession frequency $f_\mu(0)$ shown in Fig.~\ref{fig.freq}.
However, $f_\mu$ is reduced by 75\% from 28~MHz ($x$=0) to 7~Mhz ($x$=0.048(3)) but the Fe hyperfine field only by 65\% from 5.6~T to 2~T if we neglect the low temperature enhancement.
This suggests either that the muon site is slightly different in the doped materials\cite{maeter09}, that the polarization of the Ce moments by the Fe sublattice is oriented differently\cite{maeter09}, or that the Fe magnetic structure is slightly different.

Below $T$$\approx$10~K, $B_{hf}$ of $x$=0.048(3), and 0.063(2) increases to $\approx$4~T.
This increase of the Fe hyperfine field \textit{does not} cause the muon spin precession frequency $f_\mu$ in Fig.~\ref{fig.freq} to increase.
Instead the $\mu$SR transverse relaxation rate $\lambda_T$, shown in Fig.~\ref{img.ce-mos} increases simultaneously with $B_{hf}$.
This indicates that the additional magnetic moment that enhances $B_{hf}$ is static but disordered.
It could be due to an increase of the Fe magnetic moment or a transferred hyperfine field due to short range Ce order.

\begin{figure*}[htbp]
\includegraphics[width=\textwidth]{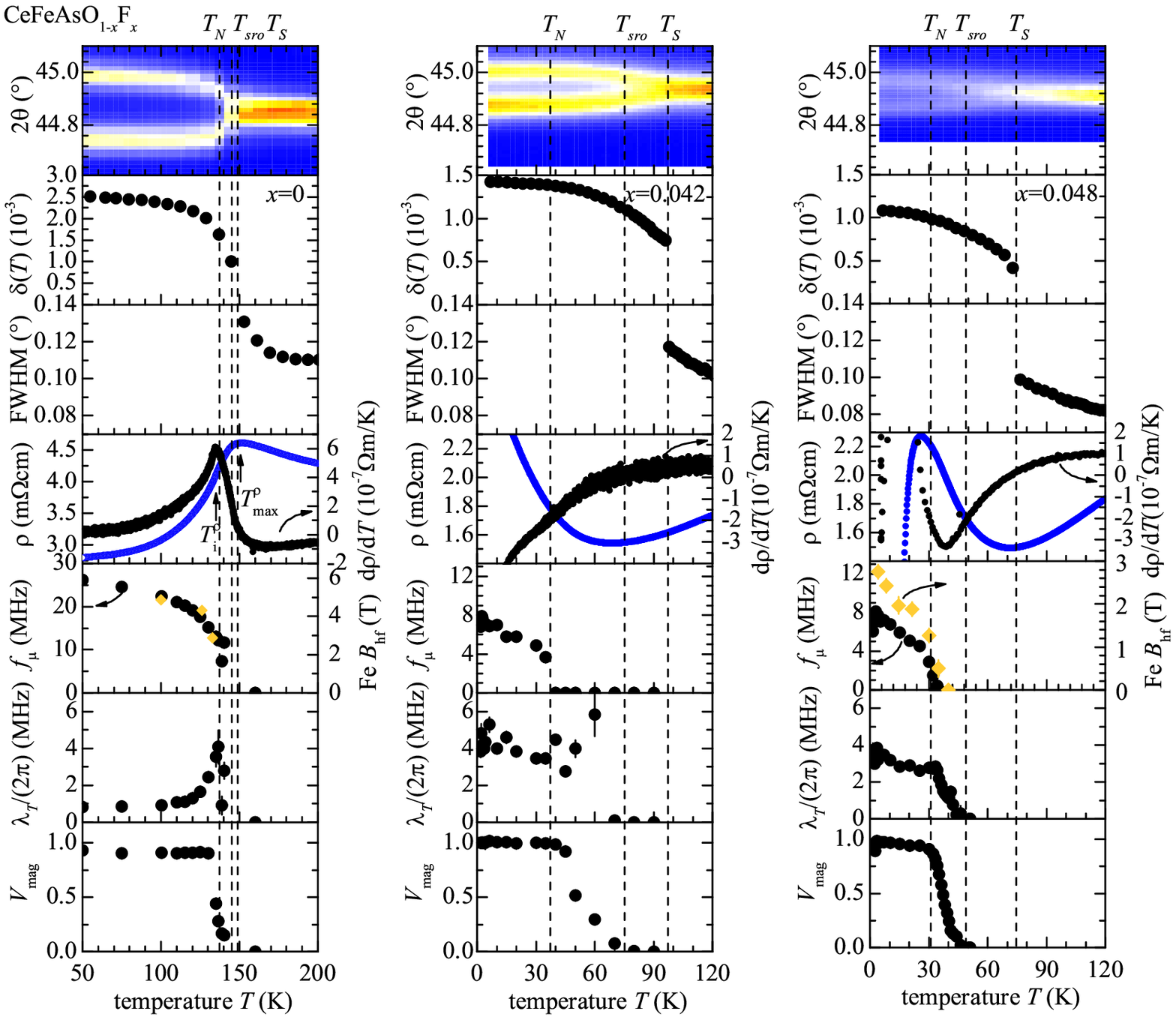}
\caption{Intensity maps of X-ray diffraction patterns that illustrate the tetragonal to orthorombic structural transition of CeFeAsO$_{1-x}$F$_x$ and its relationship to the magnetic short and long range order, and anomalies in the electrical resistivity.
In the doped materials the pronounced anomalies of the resistivity disappear and an estimation of $T_N$ and $T_S$ is not possible by the resistivity measurements.}
\label{fig:ce-all}
\end{figure*}

\subsection{Summary}
\label{sec.ce-sum}
We have studied the electronic phase diagram of CeFeAsO$_{1-x}$F$_{x}$ by detailed resistivity, synchrotron X-ray diffraction, M\"ossbauer spectroscopy, magnetic susceptibility and muon spin relaxation experiments.
All phase transition temperatures are summarized in Tab.~\ref{tab.ce}.

\paragraph*{General results} The phase diagram is shown in Fig.~\ref{fig.phase}.
We find that the structural phase transition always precedes ($T_S$$>$$T_{sro}$) short range magnetic order and is preempted by structural fluctuations.
The short range order develops over a large temperature interval and only close to the temperature where nearly 100\% of the sample volume exhibits magnetic order, we find long range magnetic order for $T$$\leq$$T_N$$<$$T_{sro}$.
F doping suppresses both phases, leading to lower transition temperatures and smaller order parameters.
Seemingly, the F doping suppresses the structural transition completely before it induces superconductivity.
The magnetic order, however, survives and we find a mixed phase of bulk superconductivity and bulk magnetic short range order for $x$=0.063(2).
For this composition the magnetic order never becomes long ranged---whether this is due to the competition with superconductivity or simply the same short range order that we find for $x$=0, and 0.042(2) remains unclear.
Higher doping levels quench the Fe magnetic order and give rise to pure superconductivity.
The magnetic penetration depth of both superconducting materials $x$=0.145(20), and 0.150(20) is consistent with a single s-wave gap without nodes.
The F doping also weakens the Ce magnetic order: As the Fe order becomes short ranged for $x$=0.063(2), also does the Ce order and for $x$=0.145(20) it is suppressed to $T_N^{Ce}$$\approx$1~K---a reduction by 75\% compared to the undoped material---and the ordered moment is drastically reduced.
This suggests that the Ce-Ce RKKY interaction is weakened either by changes of the Fermi surface, or by the smearing of the conduction bands by doping induced disorder.

\paragraph*{Mixed phase} That the magnetic order coexisting with superconductivity is rendered short ranged for $x$=0.063(2) is consistent with Ref.~\onlinecite{shiroka11}.
We find magnetic order occurs at temperatures slightly lower than the critical temperature $T_c$ of superconductivity.
This allows us to conclude bulk superconductivity close to $T_c$ from $\mu$SR.
At lower temperatures the $\mu$SR signal is dominated by magnetic order.
However, the temperature independent large diamagnetic (macroscopic) susceptibility at low temperatures indicates that the superconducting volume fraction remains unchanged, i.e.,
that it remains bulk superconductivity also at lower temperatures.
Additionally the changes of the lattice parameters below $T_c$, which are related to the pressure dependence of $T_c$, are consistent with those of the two pure superconductors studied by X-ray diffraction, which further substantiates the evidence for bulk superconductivity.

The evidence relevant for the question of (nanoscopic) phase separation vs. microscopic coexistence is therefore similar to the Sm system:
\begin{itemize}
\item bulk superconductivity (from $\mu$SR, $\chi(T)$, and XRD),
\item bulk short range magnetic order (from $\mu$SR, and M\"ossbauer spectroscopy), 
\end{itemize}

\begin{table*}[htbp]
\begin{tabular}{c|ccccccc}
$x$&$T_S$&$T_c$&$T_{sro}$&$T_{100}$&$T_N$&$T_N^{Ce}$\\
\hline
\hline
0.00&148.9+2-8&--&145(5)&133(2)&137(2)&4.4(3)\\
0.042(2)&97.0+1-7&--&75(3)&40(2)&37(1)&3.5(10)\\
0.048(3)&74.7+1-5&--&49(1)&30.0(5)&32.0(5)&3.5(10)\\
0.063(2)&--&29.5 ($\rho(T)$), 29.8(2) ($\mu$SR)&28(2)&3(1)&--&2.5(10) ($\mu$SR), 2.7(1) ($\chi(T)$)\\
0.145(20)&--&43.8 ($\rho(T)$), 39.0(9) ($\mu$SR)&--&--&--&1.0(2)\\
0.150(20)&--&43.4 ($\rho(T)$), 38.7(1) ($\mu$SR)&--&--&--&--
\end{tabular}\caption{Phase transition temperatures for the structural ($T_S$ obtained by XRD), superconducting ($T_c$ obtained by $\mu$SR), Fe magnetic long range order ($T_N$ obtained by $\mu$SR), magnetic short range order ($T_{sro}$, $T_{100}$, and $\Delta T$ obtained by $\mu$SR---see text).
$T_N^{Ce}$ is the magnetic ordering transition temperature of the Ce sublattice obtained by $\mu$SR from the fit to $\sigma_{sc}(T)$, and susceptibility measurements ($\chi(T)$).
All temperatures are given in K.}
\label{tab.ce}
\end{table*}
In sum, above evidence are as ambiguous as for the Sm system.
We cannot argue with certainty for, or against nanoscopic phase separation, or microscopic coexistence---the experimental evidence are equally consistent with both situations:
For nanoscopic phase separation\cite{sanna09,sanna10,shiroka11}, at $T$$\approx$2~K $\mu$SR would show bulk magnetic order, while probes, such as susceptibility measurements, that are insensitive to the magnetic order, would show bulk superconductivity---consistent with our data.
However, the same evidence would be expected for microscopic coexistence.
We note that our experimental data confirm and extend previously published $\mu$SR and magnetic susceptibility measurements\cite{sanna10,shiroka11}.
We merely argue, that the available evidence are not sufficient to decide this matter.

\paragraph*{Relevance for a possible QCP} The possible existence of a magnetic quantum critical point (QCP) is connected to the above discussion.
The first phase diagram of this system suggested the existence of such a QCP.
However, \citet{zhao08} inferred their data from neutron diffraction.
This technique is, in principle, not sensitive to short range order.
Taking this into consideration, Zhao's phase diagram is corrected by the $\mu$SR work of \citet{sanna10}, \citet{shiroka11}, and the present work.
These findings, as discussed above, do not clarify the existence of a magnetic QCP because it remains unclear whether the phase transition as a function of doping is a first or second order transition.

\paragraph*{Electrical resistivity \& phase transitions} In Fig.~\ref{fig:ce-all} we compare X-ray diffraction, $\mu$SR, and M\"ossbauer with resistivity data.
Fig.~\ref{fig:ce-all} shows that $T_{S}$ and $T_{N}$ of CeFeAsO can only be \textit{estimated} from the maximum and the inflection point.
However, if only the anomalies of $\rho(T)$ were used to determine the critical temperatures, $T_N$ would be underestimated and $T_S$ overestimated by $\approx$2~K.
This is different from SmFeAsO for which the resistivity maximum is 10~K above the structural transition and inflection coincides with $T_N$.
Although the cusp of $\rho(T)$ is suppressed in doped samples, we can try to correlate the resistivity with the structural, and magnetic properties.
For example, for $x$=0.042(2) the resistivity minimum is close to the onset of short range magnetic order at $T_{sro}$.
Similarly, the inflection point of $\rho(T)$ is close to $T_S$.
However, these general correlations do not seem to be generic features of $\rho(T)$ as the structural transition is close to the minimum and short range order at even lower temperatures for $x$=0.048(3).

\paragraph*{Electrical resistivity \& fluctuations} Because the maximum of $\rho(T)$ of $x$=0.00 is so close to the structural transition, we could speculate about the maximum to be caused by the structural fluctuations marked by the increase of the FWHM of the (2,2,0)$\rm_T$ Bragg peak.
However, in Sec.~\ref{sec.sm-sum}, p.~\pageref{sec.sm-sum} we already learned that for the Sm system the increase of the FWHM is not correlated with the maximum of $\rho(T)$.
Therefore it seems that $T_S$$\approx$$T^\rho_{max}$ for the Ce system with $x$=0.00 is coincidental.
The seeming insensitivity of $\rho(T)$ to structural fluctuations and the phase transition is even more apparent for $x$=0.048(3).
Here, the structure stabilizes below $T_S$, yet $\rho(T)$ increases.
Due to the lack of any systematic correlation between $\rho(T)$ and the FWHM, and the transition temperatures we have to conclude, as for the Sm system,
that the structural and magnetic fluctuations studied in this work have no simple connection to the temperature dependence of the resistivity\cite{lv09,fernandes11b}.

\section{Conclusion}
\label{sec.dis}

\begin{figure}[htb]
\includegraphics[width=8.6cm]{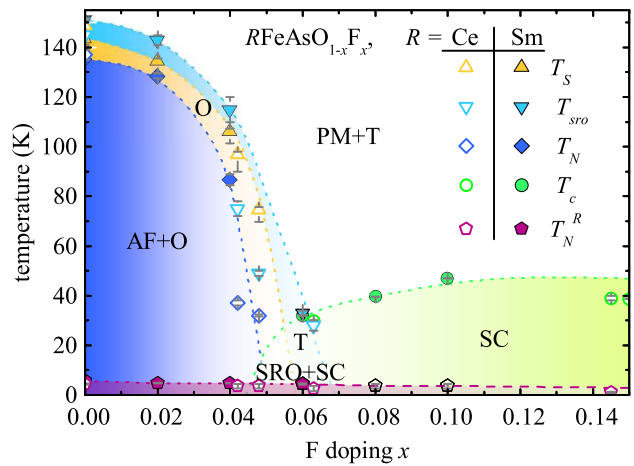}
\caption{Superposition of the phase diagrams of $R$FeAsO$_{1-x}$F$_x$ with $R$=Sm, and Ce.
$T_N$, and $T_S$ seem to follow the same trend in both systems under doping, whereas $T_{sro}$ is smaller than $T_S$ in the Ce case and larger than $T_S$ in the Sm case.
This suggests that $T_{sro}$ does \textit{not} reflect an intrinsic energy scale but rather the strength/abundance of disorder or (micro)strain in the samples.
However, the fact that short range order occurs for the two samples that show a mixed phase of superconductivity and short range magnetic order ($x$=0.06, and 0.063(2) for $R$=Sm, and Ce) suggests that the \textit{clean} systems might exhibit a low temperature \textit{long} range magnetic order in the superconducting phase.
The phase transition temperatures are listed in Tab.~\ref{tab.sm}, p.~\pageref{tab.sm}, and Tab.~\ref{tab.ce}, p.~\pageref{tab.ce}.}
\label{img.both}
\end{figure}

In Fig.~\ref{img.both} we show a superposition of the two phase diagrams.
The detailed discussions of the phase diagrams can be found on p.~\pageref{sec.sm-sum}, and p.~\pageref{sec.ce-sum}.

\paragraph*{General results} 
Both parent materials (CeFeAsO and SmFeAsO) show the well known structural, Fe magnetic, and rare-earth magnetic phase transitions.
The F doped materials also undergo these phase transitions, but the ordering temperatures and order parameters are reduced by doping.
The reduction of the phase transition temperatures by doping is accompanied by the disappearance of the resistivity anomalies.
For the studied compounds, the disappearance of the resistivity anomalies by doping is comparable in both systems.

\paragraph*{Structural transition} The structural transition temperature and order parameter are reduced by doping.
It is absent once superconductivity is induced by doping in both systems ($x$=0.06 for the Sm, and $x$=0.063(2) for the Ce system).
For the Sm system, this is supported by the absence of any additional anomaly of the temperature dependence of the specific heat.
In agreement with the X-ray diffraction study of \citet{margadonna}, our experimental evidence disprove the claims of a structural transition at $T$$\sim$170~K in superconducting Sm materials by \citet{martinelli11} We suggest that the FWHM of a structural Bragg peak should only be considered as evidence for a structural phase transition if supported by additional experiments.
Although the broadness of the structural transition (see below) hinders the determination of the critical exponents we were able to extract the following trend: Doping causes the universality class of the structural transition to change from 2D-Ising (undoped) to 3D-Ising (doped close to superconductivity).
This is similar to the reports of a universality class change of the magnetic transition\cite{wilson}.

\paragraph*{Superconducting gap symmetry} We analyzed the temperature dependence of the in-plane magnetic penetration depth of SmFeAsO$_{0.90}$F$_{0.10}$ and CeFeAsO$_{1-x}$F$_{x}$ with $x$=0.063(2), 0.145(20), and 0.150(20) and found that a single s-wave gap without nodes agrees best with the data.
However, within the experimental uncertainty a second gap cannot be excluded.
In Fig.~\ref{fig.uemura} we plot $T_c$ vs.
$1/\lambda_{ab}^2$.
The here measured superconductors roughly follow the linear relationship\cite{uemura88,*uemura91} of other doped $R$FeAsO superconductors.

\paragraph*{Mixed phase} 
Once doping induces superconductivity we do not find any evidence for a structural transition ($x$=0.06 for the Sm system, and $x$=0.063(2) for the Ce system).
But we cannot exclude that at slightly lower doping levels superconductivity can be found in the orthorhombic phase\cite{margadonna}.
Instead we find a mixed phase of superconductivity and short range magnetic order.
Both materials have in common that both phases have nearly identical transitions temperatures and that superconductivity is bulk directly at the transition but the magnetic order only becomes bulk at $\approx$5~K.
We analyzed the structural data, electrical resistivity, micro- and macroscopic magnetic susceptibility, and the magnetic volume fraction of this mixed phase.
Based on our results and the results of Refs.~\onlinecite{sanna10,sanna11,shiroka11} we conclude that, it remains unclear whether this mixed phase is nanoscopic phase separation (two thermodynamic phases) or microscopic coexistence (one thermodynamic phase).
This is mainly caused by the difficulty to detect a possible coupling of the order parameters if both ordering temperatures are very close to each other.
It is much clearer in, e.g., K doped BaFe$_2$As$_2$ for which phase separation\cite{julien09} and microscopic coexistence\cite{wiesenmayer11} are unambiguously distinguishable.


\paragraph*{Relevance for a possible QCP}
The phase diagram of the Ce system by \citet{zhao08} in principle allows a magnetic quantum critical point where $T_N$=$T_c$=0.
This feature was already clarified by the $\mu$SR works in Refs.~\onlinecite{sanna09,shiroka11} and our work confirms these results: both the Sm and the Ce system show a region in the phase diagram in which magnetic order and superconductivity occur in a mixed phase.

Our study cannot give any direct insight into the possible existence of a nematic QCP\cite{fang08}.
The nematic phase is mainly marked by electronic anisotropies (see e.g.
Ref.~\onlinecite{fisher11} and references therein) in the a-b plane and its study should require large enough single crystals.
However, the nematic order parameter and the orthorhombic distortion are directly connected\cite{fernandes11b}.
We showed that the low temperature limit of the orthorhombic distortion $\delta_0$ depends linearly on the transition temperature $T_S$ (see Fig.~\ref{img.deltats}).
This proportionality is the same for the Ce and Sm system and is also consistent with published data \cite{zhao08,margadonna}.
\citet{margadonna} reported two superconducting materials that undergo the structural phase.
$\delta_0$ of these materials is seemingly doping independent and much higher than expected from the linear extrapolation of $\delta_0(T_S)$ (while $T_S$ continues to decrease).
This suggests that the structural phase transition temperature and order parameter \textit{do not} simultaneously vanish; in contrast to the simultaneous vanishing of $T_S$ and $\delta_0$ expected at a structural QCP (or the related nematic QCP\cite{fang08}).

\begin{figure}[htb]
\includegraphics[width=8.6cm]{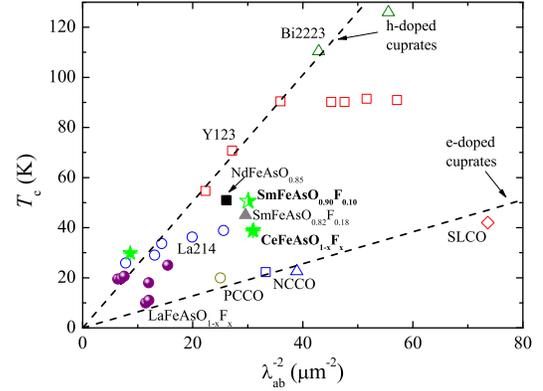}
\caption{The scaling of $T_c$ with the superfluid density $\lambda^{-2}\propto n_s$.
CeFeAsO$_{1-x}$F$_x$ and SmFeAsO$_{1-x}$F$_x$ superconductors follow a linear dependence similar to other reports.
This is an empirical relationship\cite{uemura88,*uemura91} also found for the hole and electron doped cuprate superconductors and other exotic materials.
Data for other ferropnictides and cuprates have been taken from references \cite{luetkens08,luetkens09,khasanov08,drew09,uemuraCu1,*uemuraCu2,*uemuraCu3,*uemuraCu4,*uemuraCu5}.}
\label{fig.uemura}
\end{figure}

\paragraph*{Broadness of phase transitions} Both, the magnetic and structural transitions of both systems are significantly broadened.
Evidence for the broadening of the magnetic transition is the slow increase the magnetic volume fraction.
The significant broadening of the (2,2,0)$\rm_T$ tetragonal Bragg peak and subsequent convex temperature dependence of the orthorhombic distortion indicate that also the structural transition is broadened.
The essential difference between the two systems is that in the Sm system the magnetic \emph{long range order} develops in clusters and \emph{precedes} the structural transition ($T_{sro}$$>$$T_S$), whereas in the Ce system clusters of magnetic \emph{short range order} occur after the structural transition ($T_{sro}$$<$$T_S$).
Compared to the Ce system, the Sm materials have a up to 70\% larger FWHM of the (2,2,0)$\rm_T$ tetragonal Bragg peak.
This could mask the onset of the orthorhombic order in the small magnetically ordered regions.

Reports of broad transitions exist, but sometimes are not explicitly mentioned in other studies.
Sanna, Shiroka, and co-workers\cite{sanna09,sanna10,shiroka11} reported $\mu$SR data and used a Gaussian distribution of magnetic transition temperatures with a standard deviation $\sigma$$\approx$5--10~K to quantify the transition width.
This indicates a total width of the magnetic transition from onset to saturation of $T_{100}-T_{sro}$$\approx$$4\sigma$$\approx$20--40~K (95\% of a Gaussian distribution lie within $\pm 2\sigma$ of the average).
Consistent with a broadened structural phase transition, \citet{qureshi} reported a ``precursor'' of the orthorhombic structure already above the structural phase transition of LaFeAsO$_{1-x}$F$_x$ that manifested in an increase of the FWHM of the studied Bragg peaks.
Considering the existing reports of broad structural and magnetic transitions, the broadness should be considered as a manifestation of an intrinsic property.
The spin fluctuations preceding the magnetic phase transition should, via the magneto-elastic coupling, play an important role for the broadening of the FWHM of the Bragg peak.
In addition, (random) stress, and disorder may also contribute to the broadening of both phase transitions\cite{cano12,fernandes11b,hu12,vavilov11}.


\paragraph*{Phenomenological model for the broadened magnetic transition} The broadening of the Fe magnetic transition is characterized by a power law increase of the magnetic volume fraction, a peak of the transverse relaxation rate (width of the distribution of ordered moments/local magnetic fields) and an unusual temperature dependence of the measured average ordered magnetic moment, the muon spin precession frequency.
We reproduce these features in a simple model for the Fe magnetic transition.

The model assumes that for $T$$\approx$$T_{sro}$ (the onset of the magnetic order) small, magnetically ordered regions form, that have an independent temperature dependence of the order parameter.
All regions have the same critical exponents, but a full ordered moment (at $T$=0) that depends on the local ordering temperature.
The only adjustable parameters in this model are the critical exponents.

This model fully describes the magnetic ordering of SmFeAsO as seen by $\mu$SR.
The doped Sm materials show an additional contribution to the transverse relaxation rate that is most likely caused by enhanced disorder.
The Ce system does not show muon spin precession for $T_N$$<$$T$$<$$T_{sro}$ but all other features are also present at the magnetic transition.
This suggests, that the same mechanism is present in the Ce system, albeit with a shorter magnetic correlation length.

\paragraph*{Magnetic order of the rare earth} Both systems show rare earth magnetic order.
The most accurate determination of the ordering temperature is possible with specific heat measurements.
The ordering temperature and magnetic correlation length is reduced by F doping.
$\mu$SR indicates that Sm order is present also close to optimal doping with a transition temperature that is only reduced by 1--2~K compared to the undoped material.
For the superconducting $x$=0.06 specimen, the Sm magnetic order is short ranged, just as the Fe magnetic order at this doping level.
The Ce order is affected by doping much like Sm but the transition temperature is reduced to $\approx$1~K for $x$=0.145(20).

This indicates, that the interaction between the 4$f$ moments is weakened by doping.
Considering the decrease of the lattice constants under doping (see Figs.~\ref{img.sm-lattice}, \ref{img.ce-lattice}) this excludes super exchange interaction as the main interaction because it sensitively depends on the distance between the interacting moments.
RKKY interaction between the 4$f$ moments, on the other hand, may be weakened by doping, which could be attributed to the disappearance of the Fe magnetic order, and changes of the Fermi surface topology\cite{akbari11}.
Additionally, the disorder induced by doping may smear out the conduction bands which effectively leads to a weakening of the RKKY interaction.
The smearing of the conduction bands was studied for transition metal doped BaFe$_2$As$_2$ and CaFe$_2$As$_2$\cite{wadati10,*haverkort11,*berlijn11,*levy12}.
The effect for F doping of $R$FeAsO should be weaker, but the additional electronic inhomogeneities\cite{lang10} that develop in these systems may play a similar role.

\paragraph*{Rare earth spin fluctuations} Earlier reports associated the magnetic fluctuations in the Sm system with Fe spin fluctuations preceding superconductivity\cite{drew}.
Our detailed analysis of the magnetic fluctuations indicates that this is completely coincidental and that the fluctuations seen by $\mu$SR should be associated with fluctuations of the Sm moments due to crystal electric field effects.

\begin{acknowledgments}
HM thanks I. Eremin, and R. Fernandes for helpful discussions.
We thank M. Deutschmann, S. M\"uller-Litvanyi, R. M\"uller and A. K\"ohler for experimental support in preparation and characterization of the samples.
This work was financially supported by the German Research Foundation (DFG) within the priority program SPP1458 (projects KL-1086/10-1 and GR3330/2), and the
graduate school GRK1621.
The work at the IFW Dresden has been supported by the DFG through FOR 538.
Part of this work was performed at the Swiss Muon Source (Villigen, Switzerland).
\end{acknowledgments}

\end{document}